\documentclass[preprint2]{aastex}
\usepackage{cite}
\usepackage{mathrsfs}
\usepackage{natbib}
\usepackage{amsmath}
\usepackage[normalem]{ulem}
\usepackage{morefloats}
\usepackage{bigfoot} 
\usepackage{soul}

\begin{document}

\onecolumn

\title{\emph{HERSCHEL} EVIDENCE FOR DISK FLATTENING \\
OR GAS DEPLETION IN TRANSITIONAL DISKS\altaffilmark{\dagger}}

\author{J. T. Keane\altaffilmark{1}, 
I. Pascucci\altaffilmark{1},
C. Espaillat\altaffilmark{2},
P. Woitke\altaffilmark{3},
S. Andrews\altaffilmark{4},
I. Kamp\altaffilmark{5},
W.-F. Thi\altaffilmark{6},
G. Meeus\altaffilmark{7},
W. R. F. Dent\altaffilmark{8}}

\altaffiltext{1}{Lunar and Planetary Laboratory, University of Arizona, Tucson, AZ 85721, USA}
\altaffiltext{2}{Department of Astronomy, Boston University, Boston, MA 02215, USA}
\altaffiltext{3}{SUPA, School of Physics \& Astronomy, University of St. Andrews, North Haugh, St. Andrews, KY16 9SS, UK}
\altaffiltext{4}{Harvard-Smithsonian Center for Astrophysics, Cambridge, MA 02138, USA}
\altaffiltext{5}{Kapteyn Astronomical Institute, Postbus 800, 9700 AV Groningen, The Netherlands}
\altaffiltext{6}{Universit\'{e} Joseph Fourier Grenoble-1, CNRS-INSU, Institut de Plan\'{e}tologie et dÕAstrophysique (IPAG) UMR 5274,
38041 Grenoble, France}
\altaffiltext{7}{Universidad Autonoma de Madrid, Dpt. Fisica Teorica, Campus Cantoblanco, 28049 Madrid, Spain}
\altaffiltext{8}{ALMA SCO, Alonso de Cordova 3107, Vitacura, Santiago, Chile}

\altaffiltext{$\dagger$}{{\it Herschel} is an ESA space observatory with science instruments provided by European-led Principal Investigator consortia and with important participation from NASA.}

\begin{abstract}

Transitional disks are protoplanetary disks characterized by reduced near- and mid-infrared emission, with respect to full disks.  This characteristic spectral energy distribution indicates the presence of an optically thin inner cavity within the dust disk believed to mark the disappearance of the primordial massive disk.  We present new \emph{Herschel Space Observatory} PACS spectra of [O{\sc i}] 63.18 $\mu$m for 21 transitional disks.  Our survey complements the larger \emph{Herschel} GASPS program \citep[``Gas in Protoplanetary Systems,''][]{dent13} by quadrupling the number of transitional disks observed with PACS in this wavelength.  [O{\sc i}] 63.18 $\mu$m traces material in the outer regions of the disk, beyond the inner cavity of most transitional disks.  We find that transitional disks have [O{\sc i}] 63.18 $\mu$m line luminosities $\sim2$ times fainter than their full disk counterparts.  We self consistently determine various stellar properties (e.g. bolometric luminosity, FUV excess, etc.) and disk properties (e.g. disk dust mass, etc.) that could influence the [O{\sc i}] 63.18 $\mu$m line luminosity, and we find no correlations that can explain the lower [O{\sc i}] 63.18 $\mu$m line luminosities in transitional disks.  Using a grid of thermo-chemical protoplanetary disk models, we conclude that either transitional disks are less flared than full disks, or they possess lower gas-to-dust ratios due to a depletion of gas mass.  This result suggests that transitional disks are more evolved than their full disk counterparts, possibly even at large radii.

\end{abstract}

\section{Introduction}

Protoplanetary disks (gas-rich dust disks around young stars) provide the raw building-blocks for solar systems.  While significant progress has been made in understanding the relevant evolutionary timescales of protoplanetary disks, little is known about the physical mechanisms driving the eventual dispersal of dust and gas about these young systems \citep[for review, see:][]{pascuccitachibana10}.  The goal of this paper is to gain insight into these dispersal processes by investigating a special type of protoplanetary disk that is thought to be in the process of losing its primordial dust disk: the transitional disks.

Transitional disks, like their full protoplanetary disk cousins, are often identified by their spectral energy distributions (SEDs).  While there is significant variation in the SEDs of young star systems, transitional disks appear as a distinct subgroup of protoplanetary disks: their SEDs show reduced near- and mid-infrared emission, with respect to full disks \citep{strom89}.  This characteristic SED points to the presence of an optically thin inner cavity, extending from the star out to 1 $\sim$ 20 AU.  The excavation of this cavity is believed to mark the early stages of the dispersal of the primordial, massive dust disk -- whose continuous dust disk extended as close as a few stellar radii to the central star \citep[e.g.,][]{calvet02, espaillat07}.  The existence of inner cavities has been directly confirmed for a few transitional disks via sensitive, high-resolution millimeter observations which detect reduced (or absent) dust emission from the inner disk, as a result of a deficit of millimeter size grains \citep[e.g.,][]{andrews09, brown09}.  While transitional disks may posess dust cavities, it is known that, in most cases, these dust cavities are not devoid of gas.  Transitional disks are still actively accreting \citep[e.g.,][]{n07}, and various optical emission lines (e.g. CO lines, [O{\sc i}] 6300 \AA $\,$ and 5577 \AA, etc.) indicate the presence of gas within the dust cavity region - though it may be depleted \citep[e.g. TW Hya,][]{gorti11}.

There are three leading hypotheses for the driving mechanism behind the formation of cavities in transitional disks:  \citep[for review, see:][]{espaillatPPIV}:
\begin{itemize}
\item \emph{Dust coagulation.}  As disks evolve submicron-sized dust grains coagulate into larger aggregates which have little emission at infrared wavelengths and thus reduce the disk opacity.  These larger aggregates would eventually coalesce into planetesimals and planetary embryos.  Since dynamical timescales increase with increasing radial distance from the central star, grain growth occurs inside-out and leads to the development of an expanding optically thin inner cavity, although the total mass of this inner disk region is not necessarily lower \citep[e.g.,][]{dullemonddominik2005}.
\item \emph{Photoevaporation.}  High-energy photons from the central star can drive photoevaporative winds, particularly from the outer regions of the protoplanetary disk (beyond $\sim$few AU).  
As the viscous accretion rate drops below the photoevaporation mass loss rate, a gap opens in the disk and the inner disk viscously accretes onto the star -- resulting in an inner cavity \citep[e.g.][]{alexander14}.  Direct irradiation of the cavity wall is expected to rapidly disperse the outer disk \citep{alexander06}.  Photoevaporative winds have been detected for select protoplanetary disks via blueshifted ($\sim$few km/s) [Ne {\sc ii}] 12.81 $\mu$m lines, which traces unbound winds within the inner $\lesssim$ 10's of AU \citep{pascucci09}.
\item \emph{Dynamical clearing by giant planets.} Dynamical interactions between the disk and an embedded giant planet (with masses roughly equal to that of Jupiter) can open gaps within the disk \citep[e.g.,][]{lubow99}.  Gas from the inner disk (within the planet's orbit) can continue to accrete onto the central star, while most of the gas from the outer disk (beyond the planet's orbit) accretes onto the planet, and only a small amount of gas flows past the planet into the inner disk.  In addition to the physical gap created by the planet, pressure gradients setup at the outer edge of the gap can act as a dust filter - allowing only grains below a critical size to reach the inner disk, and perhaps forming an optically thin inner cavity \citep{rice06}.
\end{itemize}
While these different mechanisms can produce qualitatively similar SEDs, they predict distinctive differences in the distribution of disk gas.  Furthermore, these different processes can, and probably do, operate simultaneously. 

In this paper, we use \emph{Herschel Space Observatory} far-infrared data to examine whether full disks and transitional disks are different in their outer disk regions, beyond 10's of AU.  We use the [O{\sc i}] 63.18 $\mu$m emission line and the nearby 63 $\mu$m continuum emission to trace the gas and dust components respectively, beyond 10$\sim$100 AU \citep[e.g.][]{aresu12}.  In addition, we use ancillary data to characterize our sample at different wavelengths.  In \emph{Section 2}, we provide a short description of our sample, the \emph{Herschel}/PACS observations and data reduction, and the ancillary stellar and disk properties used to characterize our sample.  In \emph{Section 3}, we summarize our  [O{\sc i}] 63.18 $\mu$m line 63 $\mu$m continuum results.  Most notably, we find that transitional disks possess [O{\sc i}] 63.18 $\mu$m line luminosities a factor of 2$\sim$3 lower than full disks, despite having similar 63 $\mu$m continuum luminosities { -- a trend previously identified by \citet{howard}, though expanded in this work with quadruple the number of transitional disks.}  In \emph{Section 4}, we rule out various observable stellar and disk properties (e.g. FUV and X-ray luminosity) as the potential cause for this  [O{\sc i}] 63.18 $\mu$m line luminosity difference between full disks and transitional disks.  In \emph{Section 5} we use the results of the DENT grid \citep[a grid of 300,000 thermo-chemical protoplanetary disk models, by][]{woitke10}, to examine other possible causes for the [O{\sc i}] 63.18 $\mu$m line luminosity difference.  We conclude that the lower  [O{\sc i}] 63.18 $\mu$m line luminosity of transitional disks could be due to transitional disks either being less flared, or by having lower gas-to-dust ratios.  In \emph{Section 6}, we discuss the implications of this result for disk evolution models, and potential followup observations.

\section{Observations and Data Reduction}

\subsection{Sample Description}

We selected 21 transitional disks from predominantly young (a few Myr old) and nearby ($\le$ 200 pc) star-forming regions.  Five additional transitional disks were selected from the GASPS sample \citep[``Gas in Protoplanetary Systems,''][]{dent13}, resulting in a total of 26 transitional disks.  Our sample is listed in \emph{Table 1}.  The transitional disks were identified by significant dips in their \emph{Spitzer}/IRS spectra. (For the relevant spectra used to identify each transitional disk as transitional, see: \citealt{B07, calvet02, espaillat10, furlan09, KM09, M08, M10}.)  Because only ~10\% of protoplanetary disks are transitional \citep[e.g.][]{williams11, muzerolle10} we selected targets from a number of star-forming regions, including Taurus-Auriga, Ophiuchus, Chameleon and Lupus.  Targets that have been previously modeled either with continuum radiative transfer codes, or simple prescriptions for the disk inner cavity were given preference, as were objects with archival measurements of accretion rates and infrared and millimeter observations.

For comparison with our sample of transitional disks, we selected an additional 33 protoplanetary disks from the GASPS \citep[``Gas Survey of Protoplanetary Systems,''][]{dent13} survey of the Taurus-Auriga star-forming region\footnotemark. \footnotetext{An additional full disk, SZ 50, from the Cha I star forming region, was included in order to more properly match the spectral type distribution between full disks and transitional disks.} These disks were selected as being typical protoplanetary disks, with IRS spectra close to the Taurus-Auriga mean \citep{dalessio}, and were also selected to sample similar spectral types to those of the transitional disks.  Like the transitional disk subsample, we took preference for objects with known accretion rates and millimeter observations.  Of these 33 protoplanetary disks, 15 have jets/outflows as identified by a combination of optical and near-IR spectroscopy and imaging \citep[see][and references therein]{kenyongomez08}.  We will refer to this subsample as ``outflow'' sources.  The remaining 18 disks without outflows will be referred to as ``full'' disks.  This distinction between outflow disks and full disks varies slightly from \citet{howard}, who identified outflow disks { as objects with either directly imaged jets in H$\alpha$, [O{\sc i}] $\lambda6300$, [S{\sc ii}] $\lambda6371$, being associated with Herbig-Haro objects, or having very broad [O{\sc i}] $\lambda6300$ emission line profiles.  This slight difference in definition only changes the classification of two disks (AA Tau and DL Tau).  None of our full disks were noted to have broadened or spatially-extended [O{\sc i}] 63.18 $\mu$m emission in \citet{howard}.}

Targets with binary companions represent a possible source of contamination within our samples given the large spaxel size of PACS (9.4" on a side, which corresponds to a projected separation of $\sim 1300$ AU at the distance of Taurus).  Close binary companions can interact with the primary star's disk, and produce transitional SEDs \citep[e.g. CoKu Tau/4,][]{irelandkrauss08}.  Medium and large separation binaries (projected separations $> 10$ AU) do not seem to strongly affect the first steps of planet formation (grain growth and dust settling) \citep{pascucci08}. However, binaries with separations $\le 40$ AU significantly hasten the process of disk dispersal \citep{kraus12}.  While it might be best to remove all binaries from our sample and just focus on single  stars, this approach could bias our samples and as such our results. The Taurus-Auriga star-forming region, from which we draw most of the full disks, has been well surveyed for multiplicity. However, Ophiuchus, Chameleon and the Lupus star-forming regions, from which we draw our sample of transitional disks, have not been as well studied. Thus, there are very likely undetected multiple systems within our transitional disk sample. We opted to retain full disks within multiple systems, as long as the mid-infrared flux ($\sim 10$ $\mu$m) ratio between members is large ($L_{IR, primary} / L_{IR, secondary} \gtrsim 3$), and the protoplanetary disks are not circumbinary.  Targets that do not meet this criterion are excluded from all analysis (though they are included in tables and figures, for reference).  \emph{Table 1} lists the multiplicity status for all targets, as well as the relevant references for the projected separations and mid-infrared flux ratio for binaries.

\subsection{\emph{Herschel} PACS Spectroscopy}

We obtained \emph{Herschel Space Observatory} PACS \citep{poglitsch10} spectroscopy for our sample of 21 transitional disks. The relevant \emph{Herschel} observation identification numbers (ObsIDs), exposure times, and dates of our observations are listed in \emph{Table 1}.  The five additional transitional disks (DM Tau, LkCa 15, GM Aur, TW Hya and UX Tau), and the entire sample of full disks and outflow disks were previously observed as part of the \emph{Herschel} Key Program: GASPS (PI, W. Dent).  We used the line spectroscopy mode (``PacsLineSpec'') to take spectra centered on the [O{\sc i}] 63 $\mu$m line, between 62.93 and 63.43 $\mu$m.  All of the observations were executed in "ChopNod" mode, in order to remove telescope emission and background.

We reduced our original observations and re-reduced the GASPS archival data with the \emph{Herschel} Interactive Processing Environment \citep[HIPE,][]{ott10} version 9.0.0.  We used the default ``ChopNodLineScan'' pipeline along with the most recent calibration tree (CalTree 32).  The data reduction process included: removal of saturated and overly-noisy pixels; differencing the on-source and off-source observations; spectral response function division; rebinning to the native resolution of the instrument (oversample = 2, upsample = 1); spectral flat fielding; and averaging over the two nod positions.  We extracted our spectrum from the central spaxel and accounted for diffraction losses to neighboring spaxels with an aperture correction provided in HIPE.  Since outflow targets generally can have extended emission, spanning multiple spaxels \citep{podio12,howard}, our measured fluxes for these outflow targets will generally underestimate their true fluxes.  We verified that for all of our transitional disks and full disks that there was no appreciable emission in neighboring spaxels.  This lack of extended emission also suggests that there is no significant mispointing in the observations of our transitional disks and full disks.

\subsection{Stellar and Disk Properties}

To interpret our \emph{Herschel} PACS observations of [O{\sc i}] and its relationship to the protoplanetary disk environment, we aggregated stellar properties (effective temperature, bolometric luminosity, FUV and X-ray luminosities, etc.) as well as disk properties (disk mass, disk structure, accretion rates) that, through past work, are known to affect the [O{\sc i}] 63.18 $\mu$m emission.  In this section, we explain the methods by which we derived these stellar and disk properties.  We will relate these to our \emph{Herschel} observations in \emph{Section 4}.

\subsubsection{Effective Temperature and Bolometric Luminosity}

We determined stellar effective temperatures by relating the host star's spectral type (from the literature) to the corresponding effective temperature \citep{luhman99} as listed in \emph{Table 2}.  Generally, we do not assume that the effective temperatures are accurate to more than one spectral subtype ($\sim 100$ K).  

We self-consistently derived bolometric luminosities for all targets by performing a bolometric correction on de-reddened, literature-available I-band photometry listed in \emph{Table 2}.  I-band photometry is preferential, as it is less affected by intervening dust. We de-reddened all of our I-band fluxes by relating V-band extinctions (which are more commonly reported in the literature), to I-band extinctions using relationships from \citet{mathis90}, assuming $R_V$ values typical of the ISM ($R_V = 3.1$).  The de-reddened continuum fluxes were converted to luminosities using the known distances to each different star-forming region (see \emph{Table 1} and references therein).  Finally, bolometric corrections from \citet{luhman99} were used to calculate bolometric luminosities for each target.  These effective temperatures and bolometric luminosities are listed in \emph{Table 3}.

\subsubsection{FUV Luminosities and Accretion Rates}

While ultraviolet observations of T Tauri stars would provide the most direct measurement of the FUV luminosity, these observations are notoriously difficult.  Instead, we made use of the well known correlation between accretion rate and FUV excess emission, to derive FUV luminosities from stellar accretion rates \citep[e.g.][]{dahm08, herczeg08}.  Accretion rates can be determined from a large number of other more commonly measured emission lines \citep[e.g.][]{rigliaco11}.  

We used the H$\alpha$ emission line at 6563 \AA $ $ to estimate stellar accretion rates.  H$\alpha$ is advantageous because is a very commonly reported observational diagnostic, and it correlates well with accretion luminosities (the excess luminosity arising from the infall and accretion of material onto the central star), as derived from other accretion tracers \citep[e.g.][]{rigliaco11}. To determine accretion luminosities, we calculated the H$\alpha$ line fluxes from literature H$\alpha$ equivalent widths (listed in \emph{Table 2}).  For targets with multiple H$\alpha$ equivalent widths available in the literature, we used the mean equivalent width\footnotemark.  
 \footnotetext{While H$\alpha$ is known to be variable, it has been shown that the variability does not introduce significant scatter in H$\alpha$ derived accretion rates \citep{biazzo12}. For our targets with multiple H$\alpha$ equivalent widths, using either the maximum or minimum H$\alpha$ equivalent width changes the resulting accretion rate on average only 0.12 dex.  This variation is less than the uncertainty that results from the empirical relationships used to convert H$\alpha$ luminosity to accretion luminosity \citep{fang09}.}
To determine H$\alpha$ line fluxes, we combined the H$\alpha$ equivalent width with the nearest available photometric point in the literature: R-band.  De-reddening was done similarly as for our bolometric luminosity analysis: using V-band extinctions converted to R-band extinctions via the relationships of \citet{mathis90}. The de-reddened line fluxes were converted to line luminosities using the known distances to each different star-forming region (see \emph{Table 1} and references therein).  The  H$\alpha$ line luminosities, $L_{H\alpha}$, were then converted into accretion luminosities, $L_{acc}$, with the empirical relationships of \citet{fang09}:
	\begin{equation}
	\log{(L_{acc}/L_{\sun})} = (2.27 \pm 0.23) + (1.25 \pm 0.007) \times \log{(L_{H\alpha}/L_{\sun})} 
	\end{equation}
Our derived accretion luminosities are listed in \emph{Table 3}.  From our accretion luminosities, we then used the empirical relationships of \citet{yang12} to relate accretion luminosities to FUV luminosities:
	\begin{equation}
	\log{(L_{FUV}/L_{\sun})} = -1.670 + 0.836 \times \log{(L_{acc}/L_{\sun})} 
	\end{equation}
Our derived FUV luminosities are also listed in \emph{Table 3}.  For the 12 disks shared between this study and \citet{yang12}, we found that our FUV luminosities agreed to those derived by \citet{yang12} within 0.35 dex.  We found no systematic shift between our H$\alpha$-derived FUV luminosities, and their directly measured FUV luminosities. Stellar chromospheric activity can also result in H$\alpha$ emission, so we used the spectral type dependent, equivalent width cutoffs of \citet{whitebasri03} to distinguish between chromospheric activity and accretion.  For targets where the H$\alpha$ equivalent width fell below these cutoffs, we report accretion and FUV luminosity upper limits.  

Converting accretion luminosities to accretion rates requires some physical knowledge of the system and the processes of accretion.  \citet{g98} developed a simple magnetospheric accretion model whereby the accretion luminosity is generated by the release of potential energy as gas falls from the inner edge of the disk onto the surface of the star along stellar magnetic field lines.  In this model, the accretion rate, $\dot{M}$ is related to accretion luminosity by:
	\begin{equation}
	\dot{M}=\frac{L_{acc}R_{\star}}{GM_{\star}(1-\frac{R_{\star}}{R_{in}})}
	\end{equation}
where $R_{\star}$ and $M_{\star}$ are the radius and mass of the star, $G$ is Newton's gravitational constant, and $R_{in}$ is the inner truncation radius of the disk.  $R_{in}$ is generally unknown, but is usually assumed to be $\approx 5R_{\star}$, which corresponds to the typical co-rotation distance \citep{g98, shu94}.  The stellar radius is determined from the star's effective temperature and bolometric luminosity via the Stefan-Boltzmann Law.  We used pre-main sequence evolutionary tracks from \citet{siess00} to relate the effective temperature and bolometric luminosity to specific stellar masses.  Our final accretion rates (as well as stellar masses) are listed in \emph{Table 3}.  

To test the validity of our self-consistently derived accretion rates, we compared our results with an array of other studies, including: \citet{n07}, \citet{g98}, \citet{h98}, \citet{wg01}, and \citet{heg95}.  While differences between accretion rates can develop from several factors (including the use of different accretion tracers, different estimates of extinction, different bolometric corrections, use of non-contemporaneous photometry, etc.), we find that our accretion rates generally agree with past studies to within $\sim$0.5 dex.  This level of variation between accretion rates computed from different tracers is typical, even if observations are contemporaneous \citep{rigliaco12}.  Our accretion rates are also not significantly offset from past studies of accretion rates, with the exception of \citet{heg95}, who find systematically higher accretion rates \citep[although this systematic offset from other estimates has been noted in previous studies, e.g.][]{g98}.

One of the major advantages of our study, as compared to many past studies, is that our accretion rates are self consistently derived, using all the same metric, instead of being aggregated from different literature sources which adopt different methods.

\subsubsection{Disk Structure and Disk Mass}

Many transitional disks in our sample have been previously modeled with radiative transfer codes in order to reproduce near- and mid-infrared disk spectra and resolved millimeter images. While the exact nature of these disk models can vary between papers, they all involve the creation of a simple, axisymmetric model disk with a prescribed dust and gas surface density.  Models specific to transitional disks include gas and dust cavities within a specified radius: $r_{cavity}$.  At the outer edge of this dust cavity, the frontally illuminated disk wall puffs up to a wall height of $h_{wall}$, which can significantly affect the near-infrared emission of transitional disks \citep[both due to excess emission and shadowing of the outer disk;][]{espaillat11}.  These disk models are then subject to simulated observations, and the relevant model spectra or resolved images are calculated (for some specified viewing angle) and fit to observations.  We aggregated values for the cavity size ($r_{cavity}$) and the wall height ($h_{wall}$) from the literature.  These disk properties are listed in \emph{Table 4}.  While the individual models can vary between papers, the majority of these cavity sizes and wall heights are taken from \citet{andrews11} and \citet{espaillat11}, which both use similar disk models.

In addition to looking at the cavity size and wall height, we used self consistently calculated estimates of the total disk mass, from 1.3 mm and 850 $\mu$m photometry available in the literature.  Following \citet{B90}, it is possible to invert observed millimeter flux into an apparent disk dust mass if we assume that the emission is (1) optically thin, (2) arises from an isothermal region of the disk, of known temperature, and (3) is due to material with a known opacity.  (See \citealt{B90} and \citealt{mohanty} for a more detailed explanation of this process, and the assumed dust temperatures and dust opacities.)  Using the canonical gas-to-dust ratio of 100-to-1, we then converted dust masses into total gas masses.  The resulting total disk masses are listed in \emph{Table 4}.  It is important to note that even if we disregard uncertainties in the dust temperature or opacity and the questionable gas-to-dust ratio, these disk masses are likely lower limits.  Millimeter observations are only sensitive to small dust grains, less than $\sim1$ cm in size.  It is possible that substantial mass may be in larger grains, planetesimals or even protoplanets.

\section{Results: Detection of [O{\sc i}] 63.18 $\mu$m and o-H$_{2}$O Emission}

{ We detect [O{\sc i}] 63.18 $\mu$m emission from 17 of our 21 transitional disks.  Coupling these new results with our reanalysis of select disks from the GASPS sample \citep{dent13, howard}, we report [O{\sc i}] 63.18 $\mu$m emission from 21 of 26 transitional disks, 12 of 18 full disks, and emission from all of the outflow disks.}  We fit all observed emission lines to Gaussians using an original MATLAB fitting routine.  To mitigate noise in the PACS spectrum, we fit the lines over a range of wavelength baselines (the minimum wavelength range: 63.13 - 63.23 $\mu$m; the maximum wavelength range spanned the entire PACS spectrum: 62.93 - 63.43 $\mu$m). The best-fitting spectrum was deemed as the spectrum closest to the median of all line fits for a given target.  The line flux of this best-fitting spectrum was calculated from the (continuum-subtracted) Gaussian line profile ($flux = amplitude \cdot \sigma_{Gaussian} \sqrt{2\pi}$).  For [O{\sc i}] 63.18 $\mu$m non-detections, we derive 3$\sigma$ upper limits assuming a Gaussian profile with a 3$\sigma_{RMS}$ peak height (where $\sigma_{RMS}$ is the standard deviation of the continuum linear-fit) , and a 98 km/s line width corresponding to the FWHM of an unresolved line in PACS (\emph{PACS Observer's Manual}).  Continuum fluxes at 63 $\mu$m were also found from the best fitting Gaussian line profile: as the constant baseline flux term.  Continuum emission at 63 $\mu$m was detected for all targets, with the exception of DS Tau (an upper limit of 0.037 Jy).  The [O{\sc i}] 63 $\mu$m line fluxes and 63 $\mu$m continuum fluxes are reported in \emph{Table 5}, and the spectra are provided in the appendix.  To validate our data reduction, we compared our resulting [O{\sc i}] 63 $\mu$m line fluxes and 63 $\mu$m continuum fluxes to the fluxes reported by \citet{howard}.  Despite using a more recent version of HIPE (version 9, rather than version 4), our fluxes generally agree with those of \citet{howard} to within $\sim30\%$, which is comparable to the absolute flux accuracy of PACS (which has a peak-to-peak accuracy $\backsim 30\%$, and RMS accuracy of $\backsim 10\%$; \emph{PACS Observer's Manual}).  Compared to this pipeline uncertainty, the uncertainties in our line fits are negligible.  Representative error bars for both of these types of flux calibration uncertainty are shown in \emph{Figure 1a} and \emph{Figure 1b}.

Typical line fluxes (normalized to the distance of the Taurus-Auriga star-forming region, at 140 pc) are on order $10^{-16} \backsim 10^{-17}$ W/m$^{2}$, corresponding to line luminosity of $10^{-7} \backsim 10^{-5} L_{\sun}$.  Continuum fluxes (again, normalized to 140 pc) range from  $0.1-100 $ Jy, corresponding to continuum luminosities ($L_{continuum} = f_{\nu} \nu 4 \pi d^{2}$) of $10^{-2} \backsim 1 \; L_{\sun}$.  \emph{Figure 1a} shows the [O{\sc i}] 63.18 $\mu$m line luminosity as a function of 63 $\mu$m continuum luminosity for all of our targets.  \emph{Figure 1b} shows the ratio of [O{\sc i}] 63.18 $\mu$m line luminosity to 63 $\mu$m continuum luminosity as a function of 63 $\mu$m continuum luminosity for all of our targets.

We used the Astronomy SURVival package \citep[ASURV,][]{asurvref} to perform linear regressions and correlation tests between the line and continuum luminosities for each subsample.  ASURV is particularly useful as it allows for the incorporation of censored data points (i.e. non-detection, line flux upper limits).  Compared to the other subsamples, we oversample G-type stars in transitional disks (5 G-type transitional disks; 1 G-type full disk; 0 G-type outflow disks).  Because of this oversampling and the seemingly chaotic nature of the G-type line and continuum fluxes, we have omitted them from many of our statistical tests\footnotemark.  \footnotetext{Our oversampling of G-type transitional disks is not intentional.  Due to the rarity of transitional disks we cannot discriminate transitional disks by spectral type in order to populate our transitional disk subsample.  Simultaneously, it is difficult to populate subsamples of full or outflow disks with G-type stars from the GASPS Taurus-Auriga survey, since Taurus-Auriga is a low-mass star-forming region.} Additionally, as discussed in \emph{Section 2.1}, we also exclude multiple systems where the multiplicity likely strongly affects our \emph{Herschel}/PACS observations.  \emph{Tables 6}, \emph{7}, and \emph{8} summarize the results from a variety of statistics and fitting routines that were used to characterize differences between the three subsamples.  There are a number of important trends in our [O{\sc i}] 63.18 $\mu$m line and 63 $\mu$m continuum luminosity data:

\begin{itemize}

\item $ $[O{\sc i}] 63.18 $\mu$m line luminosities are positively correlated with 63 $\mu$m continuum luminosities, both for the sample as a whole, and for each individual subsample, as shown in \emph{Table 6}.  This { correlation was previously recognized} in the \emph{Herschel}/PACS GASPS survey of Taurus-Auriga protoplanetary disks \citep{howard} and Herbig Ae/Be stars \citep{meeus12}, though our study extends this result to a significantly larger sample of transitional disks.

\item Outflow disks tend to have [O{\sc i}] 63.18 $\mu$m line luminosities and 63.18 $\mu$m continuum luminosities that differ markedly from full disks and transitional disks.  This is simply demonstrated in \emph{Table 7}, which shows that both the line and continuum luminosities of outflow disks are not likely from the same parent population as either the full disks or transitional disks.  As shown in \emph{Table 8}, outflow disks tend to have higher [O{\sc i}] 63.18 $\mu$m line luminosities (by 0.5$\backsim$1 dex), higher 63 $\mu$m continuum luminosities (by $\backsim$0.5 dex), and higher line-to-continuum luminosity ratios (by $\backsim$0.5 dex).  { This was previously recognized by \citet{podio12} and \citet{howard}. }

\item Full disks and transitional disks have similar 63 $\mu$m continuum luminosities.  This is most easily shown in \emph{Table 7}, which shows that the 63 $\mu$m continuum luminosities of transitional disks and full disks are effectively indistinguishable.

\item Given the same 63.18 $\mu$m continuum luminosity, full disks tend to have larger [O{\sc i}] 63.18 $\mu$m line luminosities than transitional disks\footnotemark \footnotetext{The one notable exception is BP Tau (C2).  BP Tau has a significantly lower [O{\sc i}] 63.18 $\mu$m line luminosity, compared to other full disks.}, by a factor of $\backsim$2.  While this is visually evident in \emph{Figure 1a}, there is sufficient scatter (and non-detections) to make this difficulty to quantify and the ASURV statistical tests point to indistinguishable line luminosities between the two subsamples (see  \emph{Table 7} and \emph{Table 8}). However, this difference between full disks and transitional disks becomes clear when we examine the ratio of the [O{\sc i}] 63.18 $\mu$m line luminosity to the 63 $\mu$m continuum luminosity, as shown in \emph{Figure 1b}.  The ASURV statistical tests indicate that the distribution of line-to-continuum ratios of transitional disks is significantly different from that of full disks (see \emph{Table 7} and \emph{Table 8}) with full disks having line-to-continuum ratios larger by a factor of $\backsim$2.  

Additionally, the best-fit linear regressions {\it in the line luminosity} for full disks and transitional disks, as shown in \emph{Table 6}, are distinct.  Transitional disks have steeper best-fit slopes and shallower best-fit intercepts than full disks; both of these effects contribute to larger differences in line luminosity at the relevant continuum luminosities.  We checked our ASURV fit results with an alternative Bayesian metric \citep[linmix\_err.pro;][]{kelly07}, and found similar differences between transitional disks and full disks.

This difference between full disks and transitional disks was previously recognized by \citet{howard} for the GASPS Taurus-Auriga sample only.  Our data extends this trend to a much larger sample of transitional disks, suggesting that this lower [O{\sc i}] 63.18 $\mu$m line emission is a characteristic property of transitional disks.

\item There is a weak trend for M-type stars to have lower line and continuum luminosities than K-type stars.  This trend is most evident in our sample of transitional disks.

\end{itemize}

Generally, the [O{\sc i}] 63.18 $\mu$m line is spectrally unresolved.  Most FWHM are within $\backsim$11 km/s (the native resolution of PACS) of the expected line width for an unresolved line for PACS ($\backsim$98 km/s at 60$\mu$m).  This result was expected, since [O{\sc i}] emission originates far out in the disk, $\backsim$AU from the central star \citep{woitke10}. 
For gas orbiting a Sun-like star, Keplerian velocities go as $V_{Keplerian}= 30 km/s \cdot(a /AU)^{-\frac{1}{2}} $.  Thus, beyond $\backsim$AU from the central star, we expect line widths on order $\backsim$10's of km/s.
 Even at these distances, Keplerian velocities dominate over thermal velocities \citep[$\sim$ 1 km/s, assuming typical  O{\sc i} 63.18 $\mu$m gas temperatures of $\sim 100$ K,][]{aresu12}, or turbulent velocities \citep[$\sim 1$ km/s,][]{hughes2011}, and are the cause of most of the line broadening.   A few objects, all outflow sources, have broader line widths, as high as 170 km/s  (e.g RW Aur).  In these sources,  [O{\sc i}] 63.18 $\mu$m emission is thought to originate from shocks along the jet and/or UV-heated gas in the outflow cavity walls \citep{podio12}.  Line widths of $\backsim$100's of km/s reflect the similarly large shock velocities.  Outflow disks can also have spatially extended [O{\sc i}] 63.18 $\mu$m emission associated with the jet, which is detectable in non-central PACS spaxels  \citep{podio12}.  We verified that [O{\sc i}] 63.18 $\mu$m emission was localized only in the central spaxel for our transitional and full disks.  For outflow disks, we only report  [O{\sc i}] 63.18 $\mu$m line and 63 $\mu$m continuum fluxes from the central spaxel \citep[for more accurate line and continuum fluxes of outflow disks, including neighboring spaxels, see][]{podio12}.

Our PACS spectral range fortuitously also includes the considerably fainter o-H$_{2}$O 63.32 $\mu$m emission line.  { We confirmed the detection o-H$_{2}$O emission in 5 full disks and outflow disks, previously identified by \citet{riviermarichalar2012}.  In addition, we report the marginal detections of o-H$_{2}$O in IQ Tau, DK Tau, and BP Tau - for which \citet{riviermarichalar2012} previously identified $3\sigma$ upper limits.  These new detections, from the same observational data, are made possible with our updated version of the Herschel HIPE pipeline and a different line-fitting algorithm.  In addition to these objects, we also report the detection of o-H$_{2}$O emission from RW Aur \citep[observed by the GASPS survey, but not included in][]{riviermarichalar2012}, and the first detection of  o-H$_{2}$O 63.32 $\mu$m emission from a transitional disk: DoAr44 (original to this study).  o-H$_{2}$O line fluxes and $3\sigma$ upper limits for all targets are reported in \emph{Table 5}.}

\section{Trends with \uline{Observable} Stellar and Disk Properties}

In this section, we compare our [O{\sc i}] 63.18 $\mu$m line flux results with stellar and disk properties summarized in \emph{Section 2.3} in order to identify the origin of the trends described in the previous section.  We performed correlation tests between [O{\sc i}] 63.18 $\mu$m and all of these disk/star parameters using the Astronomy SURVival package \citep[ASURV, ][]{asurvref}.  \emph{Tables 6}, \emph{7}, and \emph{8} summarize the results of these correlation tests.

\subsection{Effective Temperature and Bolometric Luminosity}

Since the [O{\sc i}] 63.18 $\mu$m line is generally optically thick \citep[e.g.][]{aresu12}, it would be expected that the line flux might increase for increasing stellar effective temperature.  Similarly, one might expect that the bolometric luminosity of the host star may affect the line flux.  As shown in \emph{Table 6}, we indeed find a correlation between [O{\sc i}] 63.18 $\mu$m line flux and the effective temperature and bolometric luminosity of the host star for { transitional disks}.  We also find a correlation between the 63.18 $\mu$m continuum flux and the effective temperature and bolometric luminosity.  { Curiously, we do not find either of these correlations for our sample of full disks.  This may be a result of the smaller span of effective temperature and bolometric luminosity covered by full disks, compared to transitional disks.}

{ The observed correlations between bolometric luminosity and effective temperature with  [O{\sc i}] 63.18 $\mu$m line flux and 63.18 $\mu$m continuum flux in our transitional disk sample} are expected on the basis that both the line and continuum emission are expected to be optically thick, and thus sensitive primarily to the disk temperature.  This explains why the line and continuum emission are correlated with each other, as they both increase with increasing temperature.  This correlation (though weaker) was also observed by \citet{meeus12}, for their smaller sample of Herbig Ae/Be stars.  This relationship is visually evident in \emph{Figure 1a}, where the symbol of each data point is representative of the star's spectral type; generally cooler, M-type stars have lower line and continuum fluxes than K-type stars.  What is more important, however, is that the effective temperatures and bolometric luminosities between full disks and transitional disks are not statistically different, as illustrated in \emph{Tables 7} and \emph{8}.  This similarity was expected since we attempted to uniformly sample across spectral types within each subsample.  This suggests that the effective temperature and bolometric luminosity alone are not enough to explain why full disks have systematically larger [O{\sc i}] 63.18 $\mu$m line fluxes than transitional disks.

\subsection{FUV Luminosities and Accretion Rates}

\citet{pinte10} used disk thermo-chemical models and showed that far-ultraviolet (FUV) radiation can be a significant gas-heating mechanism, and can promote [O{\sc i}] 63.18 $\mu$m line emission.  For low mass stars, where chromospheric FUV is negligible, most of the FUV luminosity is generated from the infall of disk material onto the central star.  This accretion process shocks and superheats the gas, generating FUV emission, which can then heat the surface layers of the surrounding disk.  Indeed, we find a correlation between FUV\footnotemark[5] and both [O{\sc i}]  63.18 $\mu$m line emission and 63.18 $\mu$m continuum emission in \emph{Table 6}.  However, for our sample, we find that transitional and full disks have statistically indistinguishable FUV\footnotemark[5] \footnotetext[5]{$L_{FUV}$ is directly related to $L_{acc}$ via \emph{Eq. 1} and \emph{Eq. 2}, so statistical tests for the two are identical.  In all Tables we only list $L_{acc}$.} luminosities, as shown in \emph{Tables 7} and \emph{8}.  Thus, FUV cannot be responsible for the [O{\sc i}] line flux differences between full and transitional disks.

The literature is not conclusive about any accretion rate difference between transitional disks and full disks.  Some studies find that transitional disks have accretion rates an order of magnitude lower than full disks \citep{n07, espaillat12, kim13}.  However, when the two samples are drawn from the same spectral type distribution, and the accretion rates are self-consistently derived (and not drawn from the literature), as in our study, no differences are found \citep{fang09}.  This is illustrated in \emph{Tables 7} and \emph{8}.


\subsection{X-ray Luminosities}

More recent thermo-chemical disk models by { \citet{aresu12, aresu14} }have included the effects of irradiation from stellar X-rays.  \citet{aresu12} found that X-ray irradiation tends to become a significant driver for the [O{\sc i}] 63.18 $\mu$m line emission only when $L_{X} > 10^{30}$  erg/s.   Below this limit, FUV irradiation dominates.  { As shown in \emph{Table 6} we find no correlation between $L_{X}$ and either [O{\sc i}] 63.18 $\mu$m line luminosity or 63 $\mu$m continuum luminosity.  We also find that full disks and transitional disks have statistically indistinguishable X-ray luminosities, as shown in \emph{Tables 7} and \emph{8}.}  Furthermore, the observed X-ray luminosities are generally lower than the $10^{30}$  erg/s limit suggested by \citet{aresu12}, suggesting that X-ray irradiation is not the driving mechanism for the trends between [O{\sc i}] 63.18 $\mu$m line luminosity in full disks and transitional disks. { Newer models by \citet{meijerink12} and \citet{aresu12} have predicted a correlation between [O{\sc i}] 63.18 $\mu$m line luminosity and the \emph{sum} of the X-ray luminosity and FUV luminosity, although this trend is not found in either the GASPS Taurus sample \citep{aresu14}, or in our larger sample of transitional disks.}

While X-rays may not be important for the differences between subsamples, X-ray irradiation may be important for a few of the G-type stars.  As noted previously, the G-type stars in our sample tend to have line and continuum fluxes that differ significantly from our other targets.  Many of these G-type outliers (e.g. CHX 22, CHX 7, YLW8) have X-ray luminosities at or above the $10^{30}$ erg/s $L_{X}$ limit of \citet{aresu12}.

\subsection{Disk Structure and Disk Mass}

Is the lower  [O{\sc i}] 63.18 $\mu$m line luminosity in transitional disks simply due to the \emph{lack} of gas in the inner cavity of transitional disks?  
{ \citet{kamp10} showed for a small number of thermo-chemical disk models that creating an inner cavity (out to 10 AU) \emph{completely devoid of gas}, decreased the [O{\sc i}] 63.18 $\mu$m line flux by a factor of 1.5.} \citet{bruderer13} has shown that even though most [O{\sc i}] emission originates from the outer disk (beyond 10's of AU), depleting gas within the inner cavity of transitional disks can reduce the disk's total [O{\sc i}] 63.18 $\mu$m luminosity by factors of up to several.  Given the large beam of \emph{Herschel}/PACS ($\sim1000$'s of AU, at the distance of Taurus), a reduction in the [O{\sc i}] 63.18 $\mu$m line flux by a factor of  $\sim2$ as we observe would require a depletion of gas in the inner disk by a factor $\gtrsim100$  \citep[see Fig. 18 of][]{bruderer13}.  { These scenarios proposed by \citet{kamp10} and \citet{bruderer13} seem unlikely} in view of our finding that the mass accretion rates of transitional disks in our sample are statistically indistinguishable from those of full disks in Taurus.


One might expect that the heating of the gas, and by extension the luminosity of the [O{\sc i}] 63.18 $\mu$m line, is affected by the distribution of the dust, hence a correlation between the dust cavity size or wall height and the [O{\sc i}] 63.18 $\mu$m line. We find no correlation between the cavity sizes or wall heights of transitional disks and either the [O{\sc i}] 63.18 $\mu$m line luminosites or 63.18 $\mu$m continuum luminosities.  This result is shown in \emph{Table 6} and graphically in \emph{Figure 2}.  This suggests that either [O{\sc i}] 63.18 $\mu$m is tracing material well beyond the inner cavity and/or the distribution of gas in the inner cavities of full disks and transitional disks is similar.  While this result is suggestive, { more work following \citet{kamp10} and \citet{bruderer13} needs to be done} to determine the relationship between the size of transitional disk cavities, the gas-to-dust ratio in these cavities, and  [O{\sc i}] 63.18 $\mu$m emission.

Finally, we find that the disk dust masses for full and transitional disks are statistically indistinguishable, and that there is no correlation with the dust mass and either the [O{\sc i}] 63.18 $\mu$m line luminosities or 63 $\mu$m continuum luminosities.  This is expected, given that both the [O{\sc i}] 63.18 $\mu$m line and 63 $\mu$m continuum emission are mostly optically thick.

\section{Trends with \uline{Model-Derived} Disk Properties}

Thus far, we have been unable to satisfactorily explain the difference in [O{\sc i}] 63.18 $\mu$m line flux between full and transitional disks.  To identify other possible causes for the trends we used the DENT (``Disk Evolution with Neat Theory") grid of themo-chemical models by \citet{woitke10} to look for correlations between  [O{\sc i}] 63.18 $\mu$m line emission and various disk properties, that are not directly observable.  Of the free parameters in the DENT grid (e.g. column density, disk inner/outer radius, grain sizes, inclination, etc.), there are only two parameters that can cause the observed [O{\sc i}] 63.18 $\mu$m line flux trends: disk flaring, and the disk gas-to-dust ratio.

To illustrate trends within the DENT grid, we have developed a novel approach for analyzing the large suite of DENT disk models (totaling over 300,000 unique disks).  \emph{Figure 3} illustrates an example of this technique, for the case where we investigate how [O{\sc i}] 63.18 $\mu$m line and 63.18 $\mu$m continuum emission change as a function of FUV excess luminosity (which is discussed previously, in \emph{section 4.2}).  Using an original MATLAB script, we select a randomized subsample of a few thousand\footnotemark[6] \footnotetext[6]{in general, our results are not sensitive to the number of disk models selected, as long as it is fairly large ($\gtrsim 100$).} unique disk models from the full DENT grid.  Generally, we constrain this randomized subsample to consist of low mass stars (M- and K-type), similar to our sample of transitional and full disks.  Next, for each of the selected disk models, we search the full DENT grid for all of the disk models that possess identical stellar/disk properties, \emph{except} for the quantity that we are interested in -- FUV excess in this case.  Since the DENT grid allows for two different FUV excesses (0.001 and 0.1 $L_{star}$), this results in a few thousand emph{pairs}\footnotemark[7] of disks. \footnotetext[7]{For other stellar/disk parameters where more than two values are possible, we form sets containing the same number of disk models as the number of possible values for that stellar/disk parameter.  For example, there are five possible gas-to-dust ratios within the DENT grid; thus when performing our analysis for gas-to-dust ratios, we form several thousand sets of disk models, each containing five disks that are identical with the exception of their gas-to-dust ratios.} 

From this ensemble of disk model pairs, we can perform a number of analyses.  \emph{Figure 3a} shows the ensemble of disk models in a plot of  [O{\sc i}] 63.18 $\mu$m line flux vs. 63.18 $\mu$m continuum flux\footnotemark[8]  \footnotetext[8]{It is not possible to exactly duplicate \emph{Figure 1a} with the DENT grid, as the DENT grid does not include a  63 $\mu$m photometric point.  Instead, we used the 65 $\mu$m photometric point as a proxy.  For most DENT models, there is not a significant change in the continuum luminosity between 60, 65 or 70 $\mu$m.}, similar to \emph{Figure 1a}.  In this plot, each of the disk models is represented by a colored point, with the color corresponding to its FUV excess.  The vectors connect individual disk pairs.  These vectors can be thought of as ``evolutionary tracks'' which show how one disk would change if the FUV excess changed (in this case, the arrow points in the direction of increasing FUV excess).  Due to the extreme number of disk models in the DENT grid, even the randomized subsample in \emph{Figure 3a} is dense and difficult to interpret. To simplify interpretation, \emph{Figure 3b} displays two contour intervals -- one for each FUV excess -- indicating the region that contains 67\% of the disk models for that particular FUV excess.  These contours are generated by binning the data in both continuum flux and line flux space (usually with bins 0.25 dex in size).  A small number of ``evolutionary tracks'' are included, to reinforce the concept that we are tracking disk models as a particular quantity is changed.  Lastly, \emph{Figure 3c} illustrates the \emph{mean} ``evolutionary track'' for all of our disk pairs.  To generate this figure, we take each pair and calculate the \emph{change} (signified by a ``$\Delta$'' in the figure axes) in continuum flux and \emph{change} in line flux between each pair member.  The vector displayed represents the mean change in continuum and line flux for our entire ensemble of disk pairs.  The error bars indicate the 1-$\sigma$ variations in this single step-up in FUV excess.  This last figure, \emph{Figure 3c} is particularly useful, as it shows information that is easily lost in the large apparent scatter in \emph{Figure 3a} and \emph{Figure 3b}.  For example, while it is clear in these other figures that increasing the FUV excess increases the line flux, it is not as obvious how much this line flux changes, and the relative uncertainties.  It's also not obvious in the other figures that the change in continuum flux is so consistent (represented by the very small horizontal error bars) between all the DENT models.  In the following sections, we will make use of figures similar to \emph{Figure 3b} and \emph{Figure 3c} to investigate how changing various stellar/disk parameters affect the observed [O{\sc i}] 63.18 $\mu$m line fluxes and 63.18 $\mu$m continuum fluxes.

\subsection{Disk Flaring}

One of the early predicted trends of the DENT grid was that [O{\sc i}] 63.18 $\mu$m emission may trace the flaring of the disk \citep{woitke10}.  In a flared disk, the disk surface is directly illuminated by the central star, causing higher temperatures and stronger [O{\sc i}] 63.18 $\mu$m emission.  Within the DENT model grid, geometric flaring of the gas disk is parameterized\footnotemark[9]  \footnotetext[9]{In principle, the vertical scale height of the gas disk should be self-consistently derived from hydrostatic equilibrium, given the temperature structure of the disk. { This is \emph{not} done in the DENT grid.  Using parameterized disk structures allows for a wider, unbiased exploration of disk parameter space - while still assessing the relative influence of key parameters on observable quantities.  See \emph{Section 2} of \citet{kamp11} for a discussion of the parameterized approach.}} by the value of $\beta$: the disk scale height, $h$, as a function of radial distance, $r$, can be described by:
	\begin{equation}
	h(r) = h_{0} \left(\frac{r}{r_{0}}\right)^{\beta}\
	\end{equation}
where $h_{0}$ is the disc scale height (fixed at 10 AU), and $r_{0}$ is a fixed reference distanced (fixed at 100 AU).  In the DENT grid, there are three possible flaring parameters, $\beta = 0.8$, $\beta = 1.0$, and $\beta = 1.2$.  A flaring parameter as low as $\beta = 0.8$ is more appropriate for the late stages of disk evolution (e.g. debris disks), and not for our study of young protoplanetary disks \citep{kamp11}. Thus, we have excluded models with $\beta = 0.8$ from our analysis.  It is important to note that when we refer to ``flaring,'' we are referring to the flaring of the \emph{gas} disk.  In the DENT grid, the dust is either well mixed with the gas, or settled.  The dust disk scale height, $h_{dust}$, is parameterized as:
	\begin{equation}
	h_{dust}(r,a) \propto h(r) a^{-s/2} 
	\end{equation}
where $h$ is the gas scale height (\emph{Eq. 4}), $a$ is the grain size, and $s$ describes the strength of dust settling: $s = 0$ for a well-mixed disk, and $s = 0.5$ for a settled disk.  We find that within the DENT grid, there is no \emph{systematic} difference between the [O{\sc i}] 63.18 $\mu$m line luminosity of disks with either well-mixed or settled dust disks.  Settled dust disks do have systematically lower 63 $\mu$m continuum luminosities than well-mixed disks, by $\sim 0.7$ dex.  Instead, we focus on the effects of changing the flaring of the gas disk.

\emph{Figure 4} illustrates how the [O{\sc i}] 63.18 $\mu$m line luminosity and 63 $\mu$m continuum luminosity change as disks become more flared according to the DENT grid.  From \emph{Figure 4}, we can see that increasing the disk flaring from $\beta = 1.0$ to $\beta = 1.2$ can increase the [O{\sc i}] 63.18 $\mu$m line luminosity by $\sim 0.5$ dex, while not significantly altering the 63 $\mu$m continuum luminosity.  This increase in flaring in the DENT grid results in generally warmer gas in the disk surface, resulting in larger [O{\sc i}] 63.18 $\mu$m line luminosities.  Since changing the flaring of the disks only changes the [O{\sc i}] 63.18 $\mu$m line luminosity, and not the 63 $\mu$m continuum luminosity, this may provide a natural explanation for the decreased [O{\sc i}] 63.18 $\mu$m line luminosity in transitional disks compared to full disks.  This would imply that transitional disks are \emph{less} flared than full disks, and that their lower [O{\sc i}] 63.18 $\mu$m line luminosities are the result of cooler disk gas surface layers.  If the gas in transitional disks is indeed cooler than in full disks, this might be linked to the reduction or removal of some gas heating mechanism.  \citet{AikwawaNomura06} have shown that growth and settling of larger dust grains ($\sim 10$ cm in diameter) leads to decreased photoelectric heating in the disk atmosphere and less disk flaring.  However, these large dust grains will quickly settle towards the disk midplane, resulting in reduced far-infrared emission, which we do not see in our sample.  An alternative explanation could be that the stellar FUV photons responsible for heating the  [O{\sc i}]  emitting disk surface layers \citep{aresu12} are being absorbed at a vertically extended dust inner rim.  Future SED modeling may be able to disentangle these two possibilities.

\subsection{Disk Gas-to-Dust Ratio}

A second, though less well recognized trend in the DENT grid is that [O{\sc i}] 63.18 $\mu$m emission may trace the disk gas-to-dust ratio.  From \emph{Herschel}/PACS and millimeters observations combined with dust and gas modeling, \citet{thi10} suggested that the transitional disk TW Hya possess a lower gas-to-dust ratio than the standard interstellar value of 100, though this suggestion has been disputed in recent years \citep{gorti11, bergin12}.  \citet{meeus12} has also suggested, from analysis of the DENT grid, that variations in the [O{\sc i}] 63.18 $\mu$m line luminosities of Ae/Be stars, could be a result of variations in the gas-to-dust ratio, although they do not explore this further.

\emph{Figures 5} and \emph{6} illustrate how the [O{\sc i}] 63.18 $\mu$m line luminosity and 63 $\mu$m continuum luminosity change as the gas-to-dust ratio changes within the DENT grid.  We consider two scenarios: first the effects of changing the gas-to-dust ratio while holding the \emph{dust} mass constant, as shown in \emph{Figure 5}; and second, the effects of changing the gas-to-dust ratio while holding the \emph{gas} mass constant, as shown in \emph{Figure 6}.  It is necessary to consider these two scenarios independently since identical gas-to-dust ratios can be constructed from different combinations of gas and dust mass.

As shown in \emph{Figure 5}, increasing the gas-to-dust ratio, \emph{while holding the dust mass constant} (in other words: we are increasing the gas-to-dust ratio by adding gas), results in increased [O{\sc i}] 63.18 $\mu$m line luminosities.  The increase in line luminosity is greatest for low dust masses: where changing the dust to gas ratio from  $10^{1}$ to $10^{-3}$ results in an increase in line luminosity of $\sim$ 2 dex.  At higher dust masses, the increase in line luminosity across the same range of gas-to-dust ratio results in an increase in line luminosity of $\sim$ 1 dex.  Since the [O{\sc i}] 63.18 $\mu$m line is generally optically thick \citep[e.g.][]{aresu12}, the increase in line luminosity with increasing gas mass is likely due to an increased heating rate, perhaps by H$_2$ photo-dissociation, collisional de-excitation of H$_{2}^{*}$, or photo-electric heating \citep[e.g.][]{woitke09}.  From \emph{Figure 5}, it is also clear that changing the gas-to-dust ratio, while holding the dust mass constant, does not change the 63 $\mu$m continuum luminosity.  This is not unexpected, since the continuum luminosity is tracing the dust in the disk, which in these cases, remains unchanged. 

\emph{Figure 6} shows the complicated effects of increasing the gas-to-dust ratio, \emph{while holding the gas mass constant} (in other words: we are increasing the gas-to-dust ratio by removing dust).  In general, increasing the gas-to-dust ratio by removing dust significantly decreases the 63 $\mu$m continuum luminosity by $0.2 \sim 2$ dex, depending on the gas mass. The behavior of the [O{\sc i}] 63.18 $\mu$m line luminosity as the gas-to-dust ratio changes, while holding the gas mass constant, is even more complicated.  For gas masses below $10^{-6} \, M_{\sun}$, the [O{\sc i}]  line luminosity decreases, with increasing gas-to-dust ratio.  For gas masses above $10^{-6} \, M_{\sun}$,  the [O{\sc i}]  line luminosity increases, with increasing gas-to-dust ratio - although the rate of this increase decreases with decreasing dust mass.  This decrease in [O{\sc i}] line luminosity with decreasing dust mass may indicate the signifcance of dust-driven heating processes within the disk, such as PAH heating and collisional heating \citep[e.g.][]{woitke09}.

So, could the lower  [O{\sc i}] 63.18 $\mu$m line luminosities of transitional disks be explained by changes in the gas-to-dust ratio? Given the two ways of changing the gas-to-dust ratio, the simplest possible explanation is that transitional disks have lower gas-to-dust ratios, \emph{by having less gas mass} than full disks.  As shown in \emph{Figure 5}, a decrease of gas-to-dust ratio of only $\sim$ 0.5 dex would be able to explain the factor of $\sim$ few lower [O{\sc i}] 63.18 $\mu$m line luminosities in transitional disks, while retaining similar 63 $\mu$m continuum luminosities.  While there may be specific evolutionary pathways whereby increasing the dust mass can also explain the factor of $\sim$ few lower [O{\sc i}] 63.18 $\mu$m line luminosities in transitional disks, changes in the dust mass strongly affect the 63 $\mu$m continuum luminosities, as shown in \emph{Figure 6}.  Furthermore, our estimates of dust mass from millimeter observations (see \emph{Section 4.4}), suggest that there is no statistical difference between the dust masses of full and transitional disks.

BP Tau may be an example of a more evolved full disk that is dispersing its gas, and decreasing its gas-to-dust ratio.  \citet{dutrey03} showed that BP Tau is anomalous in many regards: its CO and dust disk are small and faint; the $^{12}$CO \emph{J} = $2 \rightarrow 1$ transitional is optically thin; and that with respect to the dust, the CO is depleted by a large factor ($\sim$ 100).  One possible explanation discussed by \citet{dutrey03} is that BP Tau may be depleted in gas with respect to dust, and have a lower gas-to-dust ratio than other full disks.  As shown in \emph{Figure 1a}, BP Tau has an anomalously low [O{\sc i}] 63.18 $\mu$m line luminosity compared to other full disks.  This result confirms that BP Tau is indeed different from other full disks.  Furthermore, the low [O{\sc i}] 63.18 $\mu$m line luminosity is consistent with the hypothesis of \citet{dutrey03}, that BP Tau has a lower gas-to-dust ratio than typical full disks, by $\sim$ 1 dex.

\section{Discussion}


\subsection{Implications for Disk Evolution Models}

Photoevaporation may be a natural mechanism by which the disk gas-to-dust mass ratio is reduced with time. High-energy stellar photons heat the disk and drive a photoevaporative wind which primarily removes the gas component from the disk surface.  Amongst our sample of transitional disks, CS Cha, TW Hya, T Cha, RXJ1615.3-3255 and YLW8 have been observed with {\it VLT/VISIR} and present [Ne~II] emission lines blueshifted by several km/s, implying on-going photoevaporation \citep{pascucci09, sacco12}.  GM Aur has been observed with \emph{Gemini/TEXES}, but with insufficient S/N to precisely determine the line centroid \citep{najita09}.  While photoevaporation has been detected from these objects, the rate at which gas is lost via this mechanism is still unknown.  If [Ne {\sc ii}] is tracing the very thin EUV irradiated region, the mass loss rate is negligible ($\sim 10^{-10} M_{\sun}/yr$);  while, if [Ne~{\sc ii}] is tracing the deeper X-ray irradiated layer, the mass loss rate may be significant ($\sim 10^{-8} M_{\sun}/yr$). In the latter case, if we assume that full disks start with a mass of $\sim 22 \, M_{Jupiter}$ (the mean value derived from millimeter data; see \emph{Section 2.3.3}), they could loose half of their gas mass in just 1\,Myr via photoevaporation.

Planet-disk interactions may also provide a mechanism for reducing the gas-to-dust ratio in protoplanetary disks \citep[e.g.][]{espaillat13}.  \citet{rice06} showed that pressure gradients at the outer edge of a gap cleared by a giant planet can act as dust filters.  In such a scenario, small dust grains and gas flow across the gap and are either lost to the planet or the inner disk (and eventually the host star), while large dust grains remain trapped in the outer disk.  This has the effect of removing gas from the outer disk while retaining most of the mm- and cm-size dust, and thus decreasing the gas-to-dust ratio of the outer disk.  However, the leak of small, micron-size dust particles into the inner disk still necessitates some additional mechanism, such as dust coagulation, to explain the dust cavities in transitional disks \citep{zhu12}.  Additionally, dust filtration alone is not a realistic mechanism for a decreasing the gas-to-dust ratio by 0.5 dex, as suggested by our work.  As gas leaves the outer disk and flows into the gap formed by the planet, it will either be accreted onto the planet, or completely cross the gap into the inner disk, where it can then accrete onto the central star.  \citet{lubow06} showed that when mass flows across into these gaps formed by giant planets, $\sim 90 \%$  of the mass will be accreted onto the planet.  Thus, for dust filtration to be the driver of a low gas-to-dust ratio in the outer disk, it is at the expense of putting a large majority of the outer disks's gas mass directly into planets.  If we assume full disks start with a gas mass of $\sim 22 \, M_{Jupiter}$ (the mean value derived from millimeter data; see \emph{Section 2.3.3}), $\sim 7 \, M_{Jupiter}$ of gas would need to be lost to planet formation to result in a decrease in the gas-to-dust of 0.5 dex.  If instead, we assume that a full  protoplanetary disk can be characterized by a minimum mass solar nebula \citep[MMSN,][]{weidenschilling77, kuchner04}, then it would be necessary for the disk to lose even more mass: upwards of $\ga 20 \, M_{Jupiter}$\footnotemark[10]\footnotetext[10]{The total disk mass is calculated by integrating the surface mass density from the inner edge of the protoplanetary disk ($\sim 0.07$ AU) to the outer edge (conservatively, $\sim40$ AU).  Using the MMSN described by \citet{kuchner04} ($\Sigma = 4225 \; g/cm^{2} \; (a / 1 \, AU)^{-1.78}$) results in a total disk mass of 24 $M_{Jupiter}$.  Using the classical MMSN described by \citet{weidenschilling77} ($\Sigma = 4200 \; g/cm^{2} \; (a / 1 \, AU)^{-1.5}$) results in a total disk mass of 38 $M_{Jupiter}$.  A loss of 0.5 dex of the disk mass for these two models correspond to $17$ and $26$ $M_{Jupiter}$, respectively.  Using more liberal estimates of the outer edge of the protoplanetary disk \citep[e.g. 270 AU;][]{chianggoldreich97} results in even larger masses.}.  These simple calculations also assume that all of the dust in the outer disk is somehow protected, perhaps due to a planet-induced pressure bump.  If the loss of dust across the gap is large, these mass estimates would only be lower limits.  If all of this mass is lost to forming planets, this would suggest the formation of a large number of giant planets at large semimajor axes ($\gtrsim 10$ AU), which does not seem to agree with the current (though still debated) statistics of giant exoplanets \citep{nielsen13, fressin13, biller13}.  Lastly, while large, Jovian-mass planets can clear gaps and cause global depletions in the gas surface density of disks, they only deplete the surface density of the disk by a factor of a few \citep[e.g. Fig. 3 of ][]{lubow06}.  As discussed in \citet{bruderer13} (and in \emph{Section 4.4}) our observed factor of 2 line flux difference between transitional disks and full disks would require a drop in the surface density by a factor of $\gtrsim100$.

\subsection{Potential Followup Observations}

Direct measurement of the gas-to-dust ratio in full disks and transitional disks would break our observed degeneracy between gas-to-dust ratio and disk flaring.  While the dust mass of protoplanetary disks can be estimated with millimeter observations \citep[e.g.][]{mohanty}, the total gas mass of protoplanetary disks is difficult to directly measure.  Combining our observations of the [O{\sc i}] 63.18 $\mu$m line with low J CO rotational lines, has been suggested as a possible way to directly measure total disk gas mass.  While this method has been implemented for select, well studied disks \citep[e.g. TW Hya,][]{thi10}, its reliability is still under discussion \citep{gorti09, bergin12}.  Both low J CO and [O{\sc i}] lines are optically thick, which make them both primarily sensitive to temperature - and only weakly dependent on disk mass.  Alternatively, observations of isotopologues may provide direct estimates for disk mass.  Isotopologues (such as $^{13}$C) are minor components within the disk and can be optically thin and directly trace disk mass (modulo the assumed abundances of the relative species). With the significant ($\sim10$x) increase in sensitivity allowed by ALMA, detecting emission from minor disk components out to nearby star-forming regions (e.g. Taurus-Auriga) is now possible.


It is difficult to directly measure the flaring of \emph{gas} in protoplanetary disks.  For select nearby and edge-on disks, it may be possible to directly measure the relative vertical distribution of dust (via mm-emission) and gas (via gas emission lines, such as CO and its isotopologues) with high spatial and spectral resolution observations with ALMA \citep{rosenfeld13}.  Detailed SED modeling covering the mid-infrared, far-infrared, and millimeter wavelengths may be able to break the degeneracy between disk gas mass and disk scale height. Flared disks intercept more stellar radiation at larger semimajor axes than flatter disks.  Emission from these warm, outer disk, surface layers dominate the SED beyond $\sim$20 $\mu$m \citep{chianggoldreich97}.

\section{Summary}

We obtained \emph{Herschel}/PACS spectra of [O{\sc i}] 63.18 $\mu$m for 21 transitional disks in the Ophiuchus, Chameleon, and Lupus star forming regions.  This survey complements the larger \emph{Herschel} GASPS survey of the Taurus star forming region \citep{dent13} by quadrupling the number of transitional disks observed with PACS in this wavelength.  [O{\sc i}] 63.18 $\mu$m is significant because it traces the cool, outer regions ($\gtrsim$ 10 AU) of the protoplanetary disk, where the majority of the disk mass lies. Our primary results can be summarized as follows:

\begin{enumerate}

\item Full disks have larger  [O{\sc i}] 63.18 $\mu$m line luminosities than transitional disks, while having similar 63.18 $\mu$m continuum luminosities.  While this result was previously recognized by \citet{howard} for the GASPS Taurus-Auriga sample, our data extends this trend to a larger sample of transitional disks, suggesting that lower [O{\sc i}] 63.18 $\mu$m line emission is a characteristic property of transitional disks.

\item For all of our targets, we self-consistently derived stellar and disk parameters that have been previously shown to affect  [O{\sc i}] 63.18 $\mu$m emission.  While [O{\sc i}] 63.18 $\mu$m can correlate with these parameters, we found that transitional disks and full disks have statistically indistinguishable effective temperatures, bolometric luminosities, FUV luminosities, accretion rates, and X-ray luminosities.  Thus, these properties cannot be responsible for the lower  [O{\sc i}] 63.18 $\mu$m line luminosities  of transitional disks.

\item We found no correlation between the [O{\sc i}] 63.18 $\mu$m line luminosities of transitional disks and either their disk masses (as inferred from millimeter photometry), dust cavity sizes, or wall heights (as inferred from SED and interferometric image modeling).  This suggests that the decrease in [O{\sc i}] 63.18 $\mu$m emission is not simply due to a lack of material in the inner cavity of transitional disks, though more modeling is needed to confirm this result \citep[e.g.][]{bruderer13}.

\item Using the DENT grid of thermo-chemical protoplanetary disk models \citep{woitke10}, we determined that the lower [O{\sc i}] 63.18 $\mu$m line luminosities in transitional disks could result from either a decrease in disk flaring, or a decrease in gas-to-dust ratio via a global depletion of gas mass.  Decreasing the disk flaring results in less stellar irradiation impinging on the surface of the outer disk, thus decreasing the disk temperature and reducing [O{\sc i}] 63.18 $\mu$m emission.  Decreasing the gas-to-dust ratio by removing gas mass results in a decrease in the amount of heating from H$_2$ photo-dissociation, collisional de-excitation of H$_{2}^{*}$, and/or photo-electric heating \citep[e.g.][]{woitke09}.  Both photoevaporation, and planet formation, can result in a decrease in gas mass, although their efficiencies are still not well constrained.  While additional observations are needed to disentangle the effects of disk flaring and gas-to-dust ratio, our results show that transitional disks are more evolved than their full disk counterparts, possibly even at large radii.

\end{enumerate}

\acknowledgments
I. P., J. T. K., C. E., and S. A. acknowledge NASA/JPL for funding support.  J. T. K. and I. P. thank Elisabetta Rigliaco for helpful discussions on mass accretion rate estimates.  The authors would also like to thank the referee, Kees Dullemond, for a very constructive review.

\appendix

\section{Supplementary Figures}

\emph{Figure 7} includes all of the reduced \emph{Herschel}/PACS  [O{\sc i}] 63.18 $\mu$m spectra used in this work, and are only provided in the online version of the article.  \emph{Figures 8} through \emph{16} display null correlations of various stellar and disk parameters with the [O{\sc i}] 63.18 $\mu$m line luminosity and nearby continuum luminosity and are only provided in the online version of the article.

\clearpage


	\begin{figure}
	\figurenum{1a}
	\epsscale{1}
	\plotone{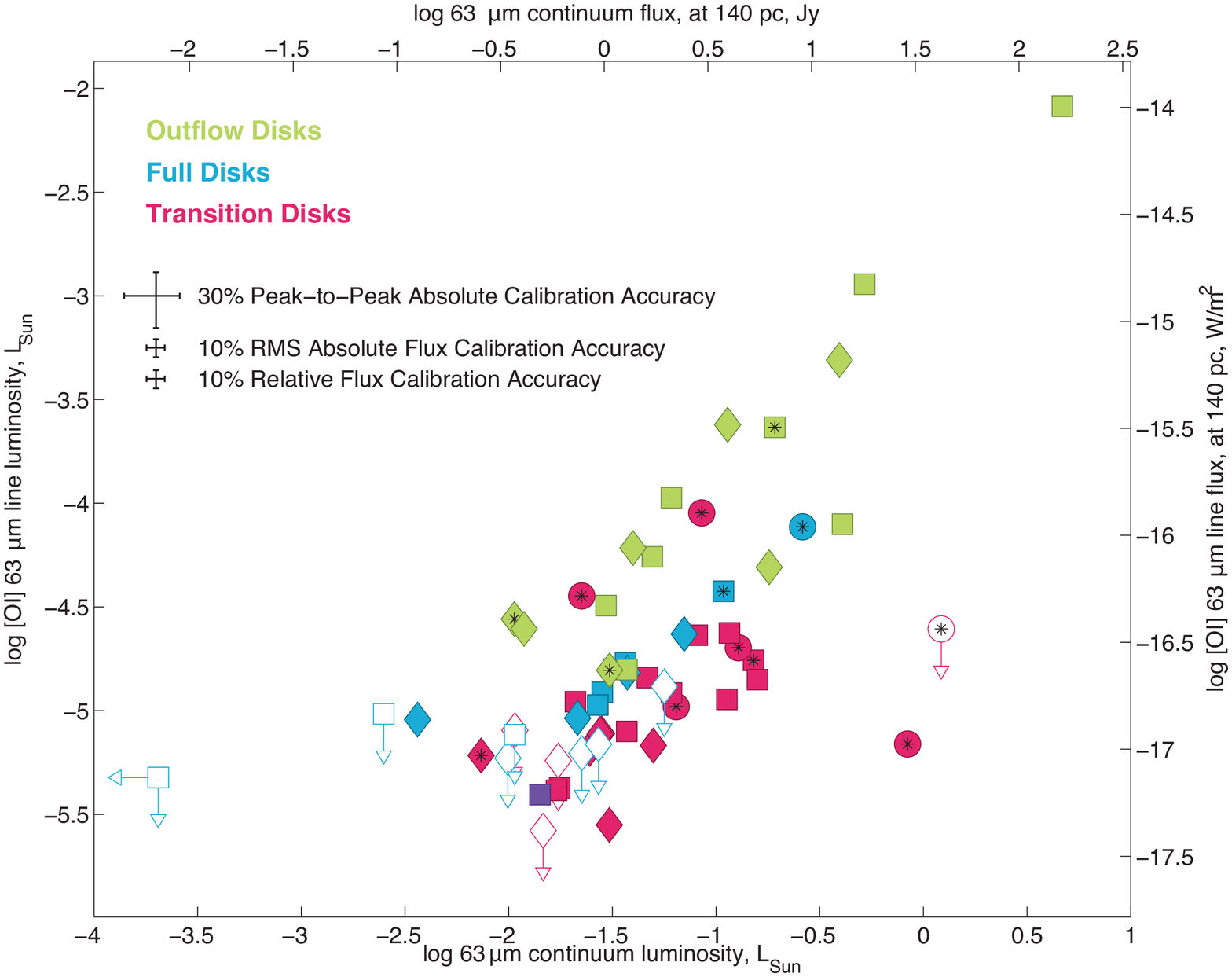}
	\caption{[O{\sc i}] 63.18 $\mu$m line luminosity as a function of 63 $\mu$m continuum luminosity for our sample of transitional disks (red), full disks (blue), and outflow disks (green). $3\sigma$ upper limits are denoted by hollow data points with arrows. Symbols correspond to stellar spectral types: circles are G-type stars (which are included in this plot, but neglected in the statistical analysis, for reasons described in the paper), squares are K-type stars, and diamonds are M-type stars. BP Tau (an evolved full disk) is indicated in purple. Targets excluded from statistical tests (for either being a binary that does not meet the criteria in Sect. 2.1, or being a G-type star) are marked by an asterisks. }
	\end{figure}
	
	\begin{figure}
	\figurenum{1b}
	\epsscale{1}
	\plotone{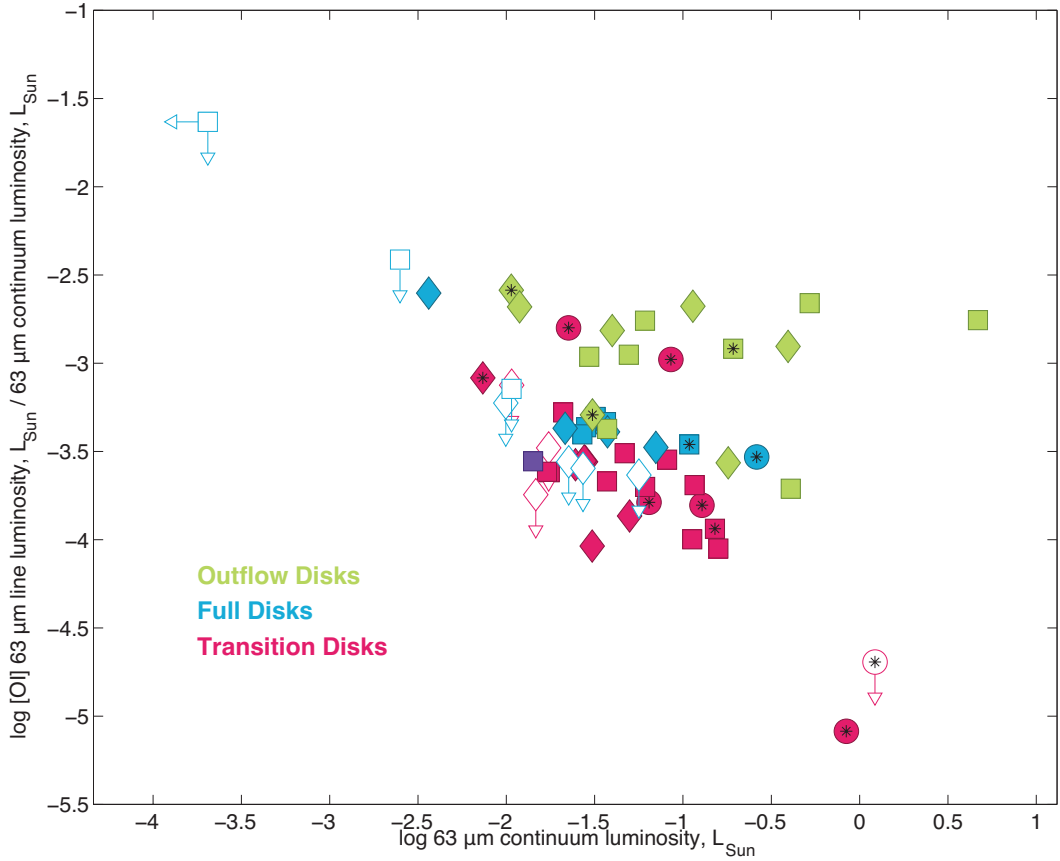}
	\caption{The ratio of [O{\sc i}] 63.18 $\mu$m line luminosity / 63 $\mu$m continuum luminosity as a function of 63 $\mu$m continuum luminosity for our sample of transitional disks (red), full disks (blue), and outflow disks (green). $3\sigma$ upper limits are denoted by hollow data points with arrows. Symbols are as in \emph{Figure 1a}. }
	\end{figure}

	
'	\begin{figure}
	\figurenum{2}
	\epsscale{1}
	\plotone{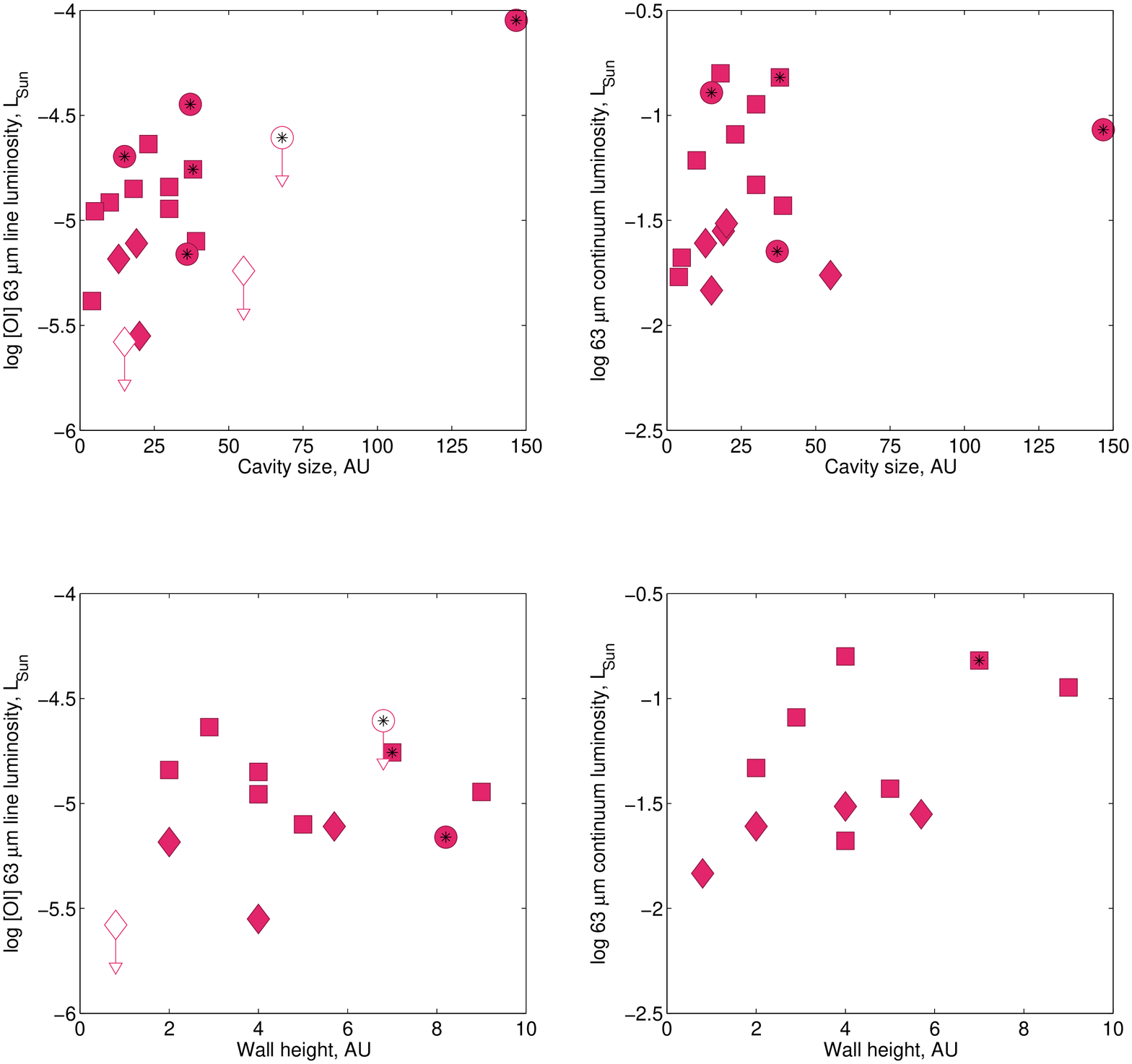}
	\caption{[O{\sc i}] 63.18 $\mu$m line luminosity and 63 $\mu$m continuum luminosity as a function of gap size and wall height for all of the transitional disks within our sample for which such measurements have been made in the past. $3\sigma$ upper limits are denoted by hollow data points with arrows. Symbols are as in \emph{Figure 1a}.}
	\end{figure}

	
	\begin{figure}
	\figurenum{3}
	\epsscale{1}
	\plotone{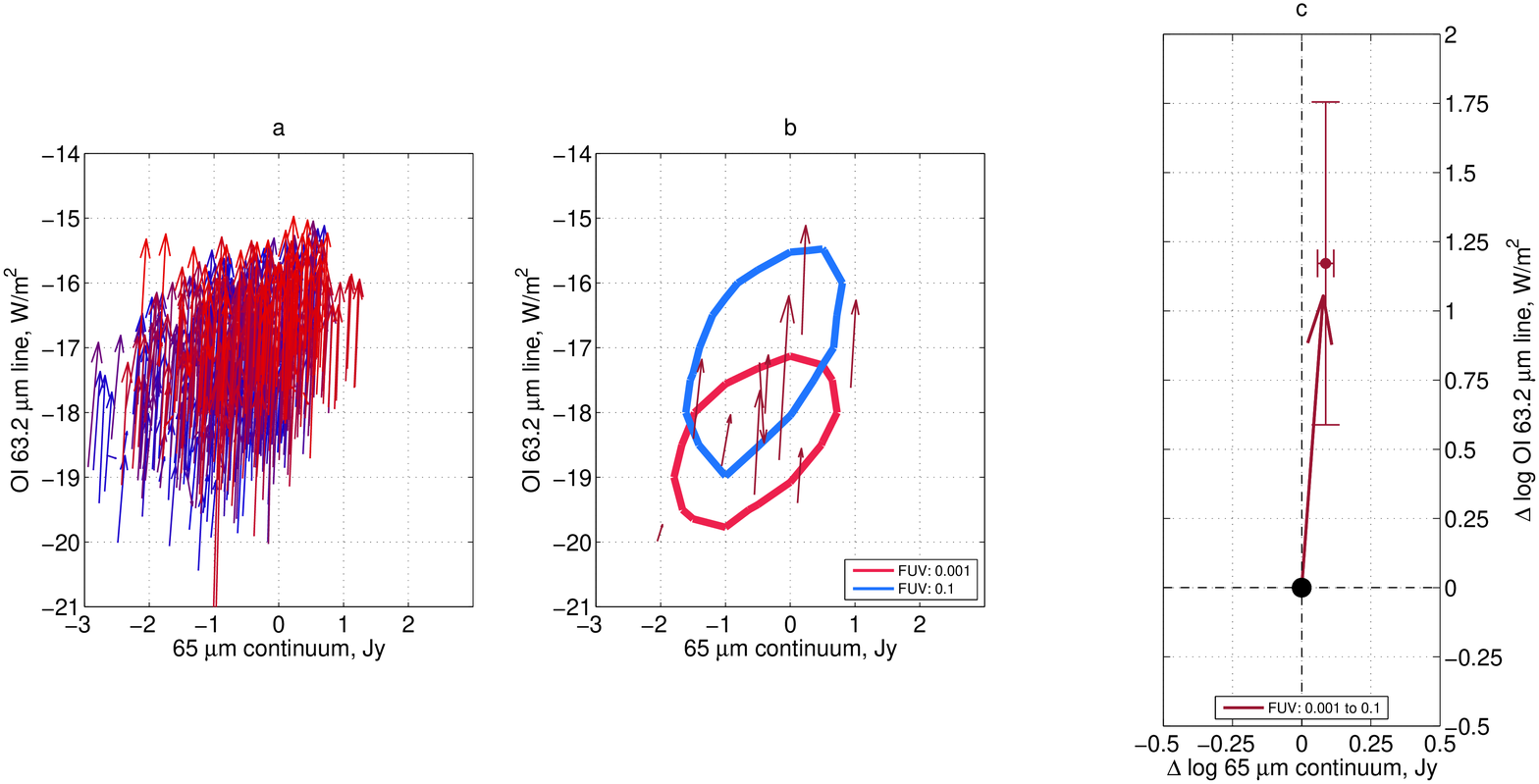}
	\caption{DENT grid predictions for how [O{\sc i}] 63.18 $\mu$m line flux and 65 $\mu$m continuum flux change, with increasing disk FUV excess (FUV = 0.001 and 0.1), for a random subsample (N $\sim$ 5000) of disks around low mass stars ($\le1$ M$_{Sun}$). The left shows all of the ``evolutionary tracks'' for this sample of disk models. The middle panel shows the regions that contain $67\%$ of the models as a function of FUV. The ``evolutionary tracks'' of 10 randomly selected disk models are included for reference. These tracks indicate the path that that particular disk would move if the FUV increased. The panel on right shows the mean change (``delta'') in  [O{\sc i}] 63.18 $\mu$m line flux and 65 $\mu$m continuum flux, with respect to an initially low FUV disk. Arrows point in the direction of increasing FUV. Error bars indicate the $1\sigma$ variations in these $\Delta$  [O{\sc i}] 63.18 $\mu$m line flux and $\Delta$ 65 $\mu$m continuum flux during each step in increasing FUV.}
	\end{figure}

	
	\begin{figure}
	\figurenum{4}
	\epsscale{1}
	\plotone{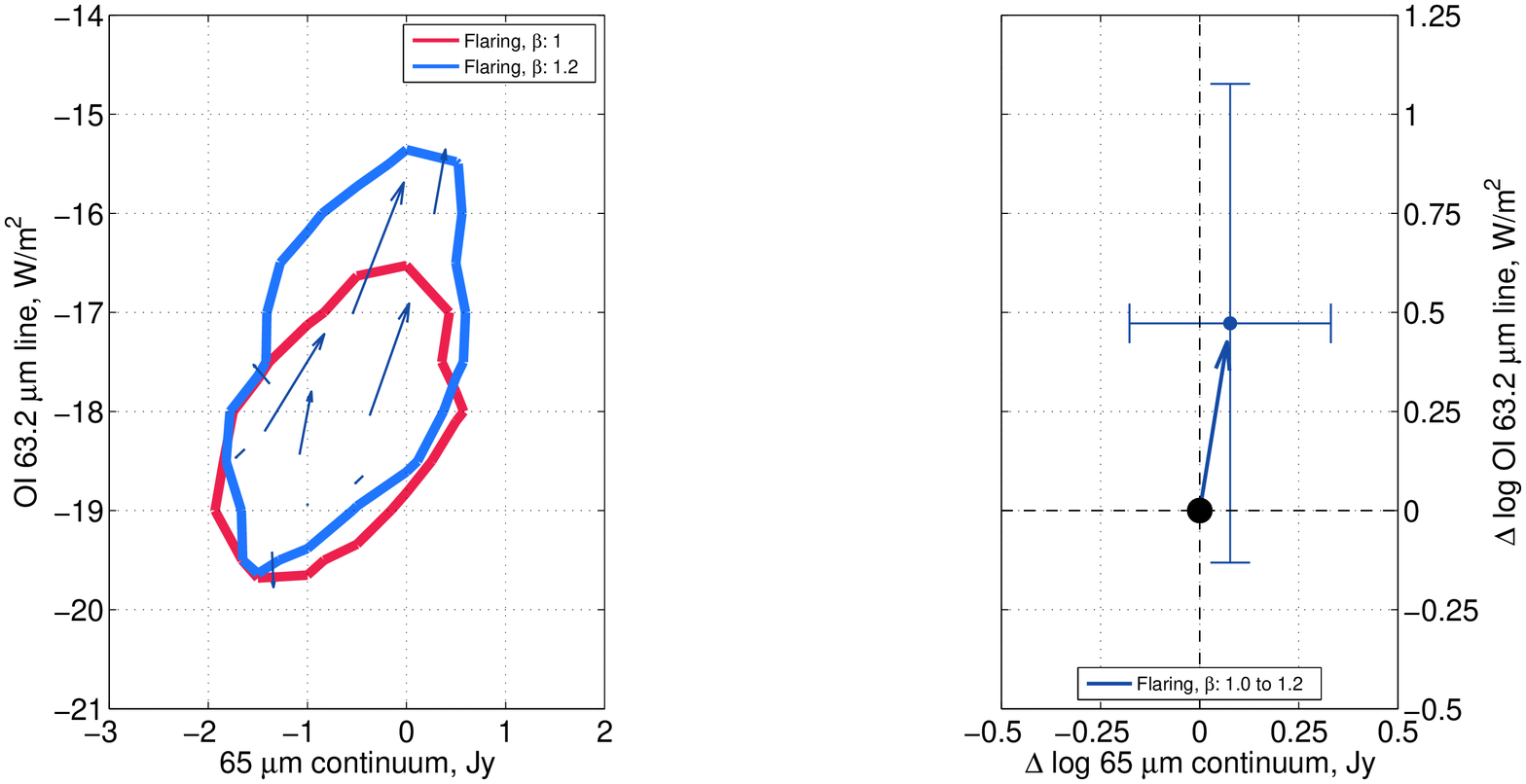}
	\caption{DENT grid predictions for how [O{\sc i}] 63.18 $\mu$m line flux and 65 $\mu$m continuum flux change, with increasing disk flaring ($\beta = 1.0$ and $\beta = 1.2$), for a random subsample (N $\sim$ 5000) of disks around low mass stars ($\le1$ M$_{Sun}$). The left panel shows the regions that contain $67\%$ of the models as a function of disk flaring. The ``evolutionary tracks'' of 10 randomly selected disk models are included for reference. These tracks indicate the path that that particular disk would move if the disk became more flared. The panel on right shows the mean change (``delta'') in  [O{\sc i}] 63.18 $\mu$m line flux and 65 $\mu$m continuum flux, with respect to an initially flatter ($\beta = 1.0$) disk. Arrows point in the direction of increasing disk flaring. Error bars indicate the $1\sigma$ variations in these $\Delta$  [O{\sc i}] 63.18 $\mu$m line flux and $\Delta$ 65 $\mu$m continuum flux during each step in increasing flaring.}
	\end{figure}


	\begin{figure}
	\figurenum{5}
	\epsscale{1}
	\plotone{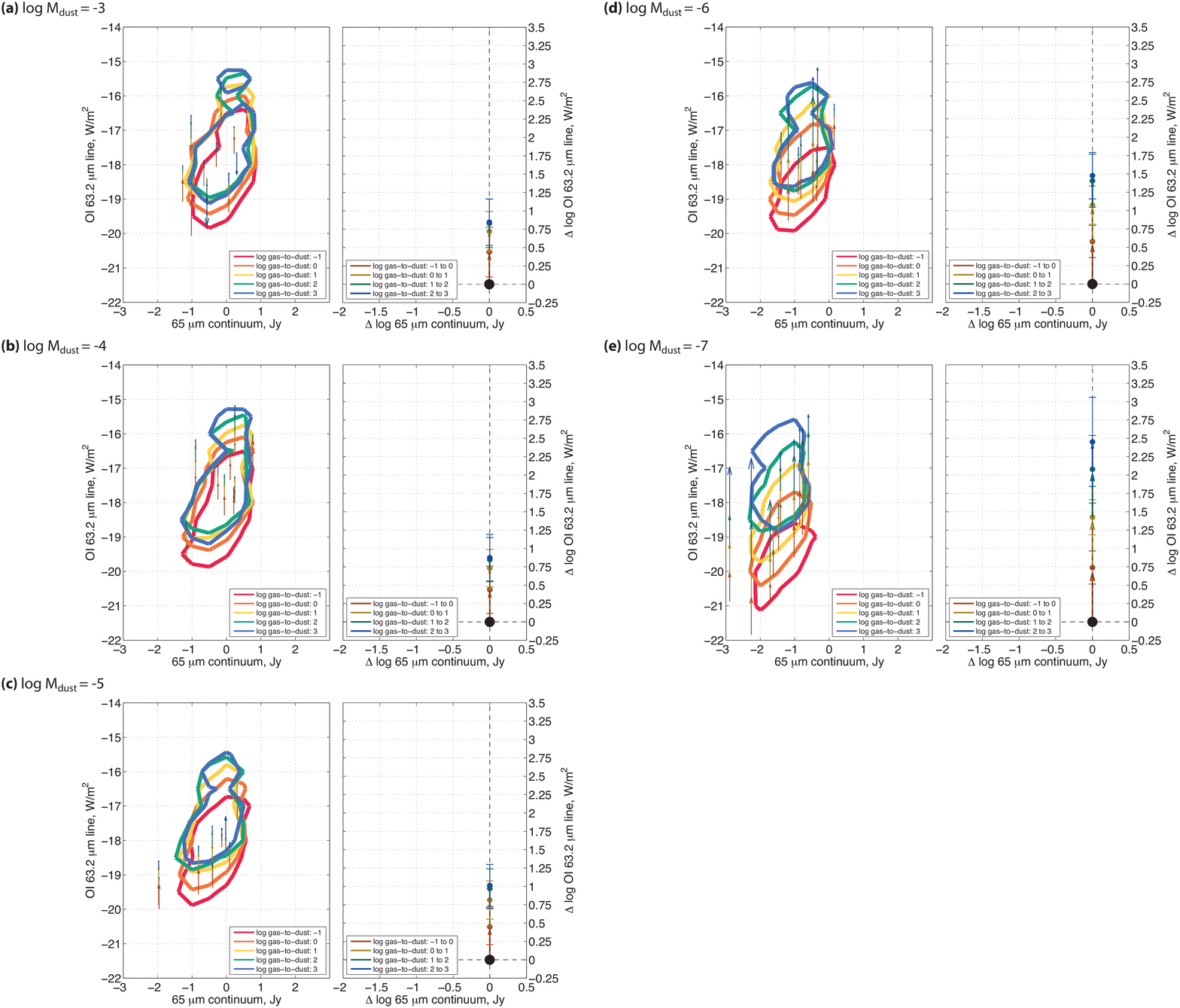}
	\caption{DENT grid predictions for how [O{\sc i}] 63.18 $\mu$m line flux and 65 $\mu$m continuum flux change, with increasing gas-to-dust ratio, \uline{while the dust mass remains fixed}.  \emph{Figures 5a} - \emph{5e} display the effect for different dust masses.  Disk models are sampled at random (N $\sim$ 5000), and consist of only low mass stars ($\le1$ M$_{Sun}$).  The panels on the left show the regions that contain $67\%$ of the models as a function of gas-to-dust ratio. The ``evolutionary tracks'' of 10 randomly selected disk models are included for reference. These tracks indicate the path that that particular disk would move if the gas-to-dust ratio increased. The panels on right show the mean change (``delta'') in  [O{\sc i}] 63.18 $\mu$m line flux and 65 $\mu$m continuum flux, with respect to an initially low gas-to-dust disk. Arrows point in the direction of increasing gas-to-dust ratio (corresponding to increasing gas in these figures). Error bars indicate the $1\sigma$ variations in these $\Delta$  [O{\sc i}] 63.18 $\mu$m line flux and $\Delta$ 65 $\mu$m continuum flux during each step in increasing gas-to-dust ratio.}
	\end{figure}


	\begin{figure}
	\figurenum{6}
	\epsscale{1}
	\plotone{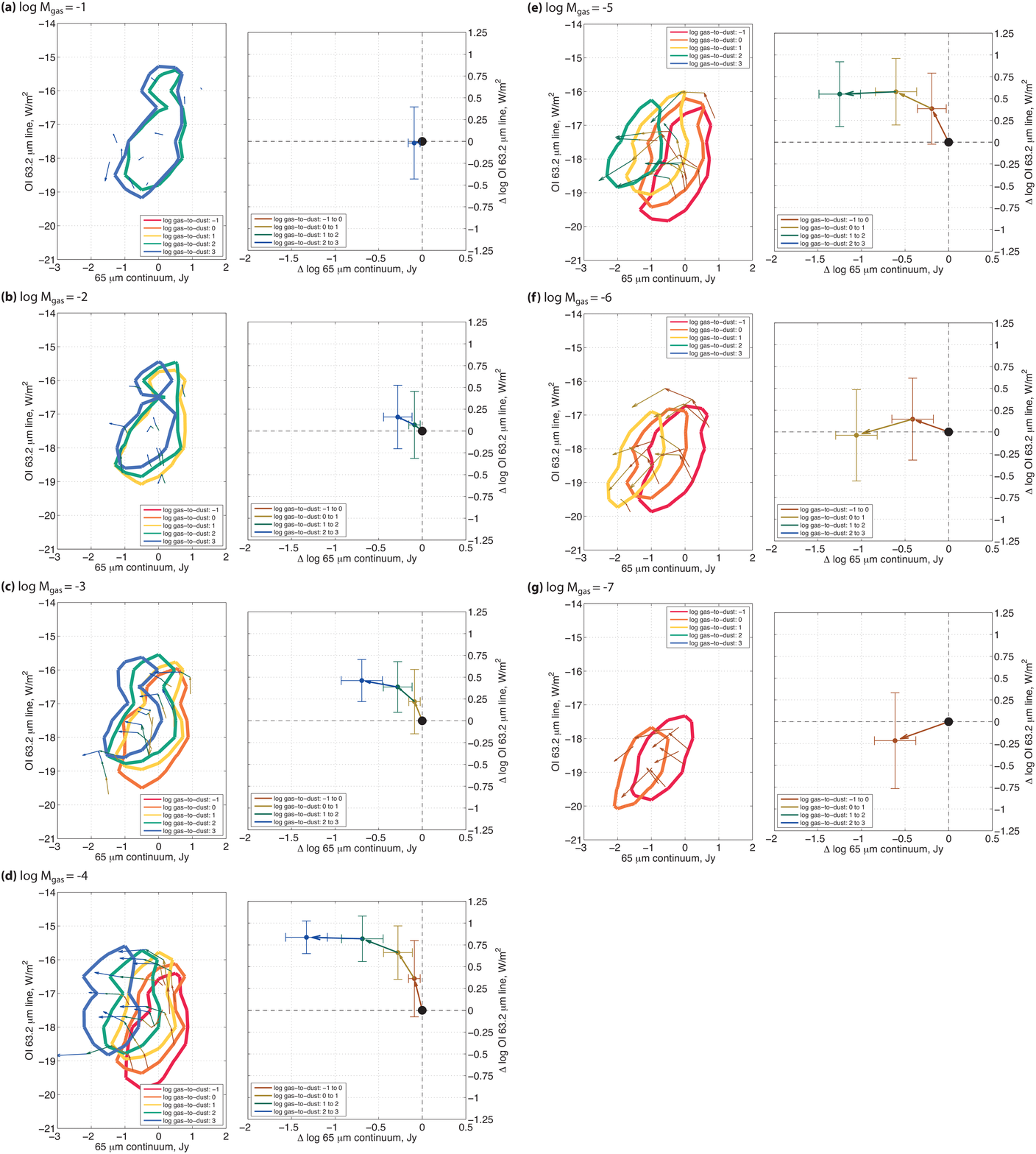}
	\caption{DENT grid predictions for how [O{\sc i}] 63.18 $\mu$m line flux and 65 $\mu$m continuum flux change, with increasing gas-to-dust ratio, \uline{while the gas mass remains fixed}.  \emph{Figures 6a} - \emph{6g} display the effect for different gas masses.  Disk models are sampled at random (N $\sim$ 5000), and consist of only low mass stars ($\le1$ M$_{Sun}$).  The panels on the left show the regions that contain $67\%$ of the models as a function of gas-to-dust ratio. The ``evolutionary tracks'' of 10 randomly selected disk models are included for reference. These tracks indicate the path that that particular disk would move if the gas-to-dust ratio increased. The panels on the right show the mean change (``delta'') in  [O{\sc i}] 63.18 $\mu$m line flux and 65 $\mu$m continuum flux, with respect to an initially low gas-to-dust disk. Arrows point in the direction of increasing gas-to-dust ratio (corresponding to decreasing dust in these figures). Error bars indicate the $1\sigma$ variations in these $\Delta$  [O{\sc i}] 63.18 $\mu$m line flux and $\Delta$ 65 $\mu$m continuum flux during each step in increasing gas-to-dust ratio.}
	\end{figure}

	
	\begin{figure}
	\figurenum{7}
	\epsscale{1}
	\plotone{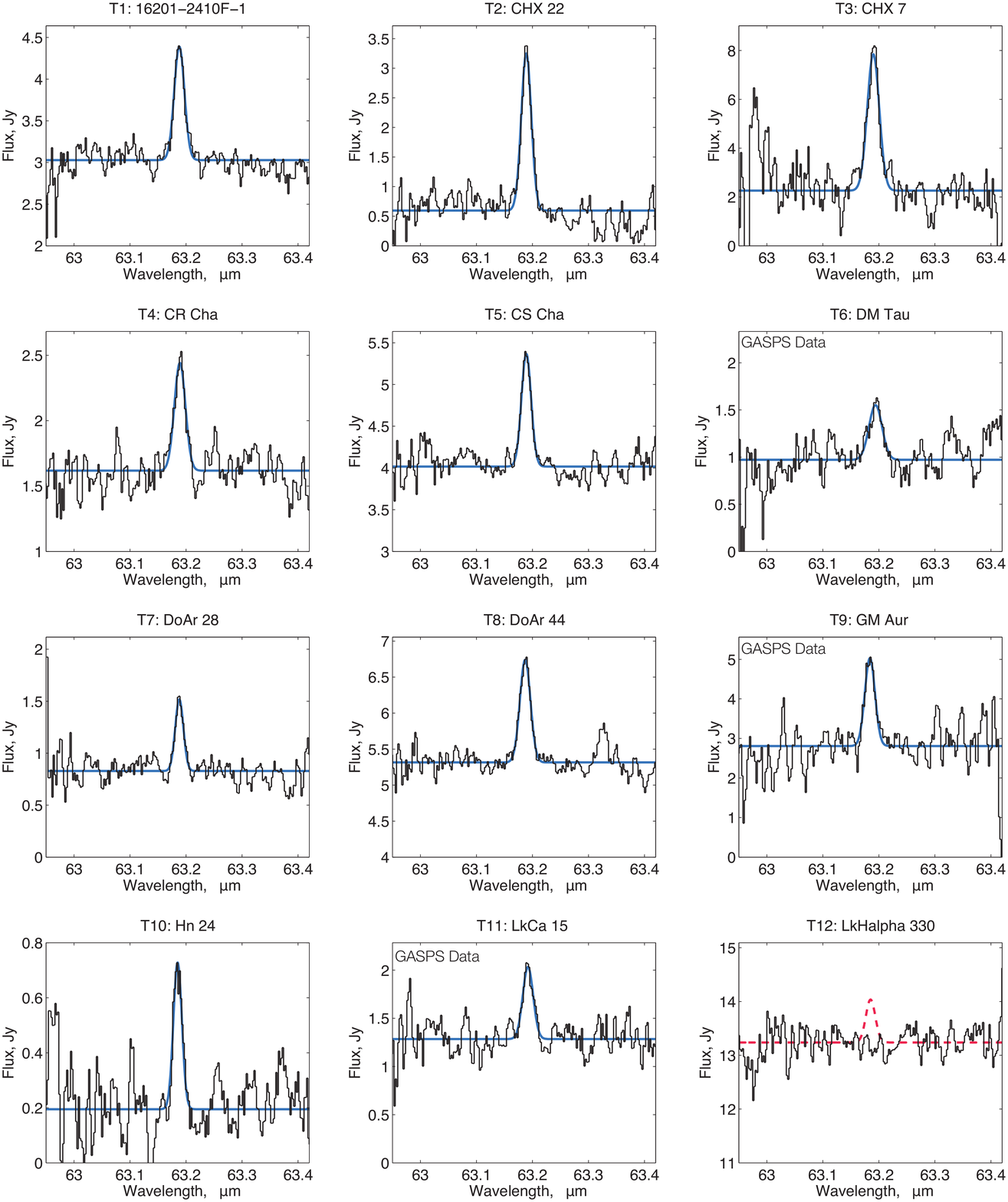}
	\caption{ONLINE ONLY.  \emph{Herschel}/PACS $63 \mu$m spectra.  Blue solid lines indicate the best fit Gaussian line profile for the [O{\sc i}] 63.18 $\mu$m line (as discussed in \emph{Section 3}).  Red dashed lines depict the hypothetical 3-sigma upper limits.  { Observations taken by the GASPS team are indicated by the annotation ``GASPS Data.''  While we re-reduced this data, these observations were previously reported in \citet{howard}, \citet{meeus12}, and \citet{podio12}}}
	\end{figure}
	
	\begin{figure}
	\figurenum{7}
	\epsscale{1}
	\plotone{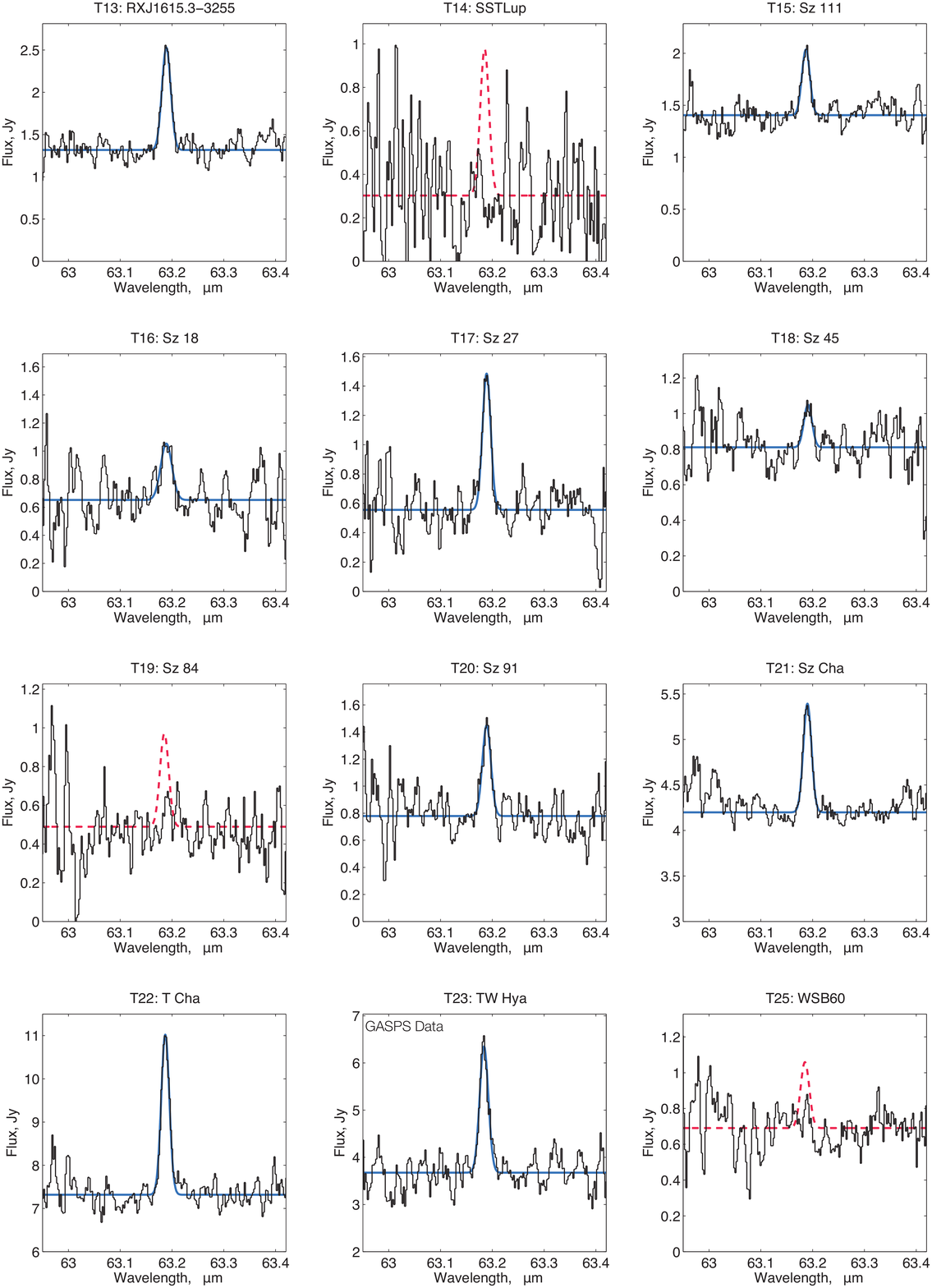}
	\caption{continued.}
	\end{figure}

	\begin{figure}
	\figurenum{7}
	\epsscale{1}
	\plotone{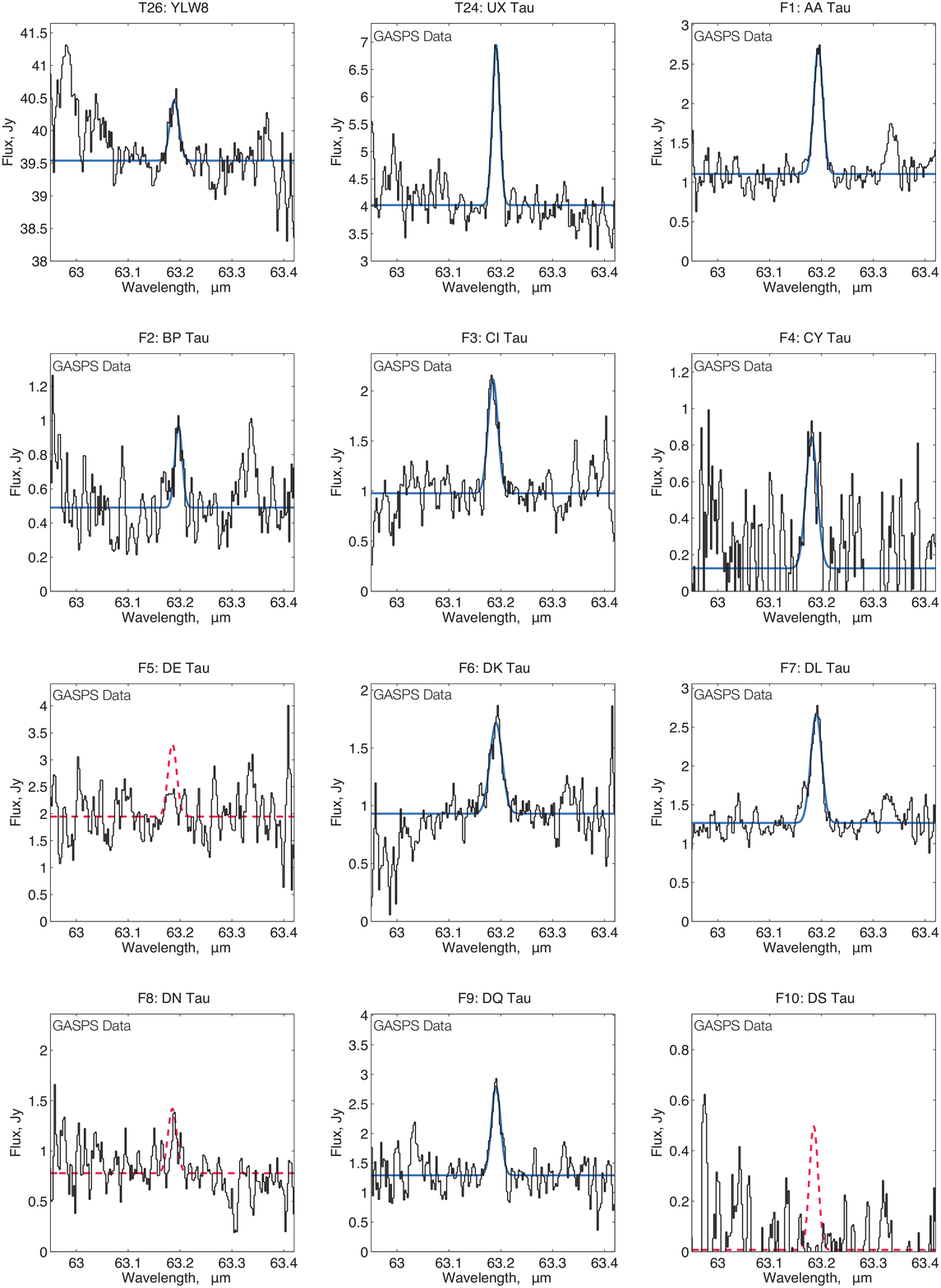}
	\caption{continued.}
	\end{figure}
	
	\begin{figure}
	\figurenum{7}
	\epsscale{1}
	\plotone{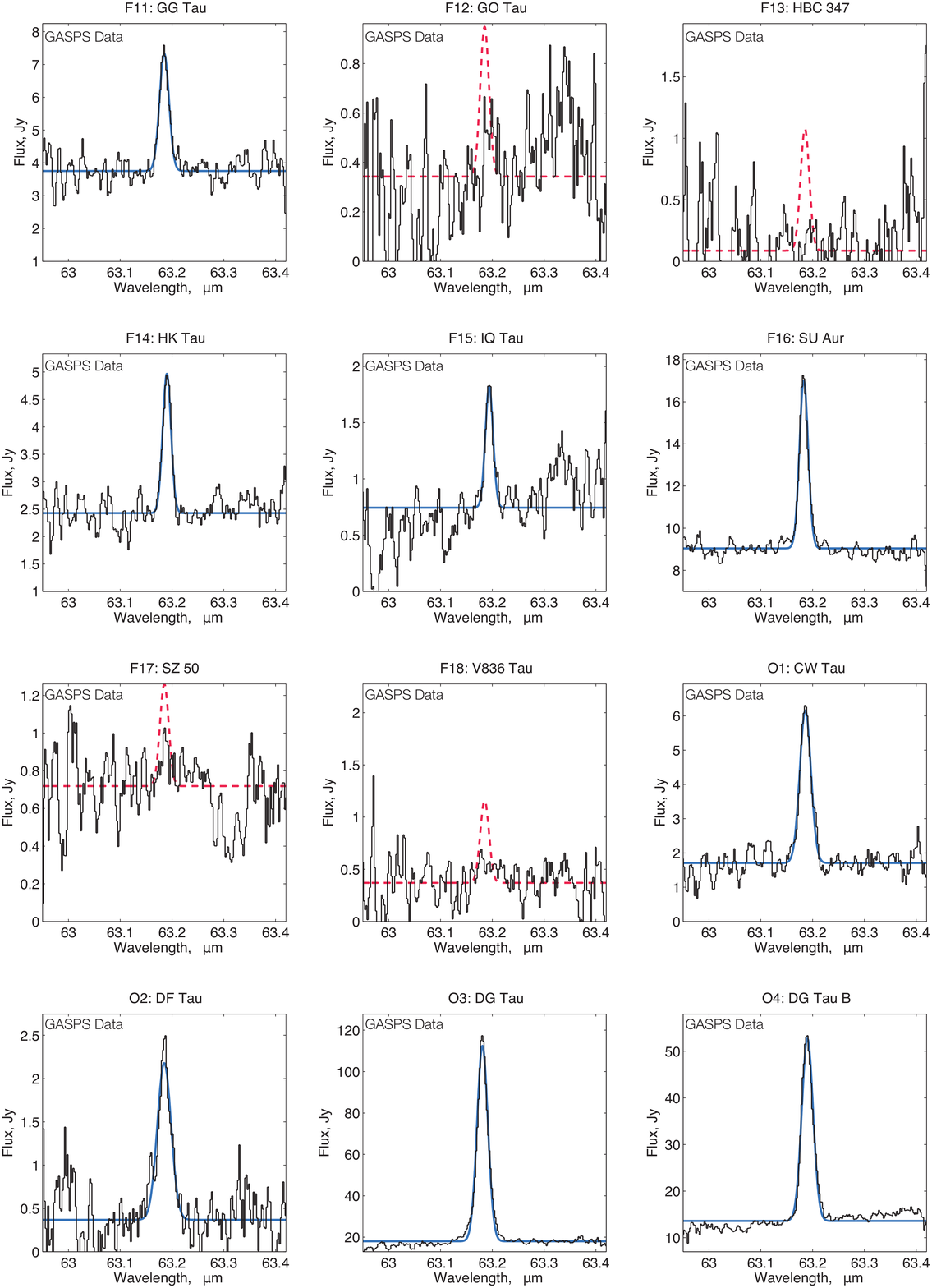}
	\caption{continued.}
	\end{figure}
	
	\begin{figure}
	\figurenum{7}
	\epsscale{1}
	\plotone{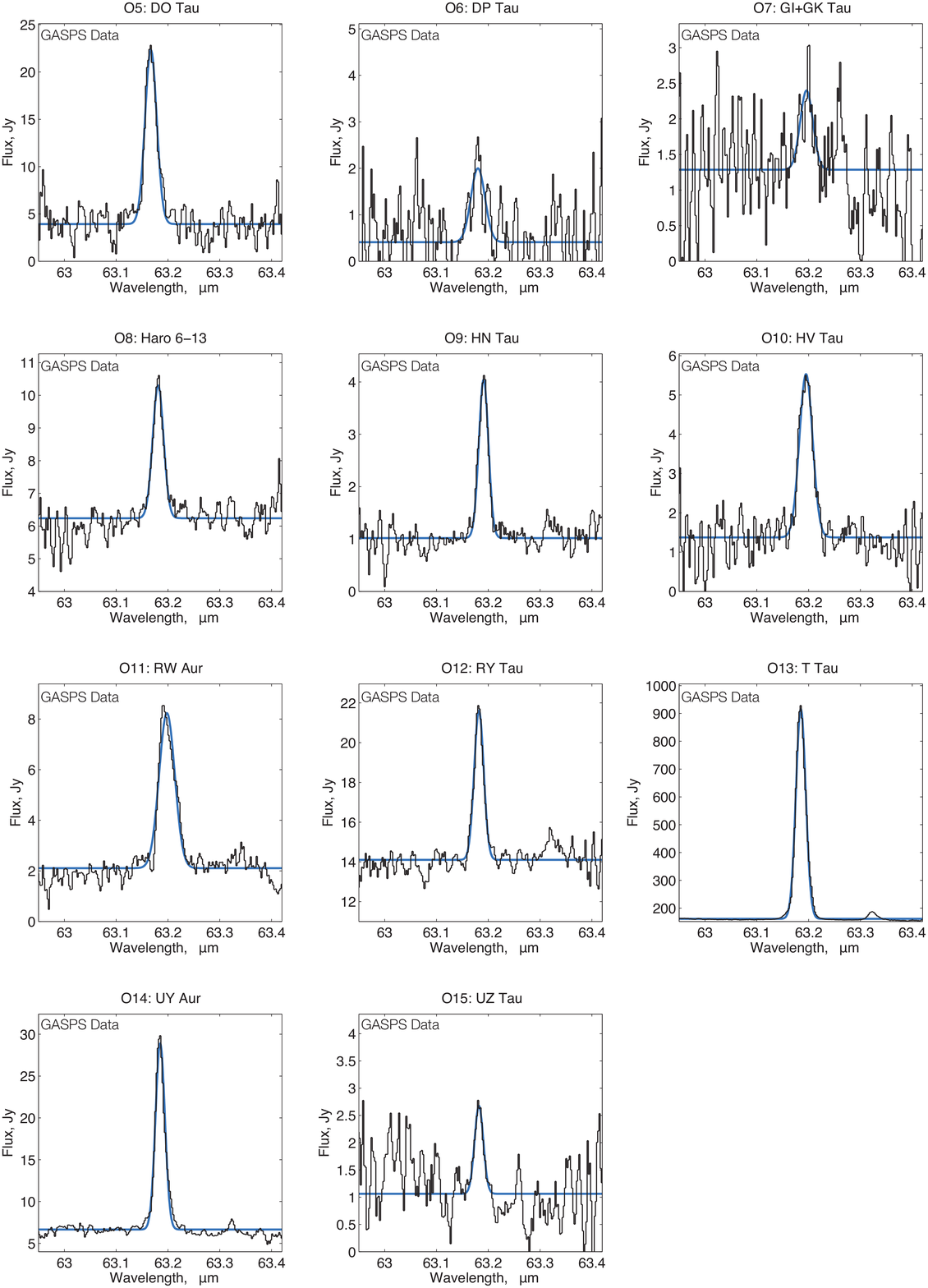}
	\caption{continued.}
	\end{figure}


	\begin{figure}
	\figurenum{8}
	\epsscale{1}
	\plotone{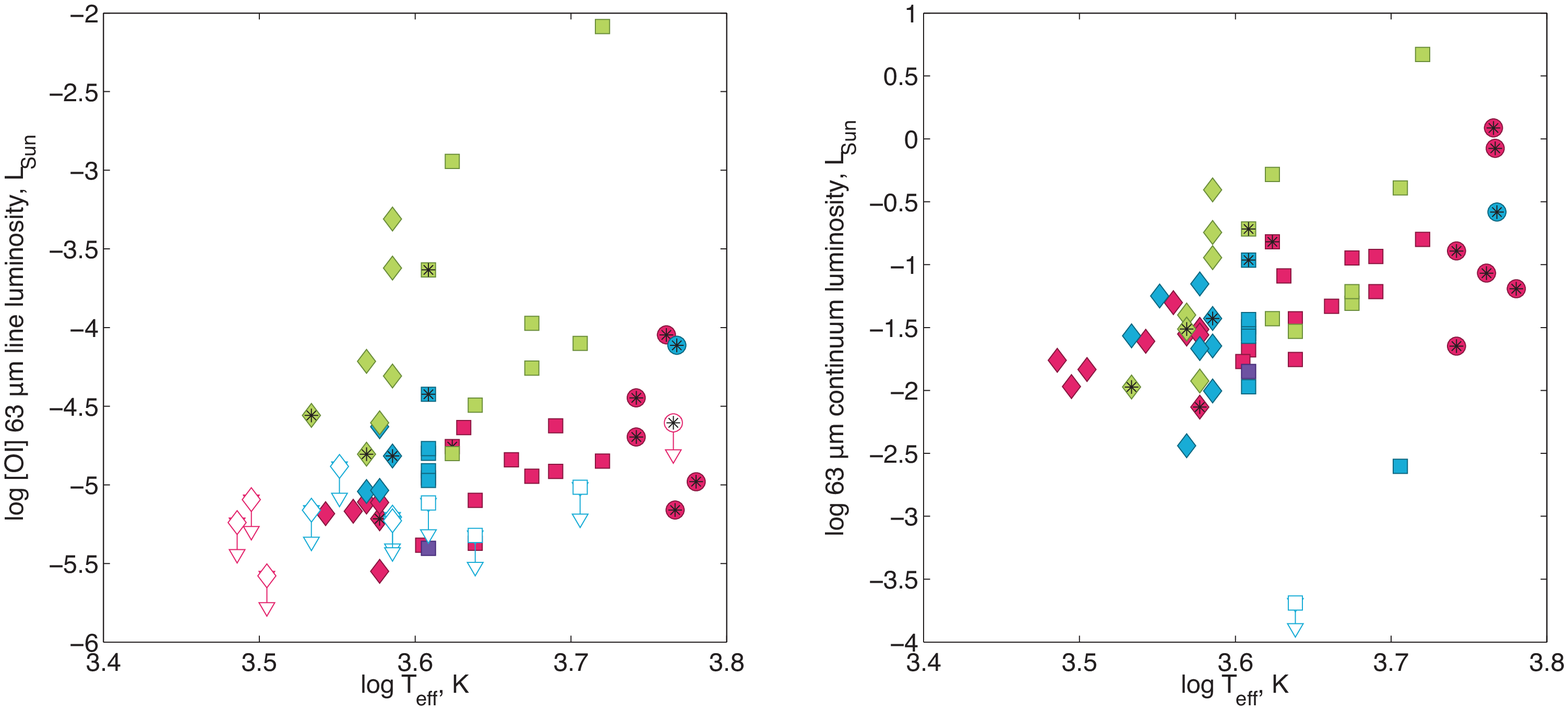}
	\caption{ONLINE ONLY.  [O{\sc i}] 63.18 $\mu$m line luminosity (left) and 63 $\mu$m continuum luminosity (right) vs effective temperature for our sample of transitional disks (red), full disks (blue), and outflow disks (green). $3\sigma$ upper limits are denoted by hollow data points with arrows. Symbols are as in \emph{Figure 1a}.}
	\end{figure}
		
	\begin{figure}
	\figurenum{9}
	\epsscale{1}
	\plotone{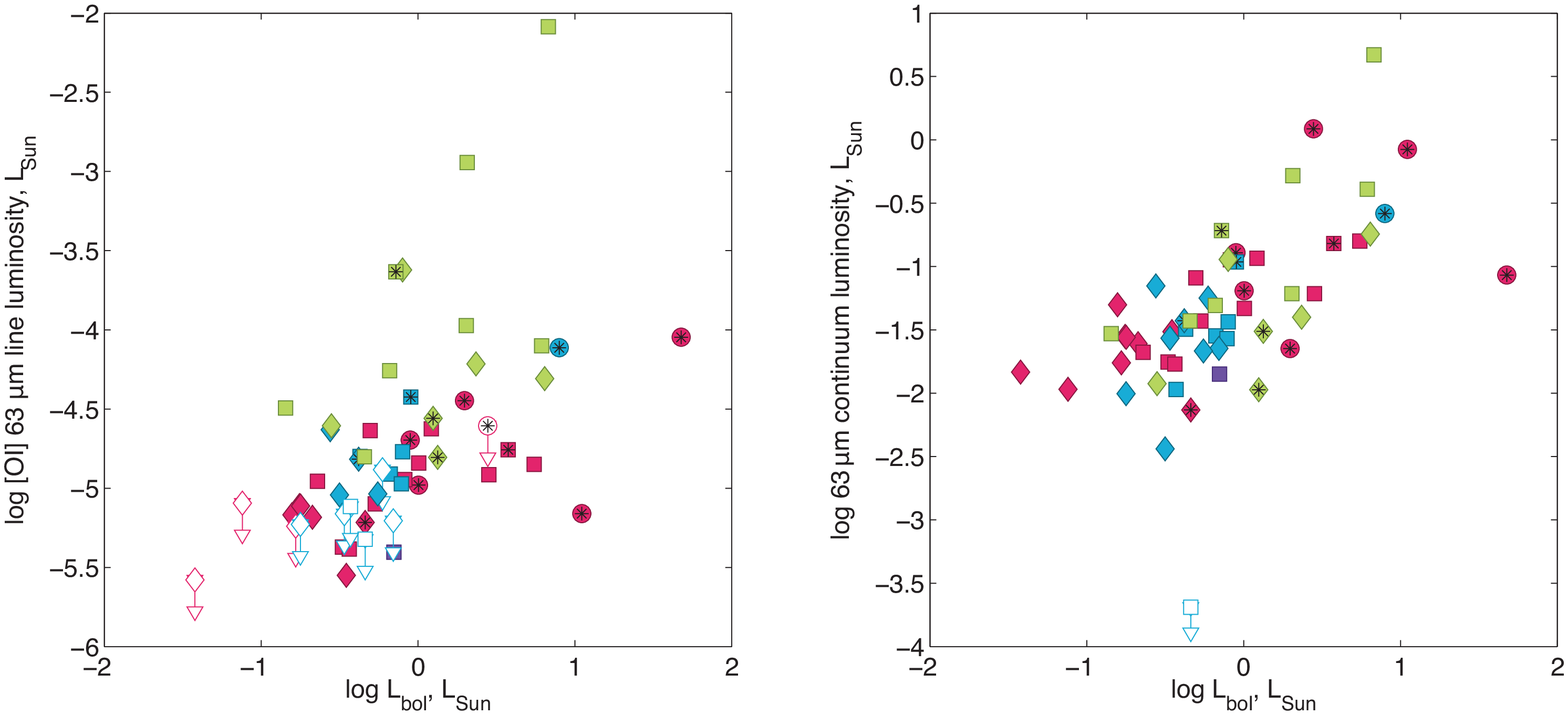}
	\caption{ONLINE ONLY.  [O{\sc i}] 63.18 $\mu$m line luminosity (left) and 63 $\mu$m continuum luminosity (right) vs bolometric luminosity for our sample of transitional disks (red), full disks (blue), and outflow disks (green). $3\sigma$ upper limits are denoted by hollow data points with arrows. Symbols are as in \emph{Figure 1a}. }
	\end{figure}
	
	\begin{figure}
	\figurenum{10}
	\epsscale{1}
	\plotone{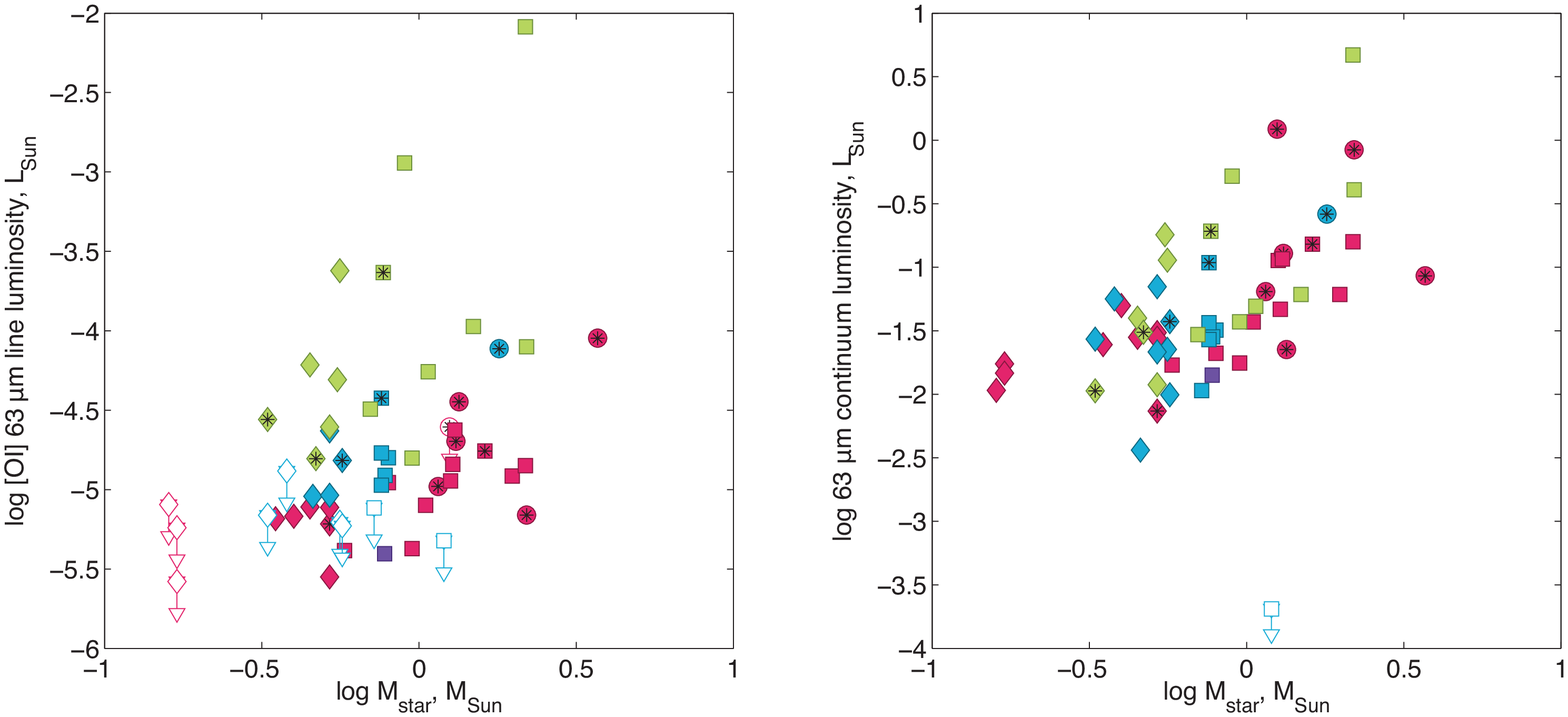}
	\caption{ONLINE ONLY.  [O{\sc i}] 63.18 $\mu$m line luminosity (left) and 63 $\mu$m continuum luminosity (right) vs stellar mass for our sample of transitional disks (red), full disks (blue), and outflow disks (green). $3\sigma$ upper limits are denoted by hollow data points with arrows. Symbols are as in \emph{Figure 1a}. }
	\end{figure}
	
	\begin{figure}
	\figurenum{11}
	\epsscale{1}
	\plotone{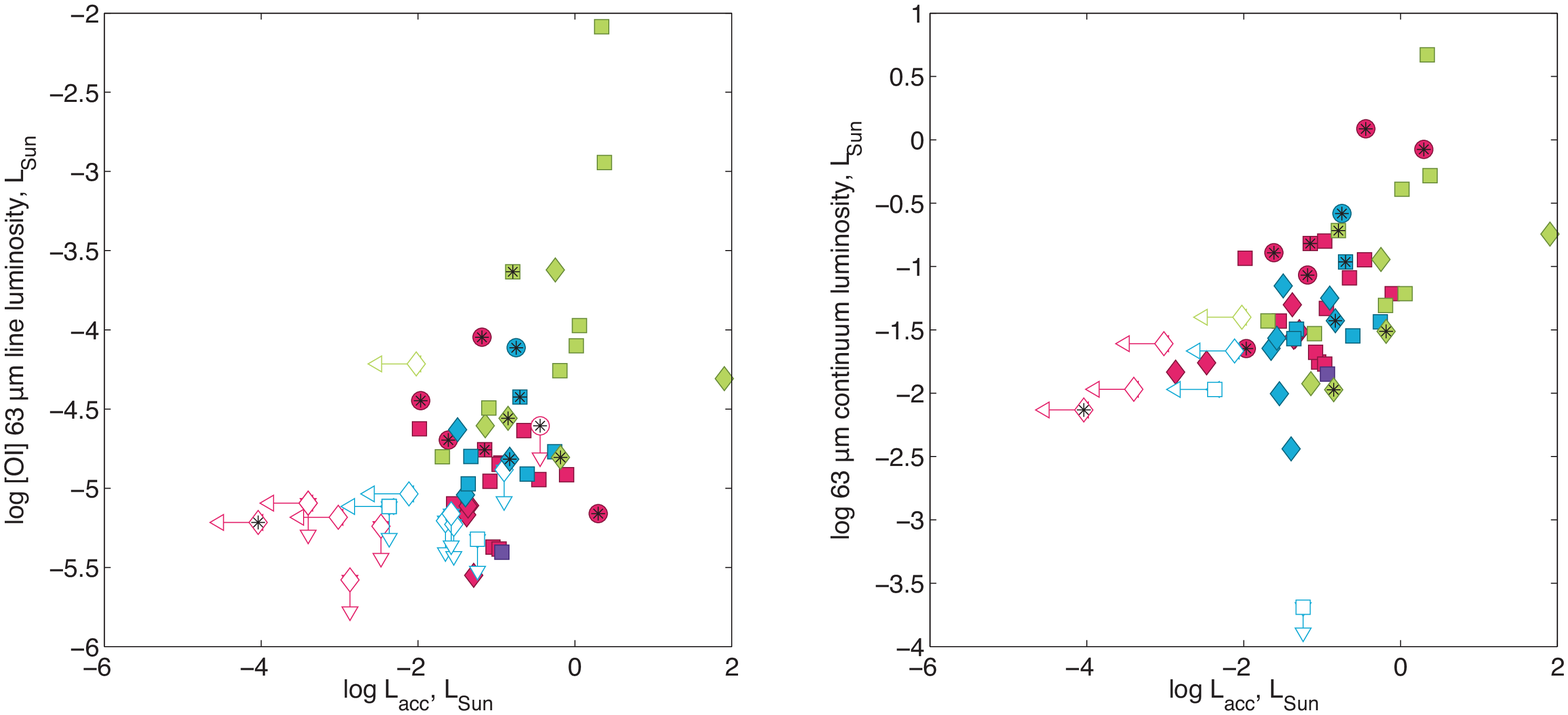}
	\caption{ONLINE ONLY.  [O{\sc i}] 63.18 $\mu$m line luminosity (left) and 63 $\mu$m continuum luminosity (right) vs accretion luminosity for our sample of transitional disks (red), full disks (blue), and outflow disks (green). $3\sigma$ upper limits are denoted by hollow data points with arrows. Symbols are as in \emph{Figure 1a}. }
	\end{figure}
	
	\begin{figure}
	\figurenum{12}
	\epsscale{1}
	\plotone{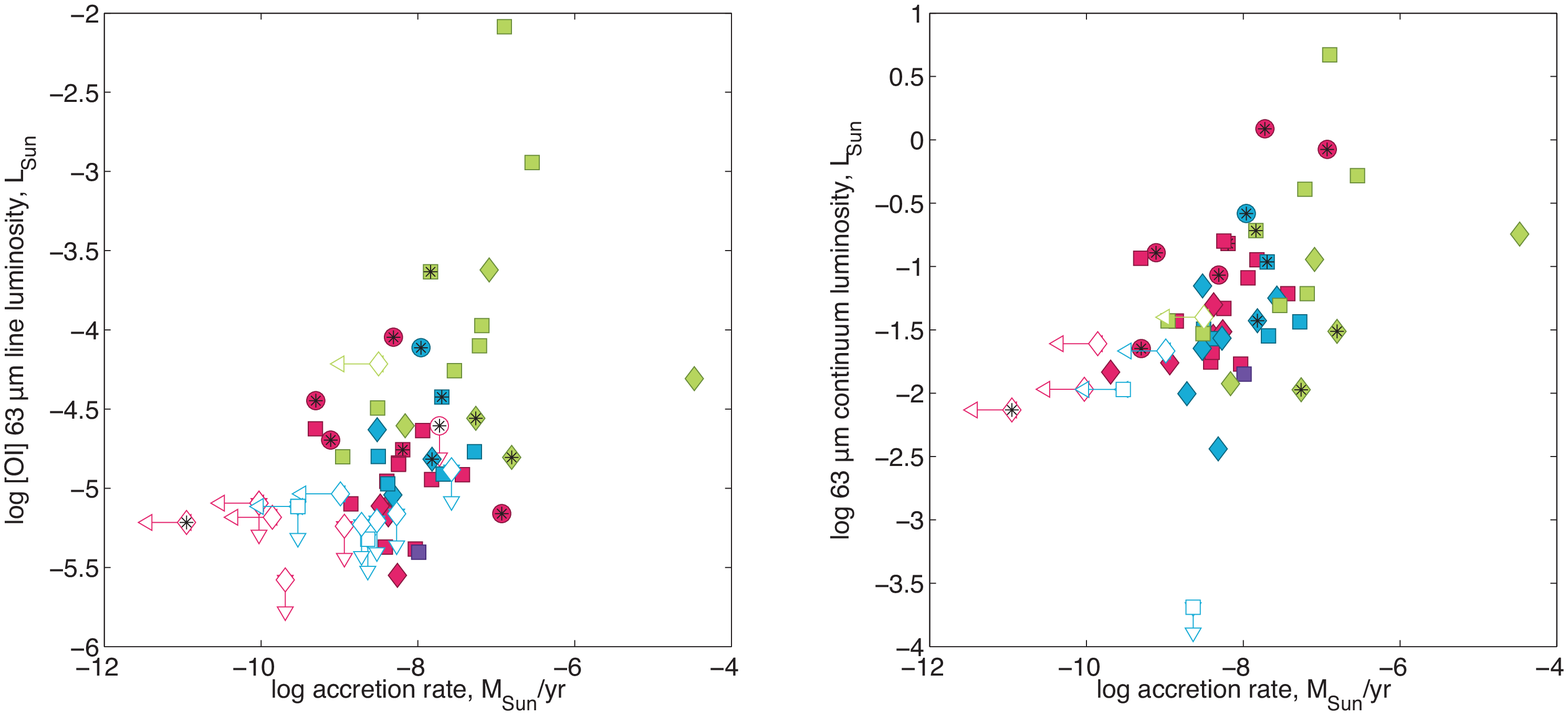}
	\caption{ONLINE ONLY.  [O{\sc i}] 63.18 $\mu$m line luminosity (left) and 63 $\mu$m continuum luminosity (right) vs accretion rate for our sample of transitional disks (red), full disks (blue), and outflow disks (green). $3\sigma$ upper limits are denoted by hollow data points with arrows. Symbols are as in \emph{Figure 1a}. }
	\end{figure}
	
	\begin{figure}
	\figurenum{13}
	\epsscale{1}
	\plotone{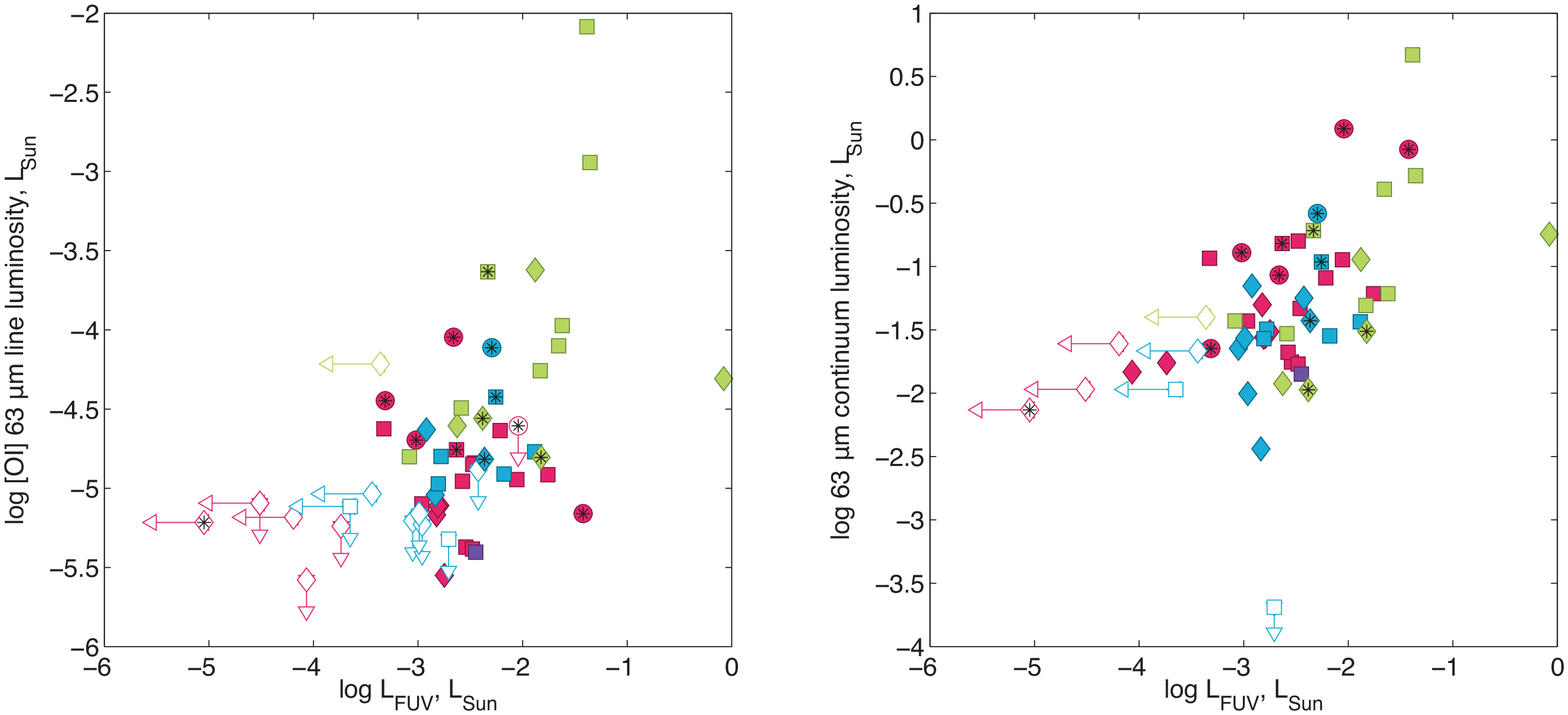}
	\caption{ONLINE ONLY.  [O{\sc i}] 63.18 $\mu$m line luminosity (left) and 63 $\mu$m continuum luminosity (right) vs FUV excess for our sample of transitional disks (red), full disks (blue), and outflow disks (green). $3\sigma$ upper limits are denoted by hollow data points with arrows. Symbols are as in \emph{Figure 1a}. }
	\end{figure}

	\begin{figure}
	\figurenum{14}
	\epsscale{1}
	\plotone{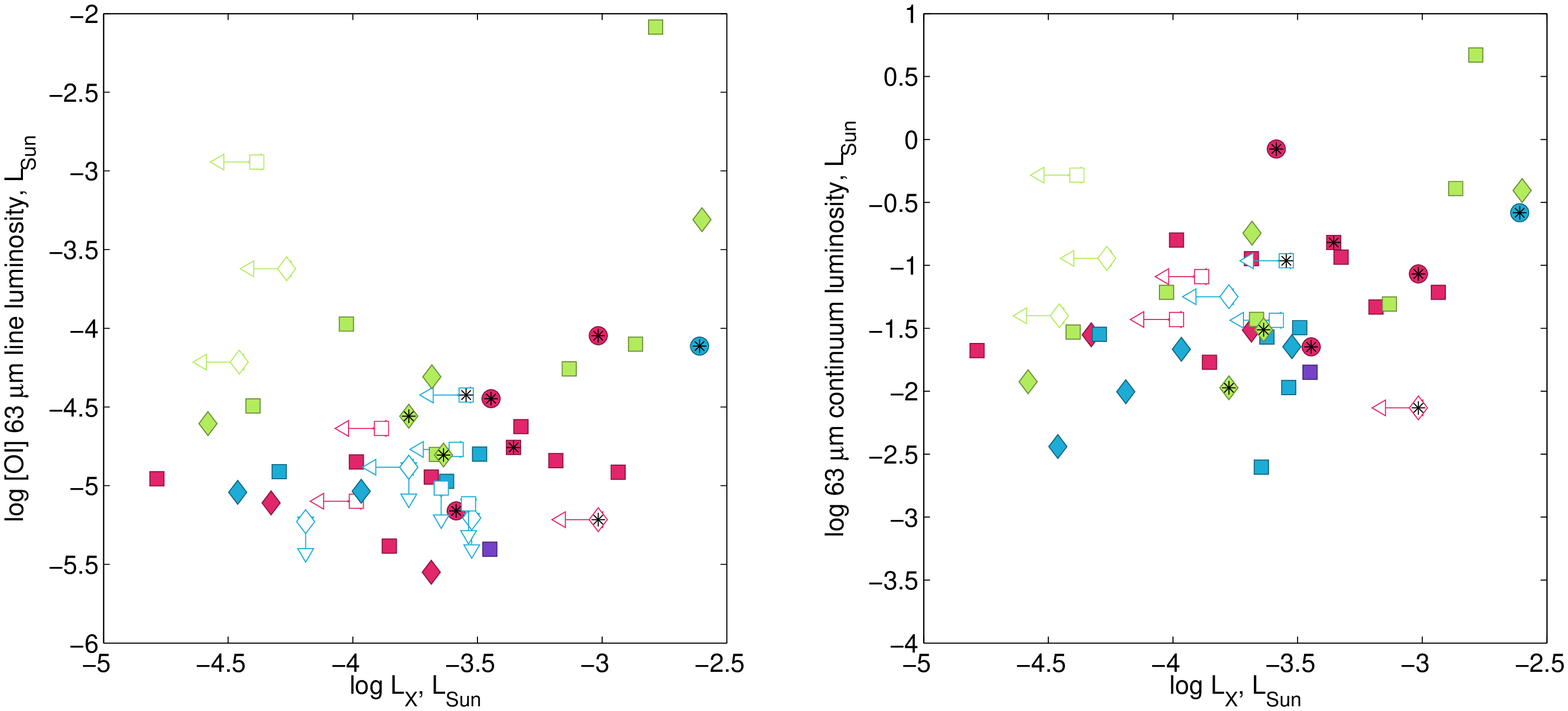}
	\caption{ONLINE ONLY.  [O{\sc i}] 63.18 $\mu$m line luminosity (left) and 63 $\mu$m continuum luminosity (right) vs X-ray luminosity for our sample of transitional disks (red), full disks (blue), and outflow disks (green). $3\sigma$ upper limits are denoted by hollow data points with arrows. Symbols are as in \emph{Figure 1a}. }
	\end{figure}

	\begin{figure}
	\figurenum{15}
	\epsscale{1}
	\plotone{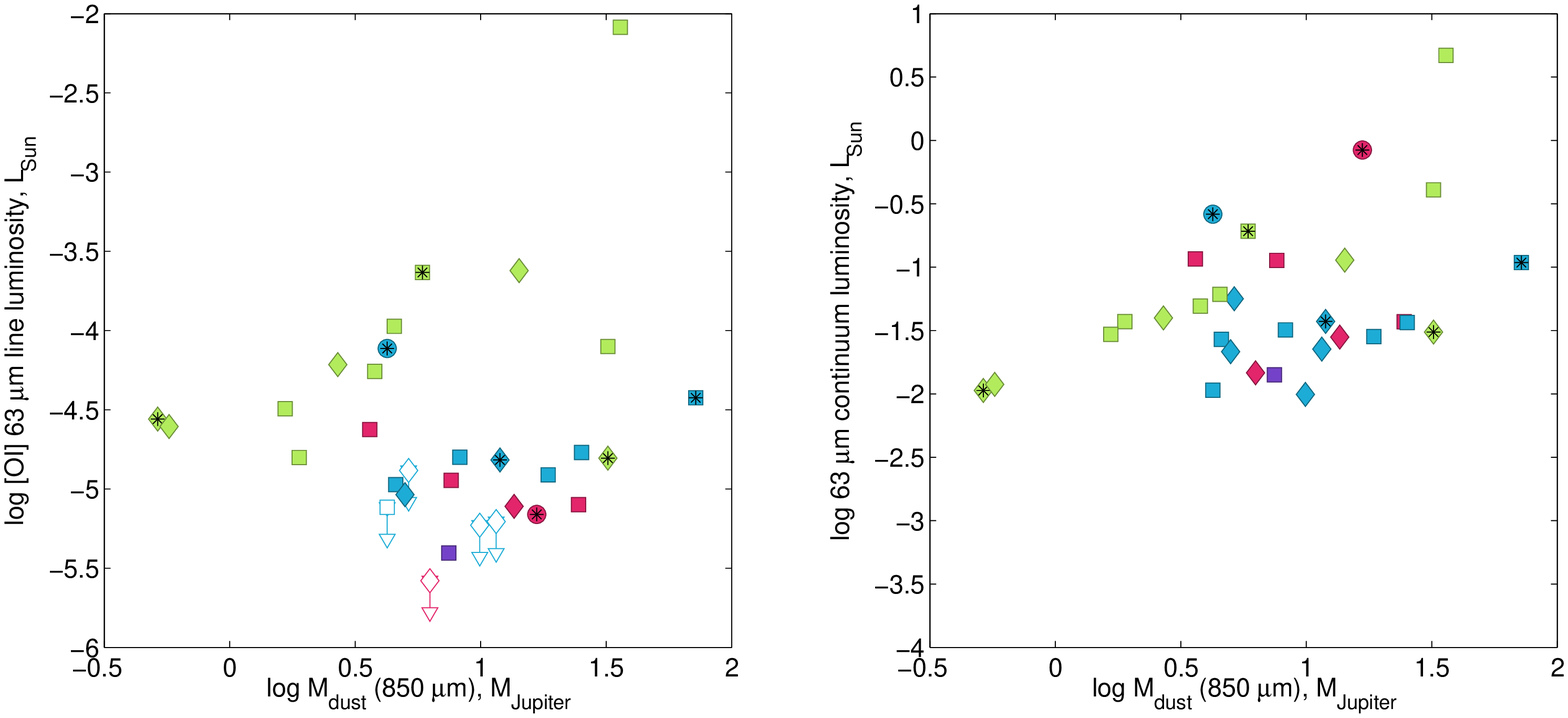}
	\caption{ONLINE ONLY.  [O{\sc i}] 63.18 $\mu$m line luminosity (left) and 63 $\mu$m continuum luminosity (right) vs dust mass (derived from 850 $\mu$m photometry) for our sample of transitional disks (red), full disks (blue), and outflow disks (green). $3\sigma$ upper limits are denoted by hollow data points with arrows. Symbols are as in \emph{Figure 1a}. }
	\end{figure}
	
	\begin{figure}
	\figurenum{16}
	\epsscale{1}
	\plotone{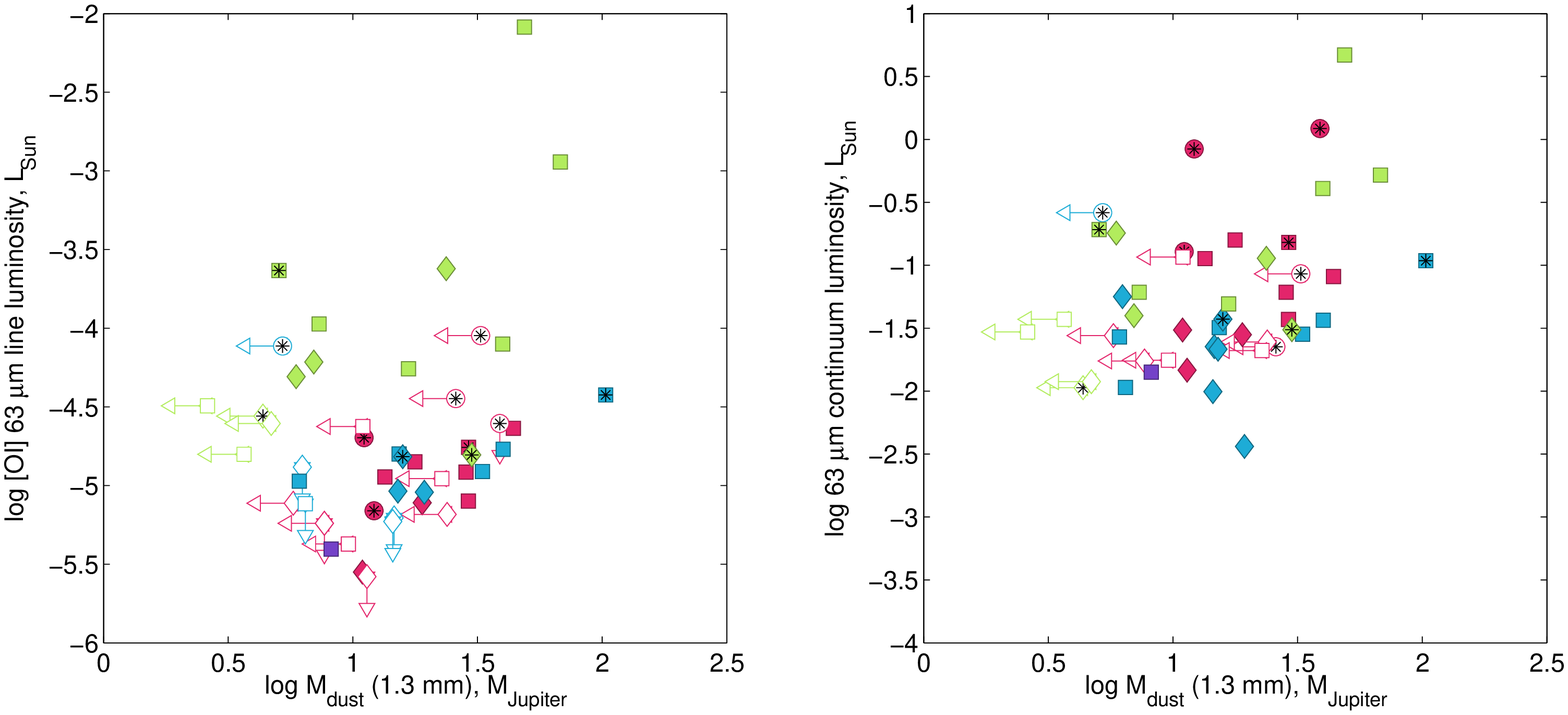}
	\caption{ONLINE ONLY.  [O{\sc i}] 63.18 $\mu$m line luminosity (left) and 63 $\mu$m continuum luminosity (right) vs dust mass (derived from 1.3 mm photometry) for our sample of transitional disks (red), full disks (blue), and outflow disks (green). $3\sigma$ upper limits are denoted by hollow data points with arrows. Symbols are as in \emph{Figure 1a}. }
	\end{figure}

\clearpage

\begin{deluxetable}{ccccccccclc}
\tabletypesize{\footnotesize}
\tablewidth{0pt}

\tablecaption{\emph{Herschel}/PACS Sample and Observations}
\rotate
\tablenum{1}

\tablehead{\colhead{ID} & \colhead{Name} & \colhead{RA} & \colhead{Dec} & \colhead{Association} & \colhead{SpTy} & \colhead{Ref} & \colhead{Multiplicity} & \colhead{Ref} & \colhead{OBSID} & \colhead{Duration [s]}} 

\startdata
 \hline
  \multicolumn{11}{c}{\emph{Transition disks}} \\
 \hline
T1 & 16201-2410F-1* & 16 23 09.23  & -24 17 04.70  & Ophiuchus & G0 & F09 & 1 & --- & 1342250127 & 8212 \\
T2 & CHX 22* & 11 12 42 69  & -77 22 23.00  & Chameleon & G8 & L07 &  2  & D13 & 1342233474 & 8212 \\
T3 & CHX 7* & 11 06 15 41  & -77 21 56.90  & Chameleon & G5 & L07 &  --- & --- & 1342233477 & 8212 \\
T4 & CR Cha & 10 59 06 99  & -77 01 40.40  & Chameleon & K2 & E11 &  2 & G07 & 1342232614 & 8212 \\
T5 & CS Cha* & 11 02 24 91  & -77 33 35.70  & Chameleon & K6 & E11 &  --- & --- & 1342233480 & 8212 \\
T6 & DM Tau  & 04 33 48.72 & +18 10 09.99 & Taurus & M1 & KH95 &  --- & --- & 1342225825$^{\star}$ & 6628 \\
T7 & DoAr 28 & 16 26 47.42  & -23 14 52.20  & Ophiuchus & K5 & M92 &  --- & --- & 1342241707 & 8212 \\
T8 & DoAr 44 & 16 31 33.46  & -24 27 37.30  & Ophiuchus & K3 & M92 &  --- & --- & 1342250578 & 8212 \\
T9 & GM Aur  & 04 55 10.99 & +30 21 59.25 & Taurus & K5.5 & E11 &  --- & --- & 1342191357$^{\star}$ & 1252 \\
T10 & Hn 24* & 13 04 55 75  & -77 39 49.50  & Chameleon & M0.5 & M10 &  2 & B96 & 1342235656 & 8212 \\
T11 & LkCa 15  & 04 39 17.80 & +22 21 03.48 & Taurus & K5 & KH95 &  --- & --- & 1342225798$^{\star}$ & 6628 \\
T12 & LkHalpha 330* & 3 45 48 28  & +32 24 11.90  & Perseus & G3 & BR07 & --- & --- & 1342238377 & 8212 \\
T13 & RXJ1615.3-3255 & 16 15 20 23  & -32 55 05.10  & Lupus & K4 & M10 &  --- & --- & 1342229825 & 8212 \\
T14 & SSTLup & 16 10 29.60  & -39 22 15.00  & Lupus & M5 & M10 & ---  & --- & 1342241709 & 8212 \\
T15 & Sz 111 & 16 08 54 69  & -39 37 43.10  & Lupus & M1.5 & H94 & --- & --- & 1342220928 & 8212 \\
T16 & Sz 18 & 11 07 19 15  & -76 03 04.80  & Chameleon & M2.5 & L07 &  --- & --- & 1342232585 & 8212 \\
T17 & Sz 27 & 11 08 39 05  & -77 16 04.20  & Chameleon & K8 & L07 & --- & --- & 1342233476 & 8212 \\
T18 & Sz 45 & 11 17 37 01  & -77 04 38.10  & Chameleon & M0.5 & L07 &  --- & --- & 1342233475 & 8212 \\
T19 & Sz 84 & 15 58 02 53  & -37 36 02.70  & Lupus & M5.5 & M10 & --- & --- &  1342229826 & 8212 \\
T20 & Sz 91 & 16 07 11 61  & -39 03 47.10  & Lupus & M0.5 & H94 &  --- & --- & 1342229827 & 8212 \\
T21 & Sz Cha & 10 58 16 77  & -77 17 17.10  & Chameleon & K0 & E11 &  2 & D13 & 1342233478 & 8212 \\
T22 & T Cha* & 11 57 13 53  & -79 21 31.50  & Chameleon & G8 & BR07 &  --- & --- & 1342232294 & 2068 \\
T23 & TW Hya & 11 01 52 03 & -34 42 18.60 & TW Hydra & K6 & R06 &  --- & --- & 1342187127$^{\star}$ & 1252 \\
T24 & UX Tau & 04 30 03.76 & +18 13 49.88 & Taurus & K2 & KH95 &  3 & M06 & 1342214357$^{\star}$ & 1252 \\
T25 & WSB60 & 16 28 16.51  & -24 36 58.00  & Ophiuchus & M4.5 & WMRG05 &  --- & --- & 1342250128 & 8212 \\
T26 & YLW8* & 16 27 10 28  & -24 19 12.70  & Ophiuchus & G2.5 & BR07 &  2 & M06 & 1342229824 & 2068 \\
 \hline
  \multicolumn{11}{c}{\emph{Full disks}} \\
 \hline
F1 & AA Tau & 04 34 55.42 & +24 28 53.16 & Taurus & K7 & KH95 &  --- & --- & 1342225758$^{\star}$ & 6628 \\
F2 & BP Tau* & 04 19 15.84 & +29 06 26.94 & Taurus & K7 & KH95 &  --- & --- & 1342225728$^{\star}$ & 3316 \\
F3 & CI Tau & 04 33 52.00 & +22 50 30.18 & Taurus & K7 & KH95 &  --- & --- & 1342192125$^{\star}$ & 1252 \\
F4 & CY Tau & 04 17 33.73 & +28 20 46.85 & Taurus & M1 & KH95 &  --- & --- & 1342192794$^{\star}$ & 1252 \\
F5 & DE Tau & 04 21 55.64 & +27 55 06.06 & Taurus & M2 & KH95 &  --- & --- & 1342192797$^{\star}$ & 1252 \\
F6 & DK Tau & 12 53 17.23 & -77 07 10.70 & Taurus & K7 & KH95 & 2 & WG01 &  1342225732$^{\star}$ & 3316 \\
F7 & DL Tau & 04 33 39.06 & +25 20 38.23 & Taurus & K7 & KH95 &  --- & --- & 1342225800$^{\star}$ & 6628 \\
F8 & DN Tau & 04 35 27.37 & +24 14 58.94 & Taurus & M0 & KH95 &  --- & --- & 1342225757$^{\star}$ & 3316 \\
F9 & DQ Tau* & 04 46 53.04 & +17 00 00.50 & Taurus & M0 & KH95 &  2 & AW05 & 1342225806$^{\star}$ & 1252 \\
F10 & DS Tau & 04 47 48.21 & 29 25 13.83 & Taurus & K5 & KH95 &  2 & AW05 & 1342225851$^{\star}$ & 3316 \\
F11 & GG Tau* & 04 32 30.35  & +17 31 40.60 & Taurus & K7 & KH95 &  2 & WG01 & 1342192121$^{\star}$ & 1252 \\
F12 & GO Tau & 04 43 03.10 & +25 20 18.75 & Taurus & M0 & KH95 &  --- & --- & 1342225826$^{\star}$ & 3316 \\
F13 & HBC 347 & 03 29 38.24 & +24 30 37.74 & Taurus & --- & --- &  --- & --- & 1342192136$^{\star}$ & 1252 \\
F14 & HK Tau & 04 31 50.67 & +24 24 17.44 & Taurus & M0.5 & KH95 &  2 & WG01 & 1342225736$^{\star}$ & 3316 \\
F15 & IQ Tau & 04 29 51.56 & +26 06 44.89 & Taurus & M0.5 & KH95 &  --- & --- & 1342225733$^{\star}$ & 3316 \\
F16 & SU Aur & 04 55 59.38 & +30 34 01.56 & Taurus & G2 & KH95 &  --- & --- & 1342217844$^{\star}$ & 3316 \\
F17 & SZ 50 & 13 00 55.36 & -77 10 22.10 & Chameleon & M3 & HH92 &  --- & --- & 1342226008$^{\star}$ & 3316 \\
F18 & V836 Tau & 05 03 06.60 & +25 23 19.71 & Taurus & K7 & KH95 & --- & --- &  1342227634$^{\star}$ & 3316 \\
 \hline
  \multicolumn{11}{c}{\emph{Outflow disks}} \\
 \hline
O1 & CW Tau & 04 14 17.00 & +28 10 57.83 & Taurus & K3 & KH95 &  --- & --- & 1342216221$^{\star}$ & 1252 \\
O2 & DF Tau* & 04 27 02.80 & +25 42 22.30 & Taurus & M3 & KH95 &  2 & P08 & 1342190359$^{\star}$ & 1252 \\
O3 & DG Tau & 04 27 04.698 & +26 06 16.31 & Taurus & K6 & KH95 &  --- & --- & 1342190382$^{\star}$ & 1252 \\
O4 & DG Tau B & 04 27 02.41 & +26 05 31.76 & Taurus & M0 & KH95 & --- & --- &  1342192798$^{\star}$ & 1252 \\
O5 & DO Tau & 04 38 28.58 & +26 10 49.44 & Taurus & M0 & KH95 & --- & --- &  1342190385$^{\star}$ & 1252 \\
O6 & DP Tau & 04 42 37.56 & +25 15 39.62 & Taurus & M0.5 & KH95 &  --- & --- & 1342191362$^{\star}$ & 1252 \\
O7 & GI/GK Tau & 04 55 10.85 & +30 22 01.69 & Taurus & K6 & KH95 &  2 & AW05 & 1342225760$^{\star}$ & 1252 \\
O8 & Haro 6-13 & 04 32 15.41 & +24 28 59.75 & Taurus & M0 & RM12 &  --- & --- & 1342192128$^{\star}$ & 1252 \\
O9 & HN Tau & 04 33 39.44 & +17 51 52.24 & Taurus & K5 & KH95 &  2 & WG01 & 1342225796$^{\star}$ & 3316 \\
O10 & HV Tau & 04 38 35.38 & +26 10 37.80 & Taurus & M1 & KH95 & 2 & WG01 &  1342225801$^{\star}$ & 3316 \\
O11 & RW Aur & 05 07 49.41 & +30 24 07.65 & Taurus & K3 & KH95 &  2 & WG01 & 1342191359$^{\star}$ & 1252 \\
O12 & RY Tau & 04 21 57.40 & +28 26 35.54 & Taurus & K1 & KH95 &  --- & --- & 1342190361$^{\star}$ & 1252 \\
O13 & T Tau & 04 21 59.30 & +19 32 08.53 & Taurus & K0 & KH95 &  2 & WG01 & 1342190353$^{\star}$ & 1252 \\
O14 & UY Aur *& 04 51 47.15 & +30 47 14.44 & Taurus & K7 & KH95 & 2 & WG01 &  1342215699$^{\star}$ & 1252 \\
O15 & UZ Tau* & 04 32 42.73 & +25 52 35.00 & Taurus & M1 & KH95 &  4 & WG01 & 1342192131$^{\star}$ & 1252 \\
\enddata

\tablecomments{Targets tagged with an asterisks were excluded from statistical tests due to either being a binary that does not meet the criteria in Sect. 2.1, or having a spectral type earlier than K-type.  BP Tau was also excluded from statistical tests, due to its nature as an ``evolved'' full disk.  { OBSIDs tagged with a star ($\star$) were observed by the GASPS team, and were previously reported in \citet{howard}, \citet{meeus12}, and \citet{podio12}, although they were re-reduced here using an updated version of the Herschel pipeline.}  Distances for each star forming association (from Reipurth et al. 2003): Chameleon, 160 pc; Lupus, 155 pc; Ophiuchus, 120 pc; Perseus, 250 pc; Taurus, 140 pc; TW Hya, 56 pc. References: Andrews and Williams 2005 (AW05), Brandner et al. 1996 (B96), Brown et al. 2007 (BR07), Daemgen et al. 2013 (D13), Espaillat et al. 2011 (E11), Furlan et al. 2009 (F09), Guenther et al. 2007 (G07), Hughes et al. 1994 (H94), Kenyon and Hartmann 1995 (KH95), Luhman 2007 (L07), Magazzu et al. 1992 (M92), McCabe et al. 2006 (M06), Merin et al. 2010 (M10), Pascucci et al. 2008 (P08), Riaz et al. 2006 (R06), White and Ghez 2001 (WG01), Wilking et al. 2005 (WMRG05).}

\end{deluxetable}

\begin{deluxetable}{ccccccccccc}

\tabletypesize{\footnotesize}
\tablewidth{0pt}

\tablecaption{Literature Data}

\tablenum{2}
\tablecomments{References: Alcala et al. 2000 (A00), Andrews et al. 2011 (A11), Antoniucci et al. 2011 (SA11), Bary et al. 2008 (B08), Beckwith et al. 1990 (B90), Bouvier and Appenzeller, 1992 (BA92), Bouvier et al. 1993 (B93), Brice\~{n}o et al. 1999 (B99), Cabrit et al. 1990 (C90), Carkner et al. 1998 (C98), Chen et al. 1995 (CMLW95), Cohen and Kuhi 1979 (CK79), Cutri et al. 2003 (C03), Damiani et al. 1995 (D95), DENIS Consortium 2005 (DENIS), Edwards et al. 1987 (E87), Edwards et al. 1994 (E94), Espaillat et al. 2011 (E11), Feigelson and Kriss 1989 (FK89), Feigelson et al. 1993 (F93), Fernandez, 1995 (F95), Furlan et al. 2009 (F09), Gauvin and Strom 1992(GS92), Gudel et al. 2007 (G07), Guenther and Emerson, 1997 (GE97), Hartigan, Edwardds, and Ghandour 1995 (HEG95), Herbst et al. 1994 (HHG94), Herczeg et al. 2007 (H07), Hughes et al. 1994 (H94), Kastner et al. 1999 (K99), Kenyon and Hartmann 1995 (KH95), Kenyon et al. 1998 (K98), Kenyon, Gomez, and Whitney, 2008 (KGW08), Kraus \& Hillenbrand 2009 (K09), Luhman 2000 (L00), Luhman 2004 (L04), Magazzu et al. 1992 (M92), McClure et al. 2010 (MFM10), Merin et al. 2010 (M10), Monet, 2003 (M03), Muzerolle, Calvet, and Hartmann 2001(MCH01), Neuhauser et al. 1995 (N95), Osterloh and Beckwith, 1995 (OB95), Prato, Greene, Simon, 2003 (PGS03), Rebull et al. 2010 (R10), Riaz et al. 2006 (R06), Salyk et al. 2009 (SB09), Schisano et al. 2009 (S09), Spezzi et al. 2007 (S07), Taguchi et al. 2009 (T09), White and Ghez 2001 (WG01), White et al. 2000 (W00), Wilking et al. 2006 (WMRG05).}

\tablehead{ \colhead{ID} &	\colhead{R$_{mag}$}	&	\colhead{Ref} 	&	\colhead{I$_{mag}$}	&	\colhead{Ref} 	&	\colhead{A$_{V}$}	&	\colhead{Ref} 	&	\colhead{H$\alpha$ EW [\AA]}	&	\colhead{Ref} 	&		\colhead{log L$_{X}$ [L$_{\sun}$]}    &     \colhead{Ref}
}
\startdata

 \hline																						
  \multicolumn{11}{c}{\emph{Transition disks}} \\																						
 \hline

T1	&	14.20	&	C03	&	12.88	&	DENIS	&	6.90	&	MFM10	&	---	&	---	&		---	&	---	\\
T2	&	10.45	&	GS92	&	10.02	&	GS92	&	1.21	&	GS92	&	1.5	&	GS92	&		-3.45	&	FK89	\\
T3	&	10.28	&	GS92	&	7.64	&	GS92	&	3.39	&	GS92	&	1.2	&	B08	&		-3.02	&	FK89	\\
T4	&	10.46	&	GS92	&	9.73	&	GS92	&	1.50	&	E11	&	38.1	&	GE97	&		-2.94	&	FK89	\\
T5	&	10.92	&	GS92	&	9.11	&	GS92	&	0.85	&	GS92	&	13.3	&	GS92	&		-3.36	&	FK89	\\
T6	&	12.92	&	KH95	&	11.77	&	KH95	&	0.00	&	KH95	&	138.7	&	CK79	&		-4.33	&	G07	\\
T7	&	12.10	&	C03	&	11.69	&	DENIS	&	2.10	&	CMLW95	&	36	&	M92	&		---	&	---	\\
T8	&	11.70	&	BA92	&	10.80	&	BA92	&	2.20	&	A11	&	68.3	&	BA92	&		-3.69	&	A11	\\
T9	&	11.20	&	B93	&	10.70	&	B93	&	0.14	&	KH95	&	96.5   109   71	&	CK79, E94, C90	&	$\leq$	-3.89	&	A11	\\
T10	&	13.00	&	C03	&	11.95	&	S07	&	2.00	&	M10	&	0.2	&	M10	&	$\leq$	-3.02	&	A00	\\
T11	&	11.58	&	KH95	&	10.79	&	KH95	&	0.62	&	KH95	&	18.05	&	SB09	&	$\leq$	-3.99	&	A11	\\
T12	&	11.20	&	C03	&	10.80	&	F95	&	1.55	&	OB95	&	16	&	SB09	&		---	&	---	\\
T13	&	11.28	&	M10	&	10.54	&	M10	&	1.00	&	M10	&	26	&	M10	&		-3.19	&	A11	\\
T14	&	15.79	&	M10	&	13.90	&	M10	&	1.00	&	M10	&	18	&	M10	&		---	&	---	\\
T15	&	13.34	&	M08	&	12.17	&	M03	&	0.10	&	H94	&	145.2	&	H94	&		---	&	---	\\
T16	&	14.15	&	GS92	&	12.69	&	GS92	&	1.60	&	E11	&	5	&	L04	&		---	&	---	\\
T17	&	14.96	&	GS92	&	13.41	&	GS92	&	3.50	&	E11	&	100	&	L04	&		-4.79	&	W00	\\
T18	&	12.57	&	GS92	&	11.59	&	GS92	&	0.60	&	E11	&	56	&	L04	&		-3.69	&	W00	\\
T19	&	14.53	&	M10	&	12.94	&	M10	&	0.50	&	M10	&	44	&	M10	&		---	&	---	\\
T20	&	14.28	&	H94	&	12.92	&	H94	&	2.00	&	H94	&	95.9	&	H94	&		---	&	---	\\
T21	&	11.21	&	GS92	&	9.25	&	GS92	&	1.88	&	GS92	&	12	&	GS92	&		-3.99	&	F93	\\
T22	&	11.07	&	S09	&	10.28	&	DENIS	&	1.70	&	S09	&	7.8	&	S09	&		---	&	---	\\
T23	&	11.40	&	DENIS	&	9.38	&	DENIS	&	1.00	&	K99	&	213.8	&	R06	&		-3.85	&	H07	\\
T24	&	10.48	&	KH95	&	9.75	&	KH95	&	0.21	&	KH95	&	3.9	&	T09	&		-3.33	&	D95	\\
T25	&	16.55	&	WMRG05	&	14.33	&	DENIS	&	2.00	&	WMRG05	&	81	&	WMRG05	&		---	&	---	\\
T26	&	12.20	&	DENIS	&	11.29	&	DENIS	&	9.00	&	PGS03	&	4	&	SB09	&		-3.59	&	A11	\\
 \hline																						
  \multicolumn{11}{c}{\emph{Full disks}} \\																						
 \hline																						
F1	&	12.06	&	KH95	&	10.99	&	HHG94	&	0.49	&	KH95	&	37.1   80   21	&	CK79, E94, C90	&		-3.49	&	G07	\\
F2	&	11.31	&	KH95	&	10.45	&	KH95	&	0.49	&	KH95	&	40.1   55   47   49.4	&	CK79, E94, C90, MCH01	&		-3.45	&	G07	\\
F3	&	12.22	&	KH95	&	11.12	&	KH95	&	1.77	&	KH95	&	102.1   64	&	CK79, C90	&		-4.30	&	G07	\\
F4	&	12.35	&	KH95	&	11.18	&	KH95	&	0.10	&	KH95	&	69.5	&	CK79	&		-4.46	&	G07	\\
F5	&	11.66	&	KH95	&	10.75	&	HHG94	&	0.59	&	KH95	&	54   76	&	CK79, C90	&	$\leq$	-3.78	&	D95	\\
F6	&	11.43	&	KH95	&	10.46	&	KH95	&	0.76	&	KH95	&	19.4   13   28	&	CK79, E94, C90	&		-3.62	&	G07	\\
F7	&	11.85	&	KH95	&	10.89	&	KH95	&	1.70	&	HEG95	&	105   111   138	&	WG01, CK79   C90	&	$\leq$	-3.59	&	D95	\\
F8	&	11.49	&	KH95	&	10.49	&	KH95	&	0.49	&	KH95	&	11.9   22   15   11.1	&	CK79, E94, C90, MCH01	&		-3.52	&	G07	\\
F9	&	12.40	&	KH95	&	11.27	&	KH95	&	0.97	&	KH95	&	112.9	&	CK79	&		---	&	---	\\
F10	&	11.56	&	KH95	&	10.80	&	KH95	&	0.31	&	KH95	&	38.5	&	K98	&		---	&	---	\\
F11	&	11.31	&	WG01	&	10.44	&	WG01	&	1.03	&	WG01	&	56   43   52	&	WG01, CK79, C90	&	$\leq$	-3.55	&	D95	\\
F12	&	13.62	&	KH95	&	12.30	&	KH95	&	1.18	&	KH95	&	80.8	&	CK79	&		-4.19	&	G07	\\
F13	&	---	&	---	&	---	&	---	&	---	&	---	&	---	&	---	&		-3.65	&	D95	\\
F14	&	13.93	&	KH95	&	12.37	&	KH95	&	2.32	&	KH95	&	53.5	&	K98	&		---	&	---	\\
F15	&	12.28	&	KH95	&	11.11	&	KH95	&	1.25	&	KH95	&	7.8	&	CK79	&		-3.97	&	G07	\\
F16	&	8.62	&	KH95	&	8.10	&	KH95	&	0.90	&	KH95	&	3.5   5	&	CK79, C90	&		-2.61	&	G07	\\
F17	&	14.30	&	HH92	&	12.50	&	HH92	&	2.14	&	HH92	&	66   46	&	SA11, CK79	&		---	&	---	\\
F18	&	12.17	&	KH95	&	11.19	&	KH95	&	0.59	&	KH95	&	9   5	&	B90, C90	&		-3.54	&	N95	\\
 \hline																						
  \multicolumn{11}{c}{\emph{Outflow disks}} \\																						
 \hline																						
O1	&	12.33	&	KH95	&	11.42	&	KH95	&	2.29	&	KH95	&	137.9	&	R10	&		-3.13	&	G07	\\
O2	&	11.07	&	KH95	&	9.87	&	KH95	&	0.21	&	KH95	&	53.9	&	CK79	&		-3.78	&	D95	\\
O3	&	11.51	&	KH95	&	10.54	&	KH95	&	3.20	&	HEG95	&	112.8   73   110	&	CK79, E94, C90	&	$\leq$	-4.39	&	---	\\
O4	&	---	&	---	&	---	&	---	&	---	&	---	&	---	&	---	&		-2.60	&	G07	\\
O5	&	12.41	&	KH95	&	11.37	&	HHG94	&	2.64	&	KH95	&	108.9   101	&	CK79, C90	&	$\leq$	-4.27	&	B99	\\
O6	&	13.09	&	KH95	&	11.95	&	KH95	&	1.46	&	KH95	&	85.4	&	CK79	&		-4.58	&	G07	\\
O7	&	12.15	&	KH95	&	11.06	&	KH95	&	0.87	&	KH95	&	22.5   17	&	K98, CK79	&		-3.66	&	G07	\\
O8	&	14.85	&	KGW08	&	13.54	&	L00	&	11.90	&	K09	&	88.2	&	CK79	&		-3.68	&	G07	\\
O9	&	12.96	&	KH95	&	12.17	&	KH95	&	0.52	&	KH95	&	158	&	E87	&		-4.40	&	G07	\\
O10	&	12.68	&	KH95	&	9.87	&	KH95	&	1.91	&	KH95	&	8.5	&	E87	&	$\leq$	-4.46	&	N95	\\
O11	&	9.95	&	KH95	&	9.34	&	KH95	&	0.50	&	F09	&	84.2	&	CK79	&		-4.03	&	D95	\\
O12	&	9.53	&	KH95	&	8.80	&	KH95	&	1.84	&	KH95	&	21	&	B90	&		-2.87	&	G07	\\
O13	&	9.19	&	KH95	&	8.50	&	KH95	&	1.39	&	KH95	&	38	&	T09	&		-2.79	&	C98	\\
O14	&	11.92	&	KH95	&	10.83	&	KH95	&	1.35	&	KH95	&	47   72.8	&	E87, CK79	&		---	&	---	\\
O15	&	11.20	&	KH95	&	10.28	&	M03	&	1.49	&	KH95	&	73.5   98.1	&	E87, CK79	&		-3.64	&	G07	\\

																																			\enddata

\end{deluxetable}

\begin{deluxetable}{cccccccc}

\tabletypesize{\footnotesize}
\tablewidth{0pt}

\tablecaption{Stellar and Accretion Properties}

\tablenum{3}

\tablehead{ 
\colhead{ID} &    \colhead{T$_{eff}$ [K]} 	&	\colhead{M$_{\star}$ [M$_{\sun}$]}	&	\colhead{L$_{bol}$ [L$_{\sun}$]}	&	\colhead{R$_{\star}$ [R$_{\sun}$]}	&		\colhead{log L$_{acc}$ [L$_{\sun}$]}	&		\colhead{log $\dot{M}$ [M$_{\sun}$/yr]}	&		\colhead{log L$_{FUV}$ [L$_{\sun}$]} 
}
\startdata
																		
 \hline																				
 \multicolumn{8}{c}{\emph{Transition disks}} \\																				
 \hline																				
T1	&	6030	&	1.15	&	1.01	&	0.92	&		---	&		---	&		---	\\
T2	&	5520	&	1.34	&	1.98	&	1.54	&		-2.0	&		-9.3	&		-3.3	\\
T3	&	5770	&	3.7	&	47.60&	6.92 &		-1.2	&		-8.3	&		-2.7	\\
T4	&	4900	&	1.98	&	2.83	&	2.34	&		-0.1	&		-7.4	&		-1.8	\\
T5	&	4205	&	1.62	&	3.76	&	3.66	&		-1.2	&		-8.2	&		-2.6	\\
T6	&	3705	&	0.45	&	0.18	&	1.02	&		-1.3	&		-8.4	&		-2.8	\\
T7	&	4350	&	0.95	&	0.33	&	1.01	&		-1.0	&		-8.4	&		-2.5	\\
T8	&	4730	&	1.26	&	0.82	&	1.35	&		-0.5	&		-7.8	&		-2.1	\\
T9	&	4277.5&	1.00	&	0.50	&	1.28	&		-0.7	&		-7.9	&		-2.2	\\
T10	&	3777.5&	0.52	&	0.46	&	1.59	&	$\leq$	-4.0	&	$\leq$	-11.0	&	$\leq$	-5.0	\\
T11	&	4350	&	1.05	&	0.53	&	1.29	&		-1.5	&		-8.9	&		-3.0	\\
T12	&	5830	&	1.25	&	2.78	&	1.64	&		-0.4	&		-7.7	&		-2.0	\\
T13	&	4590	&	1.28	&	1.01	&	1.59	&		-0.9	&		-8.2	&		-2.5	\\
T14	&	3125	&	0.16	&	0.08	&	0.94	&	$\leq$	-3.4	&	$\leq$	-10.0	&	$\leq$	-4.5	\\
T15	&	3632.5&	0.40	&	0.16	&	1.00	&		-1.4	&		-8.4	&		-2.8	\\
T16	&	3487.5&	0.35	&	0.21	&	1.26	&	$\leq$	-3.0	&	$\leq$	-9.9	&	$\leq$	-4.2	\\
T17	&	4060	&	0.80	&	0.23	&	0.97	&		-1.1	&		-8.4	&		-2.6	\\
T18	&	3777.5&	0.52	&	0.35	&	1.38	&		-1.3	&		-8.3	&		-2.7	\\
T19	&	3060	&	0.17	&	0.17	&	1.45	&		-2.5	&		-8.9	&		-3.7	\\
T20	&	3777.5&	0.52	&	0.18	&	0.99	&		-1.4	&		-8.5	&		-2.8	\\
T21	&	5250	&	2.18	&	5.50	&	2.84	&		-1.0	&		-8.2	&		-2.5	\\
T22	&	5520	&	1.31	&	0.89	&	1.04	&		-1.6	&		-9.1	&		-3.0	\\
T23	&	3850	&	0.58	&	0.39	&	1.40	&		-1.0	&		-8.0	&		-2.5	\\
T24	&	4900	&	1.3	&	1.21	&	1.53	&		-2.0	&		-9.3	&		-3.3	\\
T25	&	3197.5&	0.17	&	0.04	&	0.64	&		-2.9	&		-9.7	&		-4.1	\\
T26	&	5845	&	2.20	&	11.06	&	3.25	&		0.3	&		-6.9	&		-1.4	\\
 \hline																				
  \multicolumn{8}{c}{\emph{Full disks}} \\																				
 \hline																				
F1	&	4060	&	0.8	&	0.43	&	1.32	&		-1.3	&		-8.5	&		-2.8	\\
F2	&	4060	&	0.78	&	0.70	&	1.70	&		-0.9	&		-8.0	&		-2.4	\\
F3	&	4060	&	0.78	&	0.67	&	1.65	&		-0.6	&		-7.7	&		-2.2	\\
F4	&	3705	&	0.46	&	0.32	&	1.37	&		-1.4	&		-8.3	&		-2.8	\\
F5	&	3560	&	0.38	&	0.59	&	2.03	&		-0.9	&		-7.6	&		-2.4	\\
F6	&	4060	&	0.76	&	0.78	&	1.79	&		-1.4	&		-8.4	&		-2.8	\\
F7	&	4060	&	0.76	&	0.80	&	1.81	&		-0.3	&		-7.3	&		-1.9	\\
F8	&	3850	&	0.56	&	0.70	&	1.88	&		-1.7	&		-8.5	&		-3.0	\\
F9	&	3850	&	0.57	&	0.42	&	1.46	&		-0.8	&		-7.8	&		-2.4	\\
F10	&	4350	&	1.2	&	0.46	&	1.20	&		-1.2	&		-8.6	&		-2.7	\\
F11	&	4060	&	0.76	&	0.90	&	1.92	&		-0.7	&		-7.7	&		-2.3	\\
F12	&	3850	&	0.57	&	0.18	&	0.95	&		-1.5	&		-8.7	&		-3.0	\\
F13	&	---	&	---	&	---	&	---	&		---	&		---	&		---	\\
F14	&	3777.5	&	0.52	&	0.28	&	1.23	&		-1.5	&		-8.5	&		-2.9	\\
F15	&	3777.5	&	0.52	&	0.55	&	1.74	&		-2.1	&		-9.0	&		-3.4	\\
F16	&	5860	&	1.8	&	7.96	&	2.74	&		-0.7	&		-8.0	&		-2.3	\\
F17	&	3415	&	0.33	&	0.34	&	1.67	&		-1.6	&		-8.3	&		-3.0	\\
F18	&	4060	&	0.72	&	0.37	&	1.23	&	$\leq$	-2.4	&	$\leq$	-9.5	&	$\leq$	-3.6	\\
 \hline																				
  \multicolumn{8}{c}{\emph{Outflow disks}} \\																				
 \hline																				
O1	&	4730	&	1.07	&	0.66	&	1.21	&		-0.2	&		-7.5	&		-1.8	\\
O2	&	3415	&	0.33	&	1.25	&	3.20	&		-0.9	&		-7.3	&		-2.4	\\
O3	&	4205	&	0.9	&	2.06	&	2.71	&		0.4	&		-6.5	&		-1.4	\\
O4	&	3850	&	---	&	---	&	---	&		---	&		---	&		---	\\
O5	&	3850	&	0.56	&	0.80	&	2.01	&		-0.2	&		-7.1	&		-1.9	\\
O6	&	3777.5	&	0.52	&	0.28	&	1.24	&		-1.1	&		-8.2	&		-2.6	\\
O7	&	4205	&	0.95	&	0.46	&	1.27	&		-1.7	&		-9.0	&		-3.1	\\
O8	&	3850	&	0.55	&	6.42	&	5.71	&		1.9	&		-4.5	&		-0.1	\\
O9	&	4350	&	0.7	&	0.14	&	0.67	&		-1.1	&		-8.5	&		-2.6	\\
O10	&	3705	&	0.45	&	2.34	&	3.72	&	$\leq$	-2.0	&	$\leq$	-8.5	&	$\leq$	-3.4	\\
O11	&	4730	&	1.49	&	2.03	&	2.13	&		0.1	&		-7.2	&		-1.6	\\
O12	&	5080	&	2.2	&	6.15	&	3.21	&		0.0	&		-7.2	&		-1.7	\\
O13	&	5250	&	2.18	&	6.77	&	3.15	&		0.3	&		-6.9	&		-1.4	\\
O14	&	4060	&	0.77	&	0.72	&	1.72	&		-0.8	&		-7.8	&		-2.3	\\
O15	&	3705	&	0.47	&	1.34	&	2.81	&		-0.2	&		-6.8	&		-1.8	\\

\enddata

\end{deluxetable}

\begin{deluxetable}{cccccccccc}

\tabletypesize{\footnotesize}
\tablewidth{0pt}

\tablecaption{Disk Properties}

\tablenum{4}
\tablecomments{References: Andrews and Williams 2005, and references therein (A05), Andrews et al. 2011 (A11), Brown et al. 2007 (BR07), Espaillat et al. 2011 (E11), Guilloteau et al. 2011 (G11), Henning et al. 1993 (H93), Kim et al. 2009 (KM09), Merin et al. 2010 (M10), Mohanty et al. 2013, and references therein (M13), Nurnberger et al. 1997 (N97), Osterloh and Beckwith, 1995 (OB95), Thi et al. 2010 (T10).}

\tablehead{ 
\colhead{ID}	&	\colhead{r$_{gap}$ [AU]}	&	\colhead{Ref} 	&	\colhead{h$_{wall}$ [AU]}	&	\colhead{Ref} 	&		\colhead{$f_{850 \mu m}$ [mJy]}	&		\colhead{$f_{1.3 mm}$ [mJy]}	&	\colhead{Ref} 	&		\colhead{M$_{850 \mu m}$ [M$_{Jup}$]}	&		\colhead{M$_{1.3 mm}$ [M$_{Jup}$]}	
}
\startdata

 \hline																							
  \multicolumn{10}{c}{\emph{Transition disks}} \\																							
 \hline																							
T1	&		&		&	---	&	---	&		---	&		---	&	---	&		---	&		---	\\
T2	&	37.1	&	KM09	&	---	&	---	&		---	&	$\leq$	118	&	H93	&		---	&	$\leq$	25.8	\\
T3	&	146.7	&	KM09	&	---	&	---	&		---	&	$\leq$	143	&	H93	&		---	&	$\leq$	32.5	\\
T4	&	10	&	E11	&	---	&	---	&		---	&		124.9	&	H93	&		---	&		28.4	\\
T5	&	38	&	E11	&	7	&	E11	&		---	&		128.4	&	H93	&		---	&		29.1	\\
T6	&	19	&	A11	&	5.7	&	A11	&		237	&		109	&	M13	&		13.6	&		19.0	\\
T7	&	---	&	---	&	---	&	---	&		---	&	$\leq$	75	&	M13	&		---	&	$\leq$	9.6	\\
T8	&	30	&	A11	&	9	&	A11	&		181	&		105	&	M13	&		7.6	&		13.4	\\
T9	&	23	&	E11	&	2.9	&	E11	&		---	&		253	&	M13	&		---	&		44.1	\\
T10	&	---	&	---	&	---	&	---	&		---	&		---	&	---	&		---	&		---	\\
T11	&	39	&	E11	&	5	&	E11	&		428	&		167	&	M13	&		24.5	&		29.1	\\
T12	&	68	&	A11	&	6.8	&	A11	&		---	&		70	&	OB95	&		---	&		38.9	\\
T13	&	30	&	A11	&	2	&	A11	&		---	&		---	&	---	&		---	&		---	\\
T14	&	---	&	---	&	---	&	---	&		---	&		---	&	---	&		---	&		---	\\
T15	&	---	&	---	&	---	&	---	&		---	&		---	&	---	&		---	&		---	\\
T16	&	13	&	E11	&	2	&	E11	&		---	&		105	&	H93	&		---	&		23.9	\\
T17	&	5	&	E11	&	4	&	E11	&		---	&		100	&	H93	&		---	&		22.8	\\
T18	&	20	&	E11	&	4	&	E11	&		---	&		47.8	&	H93	&		---	&		10.9	\\
T19	&	55	&	M10	&	---	&	---	&		---	&	$\leq$	36	&	N97	&		---	&	$\leq$	7.7	\\
T20	&	---	&	---	&	---	&	---	&		---	&	$\leq$	27	&	N97	&		---	&	$\leq$	5.8	\\
T21	&	18	&	E11	&	4	&	E11	&		---	&		77.5	&	H93	&		---	&		17.7	\\
T22	&	15	&	BR07	&	---	&	---	&		---	&		105.2	&	H93	&		---	&		11.1	\\
T23	&	4	&	T10	&	---	&	---	&		---	&		---	&	---	&		---	&		---	\\
T24	&	---	&	---	&	---	&	---	&	$\leq$	173	&	$\leq$	63	&	M13	&	$\leq$	3.6	&	$\leq$	11.0	\\
T25	&	15	&	A11	&	0.8	&	A11	&		149	&		89	&	M13	&		6.3	&		11.4	\\
T26	&	36	&	A11	&	8.2	&	A11	&		397	&		95	&	M13	&		16.7	&		12.2	\\
 \hline																							
  \multicolumn{10}{c}{\emph{Full disks}} \\																							
 \hline																							
F1	&	---	&	---	&	---	&	---	&		144	&		88	&	M13	&		8.3	&		15.3	\\
F2	&	---	&	---	&	---	&	---	&		130	&		47	&	M13	&		7.5	&		8.2	\\
F3	&	---	&	---	&	---	&	---	&		324	&		190	&	M13	&		18.6	&		33.1	\\
F4	&	---	&	---	&	---	&	---	&		---	&		111	&	G11	&			&		19.3	\\
F5	&	---	&	---	&	---	&	---	&		90	&		36	&	M13	&		5.2	&		6.3	\\
F6	&	---	&	---	&	---	&	---	&		80	&		35	&	M13	&		4.6	&		6.1	\\
F7	&	---	&	---	&	---	&	---	&		440	&		230	&	M13	&		25.2	&		40.1	\\
F8	&	---	&	---	&	---	&	---	&		201	&		84	&	M13	&		11.5	&		14.6	\\
F9	&	---	&	---	&	---	&	---	&		208	&		91	&	M13	&		11.9	&		15.9	\\
F10	&	---	&	---	&	---	&	---	&		---	&		---	&	---	&		---	&		---	\\
F11	&	---	&	---	&	---	&	---	&		1255	&		593	&	M13	&		72.0	&		103.3	\\
F12	&	---	&	---	&	---	&	---	&		173	&		83	&	M13	&		9.9	&		14.5	\\
F13	&	---	&	---	&	---	&	---	&		---	&		---	&	---	&		---	&		---	\\
F14	&	---	&	---	&	---	&	---	&		---	&		---	&	---	&		---	&		---	\\
F15	&	---	&	---	&	---	&	---	&		178	&		87	&	M13	&		5.0	&		15.2	\\
F16	&	---	&	---	&	---	&	---	&		74	&	$\leq$	30	&	M13	&		4.2	&	$\leq$	5.2	\\
F17	&	---	&	---	&	---	&	---	&		---	&		---	&	---	&		---	&		---	\\
F18	&	---	&	---	&	---	&	---	&		74	&		37	&	M13	&		4.2	&		6.4	\\
 \hline																							
  \multicolumn{10}{c}{\emph{Outflow disks}} \\																							
 \hline																							
O1	&	---	&	---	&	---	&	---	&		66	&		96	&	M13	&		3.8	&		16.7	\\
O2	&	---	&	---	&	---	&	---	&		8.8	&	$\leq$	25	&	M13	&		0.5	&	$\leq$	4.4	\\
O3	&	---	&	---	&	---	&	---	&		---	&		389.9	&	G11	&		---	&		67.9	\\
O4	&	---	&	---	&	---	&	---	&		---	&		---	&	---	&		---	&		---	\\
O5	&	---	&	---	&	---	&	---	&		248	&		136	&	M13	&		14.2	&		23.7	\\
O6	&	---	&	---	&	---	&	---	&	$\leq$	10	&	$\leq$	27	&	M13	&	$\leq$	0.6	&	$\leq$	4.7	\\
O7	&	---	&	---	&	---	&	---	&		33	&	$\leq$	21	&	M13	&		1.9	&	$\leq$	3.7	\\
O8	&	---	&	---	&	---	&	---	&		---	&		34.2	&	G11	&		---	&		5.9	\\
O9	&	---	&	---	&	---	&	---	&		29	&	$\leq$	15	&	M13	&		1.7	&	$\leq$	2.6	\\
O10	&	---	&	---	&	---	&	---	&		47	&		40	&	A05	&		2.7	&		7.0	\\
O11	&	---	&	---	&	---	&	---	&		79	&		42	&	M13	&		4.5	&		7.3	\\
O12	&	---	&	---	&	---	&	---	&		560	&		229	&	M13	&		32.1	&		39.9	\\
O13	&	---	&	---	&	---	&	---	&		628	&		280	&	M13	&		36.0	&		48.8	\\
O14	&	---	&	---	&	---	&	---	&		102	&		29	&	M13	&		5.9	&		5.1	\\
O15	&	---	&	---	&	---	&	---	&		560	&		172	&	M13	&		32.1	&		30.0	\\

\enddata

\end{deluxetable}

\begin{deluxetable}{cccccccccc}
\tabletypesize{\footnotesize}
\tablewidth{0pt}
\tablecaption{\emph{Herschel} PACS Results}
\tablenum{5}

\tablehead{\colhead{ID} & \colhead{[O{\sc i}] 63.18 $\mu$m line flux} & \colhead{o-H$_{2}$O 63.32 $\mu$m line flux} & \colhead{63 $\mu$m continuum flux}\\
		  \colhead{} & \colhead{(10$^{-17}$ W/m$^{2}$)} &\colhead{(10$^{-17}$ W/m$^{2}$)}  & \colhead{(Jy)} } 

\startdata
 \hline
  \multicolumn{4}{c}{\emph{Transition disks}} \\
 \hline
T1 & 2.340 $\pm$ 0.143 & $\leq $0.716 & 3.029 $\pm$ 0.010 \\
T2 & 4.492 $\pm$ 0.200 & $\leq$ 1.192 & 0.596 $\pm$ 0.020 \\
T3 & 11.284 $\pm$ 0.851 & $\leq$ 5.325 & 2.262 $\pm$ 0.055 \\
T4 & 1.530 $\pm$ 0.150 & $\leq$ 0.658 & 1.618 $\pm$ 0.015 \\
T5 & 2.200 $\pm$ 0.134 & $\leq$ 0.702 & 4.017 $\pm$ 0.010 \\
T6 & 1.276 $\pm$ 0.305 & $\leq$ 0.960 & 0.972 $\pm$ 0.016 \\
T7 & 0.949 $\pm$ 0.076 & $\leq$ 0.541 & 0.829 $\pm$ 0.013 \\
T8 & 2.536 $\pm$ 0.189 & 1.014 $\pm$ 0.145 & 5.315 $\pm$ 0.012 \\
T9 & 3.793 $\pm$ 0.513 & $\leq$ 2.435 & 2.810 $\pm$ 0.036 \\
T10 & 0.764 $\pm$ 0.101 & $\leq$ 0.523 & 0.195 $\pm$ 0.008 \\
T11 & 1.306 $\pm$ 0.165 & $\leq$ 0.770 & 1.285 $\pm$ 0.012 \\
T12 & $\leq$ 1.276 & $\leq$ 1.276 & 13.239 $\pm$ 0.019 \\
T13 & 1.934 $\pm$ 0.118 & $\leq$ 0.534 & 1.317 $\pm$ 0.009 \\
T14 & $\leq$ 1.078 & $\leq$ 1.078 & 0.303 $\pm$ 0.016 \\
T15 & 0.910 $\pm$ 0.112 & $\leq$ 0.540 & 1.406 $\pm$ 0.008 \\
T16 & 0.823 $\pm$ 0.140 & $\leq$ 0.817 & 0.652 $\pm$ 0.012 \\
T17 & 1.390 $\pm$ 0.124 & $\leq$ 0.755 & 0.556 $\pm$ 0.011 \\
T18 & 0.354 $\pm$ 0.097 & $\leq$ 0.637 & 0.810 $\pm$ 0.008 \\
T19 & $\leq$ 0.770 & $\leq$ 0.770 & 0.490 $\pm$ 0.013 \\
T20 & 1.035 $\pm$ 0.112 & $\leq$ 0.730 & 0.779 $\pm$ 0.016 \\
T21 & 1.777 $\pm$ 0.100 & $\leq$ 0.793 & 4.199 $\pm$ 0.008 \\
T22 & 5.455 $\pm$ 0.291 & $\leq$ 1.493 & 7.318 $\pm$ 0.021 \\
T23 & 4.239 $\pm$ 0.345 & $\leq$ 1.556 & 3.675 $\pm$ 0.024 \\
T24 & 3.894 $\pm$ 0.266 & $\leq$ 1.671 & 4.021 $\pm$ 0.024 \\
T25 & $\leq$ 0.590 & $\leq$ 0.590 & 0.690 $\pm$ 0.009 \\
T26 & 1.543 $\pm$ 0.296 & $\leq$ 2.534 & 39.540 $\pm$ 0.023 \\
 \hline
  \multicolumn{4}{c}{\emph{Full disks}} \\
 \hline
F1 & 2.606 $\pm$ 0.109 & 0.956 $\pm$ 0.131 & 1.106 $\pm$ 0.016 \\
F2 & 0.647 $\pm$ 0.146 & 0.898 $\pm$ 0.173 & 0.490 $\pm$ 0.024 \\
F3 & 2.016 $\pm$ 0.183 & $\leq$ 0.921 & 0.979 $\pm$ 0.014 \\
F4 & 1.489 $\pm$ 0.416 & $\leq$ 1.573 & 0.126 $\pm$ 0.027 \\
F5 & 0.712 $\pm$ 0.384 & $\leq$ 2.150 & 1.946 $\pm$ 0.039 \\
F6 & 1.748 $\pm$ 0.155 & 0.455 $\pm$ 0.157 & 0.932 $\pm$ 0.012 \\
F7 & 2.792 $\pm$ 0.208 & 0.640 $\pm$ 0.176 & 1.268 $\pm$ 0.012 \\
F8 & $\leq$ 1.023 & $\leq$ 1.023 & 0.780 $\pm$ 0.015 \\
F9 & 2.505 $\pm$ 0.351 & $\leq$ 1.413 & 1.292 $\pm$ 0.021 \\
F10 & $\leq$ 0.782 & $\leq$ 0.782 & $\leq$ 0.037 \\
F11 & 6.185 $\pm$ 0.381 & $\leq$ 1.850 & 3.756 $\pm$ 0.028 \\
F12 & $\leq$ 0.969 & $\leq$ 0.969 & 0.343 $\pm$ 0.019 \\
F13 & $\leq$ 1.585 & $\leq$ 1.585 & 0.086 $\pm$ 0.024 \\
F14 & 3.847 $\pm$ 0.262 & $\leq$ 1.234 & 2.428 $\pm$ 0.023 \\
F15 & 1.512 $\pm$ 0.256 & 0.968 $\pm$ 0.294 & 0.744 $\pm$ 0.019 \\
F16 & 12.650 $\pm$ 0.336 & $\leq$ 1.474 & 9.043 $\pm$ 0.027 \\
F17 & $\leq$ 0.866 & $\leq$ 0.866 & 0.719 $\pm$ 0.014 \\
F18 & $\leq$ 1.259 & $\leq$ 1.259 & 0.370 $\pm$ 0.015 \\
 \hline
  \multicolumn{4}{c}{\emph{Outflow disks}} \\
 \hline
O1 & 9.061 $\pm$ 0.364 & $\leq$ 1.601 & 1.707 $\pm$ 0.024 \\
O2 & 4.541 $\pm$ 0.333 & $\leq$ 1.410 & 0.369 $\pm$ 0.035 \\
O3 & 187.160 $\pm$ 3.075 & $\leq$ 10.125 & 18.015 $\pm$ 0.224 \\
O4 & 80.446 $\pm$ 1.891 & $\leq$ 6.355 & 13.600 $\pm$ 0.113 \\
O5 & 39.268 $\pm$ 1.971 & $\leq$ 7.357 & 3.932 $\pm$ 0.129 \\
O6 & 4.069 $\pm$ 1.085 & $\leq$ 3.880 & 0.411 $\pm$ 0.060 \\
O7 & 2.592 $\pm$ 0.803 & $\leq$ 4.039 & 1.286 $\pm$ 0.057 \\
O8 & 8.075 $\pm$ 0.397 & $\leq$ 2.109 & 6.240 $\pm$ 0.030 \\
O9 & 5.272 $\pm$ 0.221 & $\leq$ 1.130 & 1.020 $\pm$ 0.017 \\
O10 & 10.000 $\pm$ 0.548 & $\leq$ 2.385 & 1.376 $\pm$ 0.034 \\
O11 & 17.452 $\pm$ 0.701 & 0.947 $\pm$ 0.320 & 2.109 $\pm$ 0.043 \\
O12 & 13.020 $\pm$ 0.640 & 2.481 $\pm$ 0.740 & 14.103 $\pm$ 0.049 \\
O13 & 1348.400 $\pm$ 14.564 & 46.876 $\pm$ 1.506 & 161.870 $\pm$ 0.966 \\
O14 & 38.197 $\pm$ 0.828 & 1.660 $\pm$ 0.389 & 6.650 $\pm$ 0.058 \\
O15 & 2.573 $\pm$ 0.554 & $\leq$ 2.789 & 1.062 $\pm$ 0.046 \\
\enddata

\tablecomments{Detections are listed with $\pm$ 1-sigma uncertainties; 3-sigma upper limits are reported for non-detections.}

\end{deluxetable}

\begin{deluxetable}{lrcccccc}
\tabletypesize{\footnotesize}
\tablewidth{0pt}
\tablecaption{Subsample Correlation Tests}
\tablenum{6}
\tablecomments{P is the probability that the correlation between the two listed parameters is obtained by chance; low P values indicate a correlation.  The different statistical tests used are: (1) Cox Hazard; (2) Kendall Tau; (3) Spearman Rho.  If the average of the three statistical tests is less than 5\%, they are listed as ``correlated,'' in boldface.  A linear regression (using the EM method) was performed for all combinations, fitting the log of the quantities listed, where the first parameter listed for each pair is the independent variable.  If no correlation is detected, the linear regression may not be significant.}																

\tablehead{
\colhead{Correlation} &  \colhead{Subsample} & \multicolumn{3}{c}{Correlation Tests} & \colhead{Correlated?} & \multicolumn{2}{c}{Linear Regression} \\ 
\colhead{Test} & \colhead{Being Tested} &  \colhead{P(1)} & \colhead{P(2)} & \colhead{P(3)} & \colhead{} & \colhead{Intercept} & \colhead{Slope} } 

\startdata
L 63 $\mu$m v. L [OI] 63 $\mu$m	&	All Objects	&	0.0\%	&	0.0\%	&	0.0\%	&	\textbf{Correlated}	&	-3.24$\pm$	0.18	&	1.15	$\pm$	0.13	\\
	&	Transitional Disks Only	&	0.0\%	&	0.0\%	&	0.1\%	&	\textbf{Correlated}	&	-4.07	$\pm$	0.21	&	0.74	$\pm$	0.15	\\
	&	Full Disks Only	&	10.0\%	&	0.7\%	&	3.3\%	&	\textbf{Correlated}	&	-4.42	$\pm$	0.31	&	0.38	$\pm$	0.18	\\
	&	Outflow Disks Only	&	0.2\%	&	0.2\%	&	0.5\%	&	\textbf{Correlated}	&	-3.01	$\pm$	0.17	&	0.97	$\pm$	0.15	\\
 \hline																			
T$_{eff}$ v. L [OI] 63 $\mu$m	&	All Objects	&	0.1\%	&	0.3\%	&	0.4\%	&	\textbf{Correlated}	&	-7.66	$\pm$	0.93	&	0.0007	$\pm$	0.0002	\\
	&	Transitional Disks Only	&	0.0\%	&	0.1\%	&	0.2\%	&	\textbf{Correlated}	&	-6.67	$\pm$	0.39	&	0.0004	$\pm$	0.0001	\\
	&	Full Disks Only	&	82.7\%	&	84.7\%	&	87.3\%	&	Not Correlated	&	-4.68	$\pm$	1.15	&	-0.0001	$\pm$	0.0003	\\
	&	Outflow Disks Only	&	24.7\%	&	29.7\%	&	33.0\%	&	Not Correlated	&	-6.22	$\pm$	1.70	&	0.0005	$\pm$	0.0004	\\
 \hline																			
T$_{eff}$ v. L 63 $\mu$m	&	All Objects	&	10.9\%	&	0.4\%	&	1\%	&	Not Correlated	&	-3.09	$\pm$	0.73	&	0.0004	$\pm$	0.0002	\\
	&	Transitional Disks Only	&	0.0\%	&	0.1\%	&	0.2\%	&	\textbf{Correlated}	&	-3.17	$\pm$	0.30	&	0.0004	$\pm$	0.0001	\\
	&	Full Disks Only	&	10.4\%	&	28.4\%	&	18.0\%	&	Not Correlated	&	1.81	$\pm$	1.63	&	-0.0009	$\pm$	0.0004	\\
	&	Outflow Disks Only	&	11.8\%	&	14.4\%	&	15.6\%	&	Not Correlated	&	-3.79	$\pm$	1.45	&	0.0007	$\pm$	0.0003	\\
 \hline																			
L$_{bol}$ v. L [OI] 63 $\mu$m	&	All Objects	&	0.0\%	&	0.0\%	&	0.0\%	&	\textbf{Correlated}	&	-4.61	$\pm$	0.10	&	1.02	$\pm$	0.19\\
	&	Transitional Disks Only	&	0.1\%	&	0.3\%	&	0.3\%	&	\textbf{Correlated}	&	-4.96	$\pm$	0.07	&	0.41	$\pm$	0.11	\\
	&	Full Disks Only	&	28.3\%	&	45.7\%	&	38.3\%	&	Not Correlated	&	-4.94	$\pm$	0.19	&	0.39	$\pm$	0.50	\\
	&	Outflow Disks Only	&	3.0\%	&	7.3\%	&	4.7\%	&	\textbf{Correlated}	&	-4.04	$\pm$	0.20	&	0.75	$\pm$	0.35	\\
 \hline																			
L$_{bol}$ v. L 63 $\mu$m	&	All Objects	&	0.0\%	&	0.0\%	&	0.0\%	&	\textbf{Correlated}	&	-1.23$\pm$	0.09	&	0.78	$\pm$	0.15	\\
	&	Transitional Disks Only	&	0.0\%	&	0.1\%	&	0.2\%	&	\textbf{Correlated}	&	-1.23	$\pm$	0.06	&	0.52	$\pm$	0.09	\\
	&	Full Disks Only	&	29.0\%	&	32.9\%	&	36.9\%	&	Not Correlated	&	-1.60	$\pm$	0.39	&	0.64	$\pm$	0.99	\\
	&	Outflow Disks Only	&	0.2\%	&	0.4\%	&	0.8\%	&	\textbf{Correlated}	&	-1.08$\pm$	0.14	&	0.97	$\pm$	0.25	\\
 \hline																			
L$_{X}$ v. L [OI] 63 $\mu$m	&	All Objects	&	5.9\%	&	59.0\%	&	80.7\%	&	Not Correlated	&	-5.59	$\pm$	0.50	&	-0.18	$\pm$	0.13	\\
	&	Transitional Disks Only	&	6.4\%	&	56.9\%	&	53.7\%	&	Not Correlated	&	-5.62	$\pm$	0.29	&	-0.14	$\pm$	0.08	\\
	&	Full Disks Only	&	65.5\%	&	92.0\%	&	---	&	Not Correlated	&	-5.35	$\pm$	0.79	&	-0.06	$\pm$	0.20	\\
	&	Outflow Disks Only	&	9.1\%	&	39.5\%	&	54.8\%	&	Not Correlated	&	-2.69	$\pm$	0.64	&	0.40	$\pm$	0.18	\\
 \hline																			
L$_{X}$ v. L 63 $\mu$m	&	All Objects	&	3.0\%	&	69.6\%	&	13.9\%	&	Not Correlated	&	-2.21	$\pm$	0.36	&	-0.18	$\pm$	0.10	\\
	&	Transitional Disks Only	&	0.3\%	&	37.1\%	&	27.6\%	&	Not Correlated	&	-2.32	$\pm$	0.34	&	-0.24	$\pm$	0.09	\\
	&	Full Disks Only	&	86.4\%	&	73.2\%	&	80.9\%	&	Not Correlated	&	-2.03	$\pm$	0.58	&	-0.03	$\pm$	0.15	\\
	&	Outflow Disks Only	&	0.8\%	&	8.9\%	&	18.5\%	&	Not Correlated	&	0.59	$\pm$	0.53	&	0.47	$\pm$	0.15	\\
 \hline																			
L$_{acc}$ v. L [OI] 63 $\mu$m	&	All Objects	&	0.0\%	&	0.0\%	&	0.0\%	&	\textbf{Correlated}	&	-4.22$\pm$	0.15	&	0.66	$\pm$	0.12	\\
	&	Transitional Disks Only	&	1.0\%	&	4.7\%	&	4.9\%	&	\textbf{Correlated}	&	-4.76	$\pm$	0.15	&	0.28	$\pm$	0.11	\\
	&	Full Disks Only	&	3.5\%	&	17.6\%	&	16.1\%	&	Not Correlated	&	-4.69	$\pm$	0.27	&	0.34	$\pm$	0.22	\\
	&	Outflow Disks Only	&	5.7\%	&	5.2\%	&	8.6\%	&	Not Correlated &	-3.88	$\pm$	0.23	&	0.41	$\pm$	0.22	\\
 \hline																			
L$_{acc}$ v. L 63 $\mu$m	&	All Objects	&	0.0\%	&	0.0\%	&	0.0\%	&	\textbf{Correlated}	&	-0.97$\pm$	0.14	&	0.47	$\pm$	0.10	\\
	&	Transitional Disks Only	&	0.3\%	&	4.1\%	&	2.8\%	&	\textbf{Correlated}	&	-1.06	$\pm$	0.14	&	0.29	$\pm$	0.09	\\
	&	Full Disks Only	&	3.6\%	&	19.9\%	&	18.2\%	&	Not correlated	&	-1.21	$\pm$	0.61	&	0.54	$\pm$	0.44	\\
	&	Outflow Disks Only	&	0.7\%	&	1.6\%	&	1.2\%	&	\textbf{Correlated}	&	-0.85	$\pm$	0.19	&	0.48	$\pm$	0.18	\\
 \hline																			
$\dot{M}$ v. L [OI] 63 $\mu$m	&	All Objects	&	0.0\%	&	0.1\%	&	0.1\%	&	\textbf{Correlated}	&	-0.44 	$\pm$	0.94	&	0.55	$\pm$	0.12	\\
	&	Transitional Disks Only	&	4.0\%	&	13.5\%	&	12.1\%	&	Not Correlated	&	-3.03	$\pm$	1.15	&	0.25	$\pm$	0.14	\\
	&	Full Disks Only	&	7.0\%	&	36.8\%	&	25.2\%	&	Not Correlated	&	-2.68	$\pm$	1.66	&	0.30	$\pm$	0.20	\\
	&	Outflow Disks Only	&	10.8\%	&	1.5\%	&	4.5\%	&	\textbf{Correlated}	&	-1.91	$\pm$	1.44	&	0.28	$\pm$	0.19	\\
 \hline																			
$\dot{M}$ v. L 63 $\mu$m	&	All Objects	&	0.0\%	&	0.0\%	&	0.0\%	&	\textbf{Correlated}	&	2.12	$\pm$	0.81	&	0.44	$\pm$	0.10	\\
	&	Transitional Disks Only	&	0.6\%	&	1.2\%	&	2.5\%	&	\textbf{Correlated}	&	1.24	$\pm$	1.01	&	0.32	$\pm$	0.12	\\
	&	Full Disks Only	&	0.7\%	&	4.4\%	&	4.3\%	&	\textbf{Correlated}	&	4.50	$\pm$	3.29	&	0.77	$\pm$	0.40	\\
	&	Outflow Disks Only	&	1.2\%	&	1.0\%	&	1.0\%	&	\textbf{Correlated}	&	1.56	$\pm$	1.21	&	0.35	$\pm$	0.16	\\
 \hline																			
m$_{disk}$ v. L [OI] 63 $\mu$m	&	All Objects	&	0.0\%	&	26.5\%	&	43.6\%	&	Not Correlated	&	-5.79	$\pm$	0.25	&	0.04	$\pm$	0.01	\\
	&	Transitional Disks Only	&	0.3\%	&	26.6\%	&	20.0\%	&	Not Correlated	&	-4.72	$\pm$	0.18	&	0.03	$\pm$	0.01	\\
	&	Full Disks Only	&	1.9\%	&	44.7\%	&	8.2\%	&	Not Correlated	&	-5.25	$\pm$	0.13	&	0.01	$\pm$	0.01	\\
	&	Outflow Disks Only	&	0.0\%	&	4.4\%	&	2.3\%	&	\textbf{Correlated}	&	-4.78	$\pm$	0.26	&	0.03	$\pm$	0.01	\\
 \hline																			
m$_{disk}$ v. L 63 $\mu$m	&	All Objects	&	0.0\%	&	5.7\%	&	7.9\%	&	\textbf{Correlated}	&	-2.14	$\pm$	0.16	&	0.03	$\pm$	0.01	\\
	&	Transitional Disks Only	&	0.6\%	&	7.7\%	&	5.6\%	&	\textbf{Correlated}	&	-2.13	$\pm$	0.25	&	0.03	$\pm$	0.01	\\
	&	Full Disks Only	&	6.6\%	&	95.1\%	&	86.7\%	&	Not Correlated	&	-2.05	$\pm$	0.20	&	0.02	$\pm$	0.01	\\
	&	Outflow Disks Only	&	0.0\%	&	4.4\%	&	13.0\%	&	\textbf{Correlated}	&	-1.71	$\pm$	0.24	&	0.03	$\pm$	0.01	\\
 \hline																			
a$_{cavity}$ v. L [OI] 63 $\mu$m	&	Transitional Disks Only	&	71.1\%	&	77.0\%	&	69.5\%	&	Not Correlated	&	-5.10	$\pm$	0.17	&	-0.0012	$\pm$	0.01	\\
 \hline																			
a$_{cavity}$ v. L 63 $\mu$m	&	Transitional Disks Only	&	75.0\%	&	24.6\%	&	26.5\%	&	Not Correlated	&	-1.46	$\pm$	0.16	&	0.00	$\pm$	0.01	\\
 \hline																			
h$_{wall}$ v. L [OI] 63 $\mu$m	&	Transitional Disks Only	&	23.6\%	&	78.5\%	&	64.7\%	&	Not Correlated	&	-5.25	$\pm$	0.20	&	0.04	$\pm$	0.04	\\
 \hline																			
h$_{wall}$ v. L 63 $\mu$m	&	Transitional Disks Only	&	31.0\%	&	23.6\%	&	24.5\%	&	Not Correlated	&	-1.65	$\pm$	0.18	&	0.07	$\pm$	0.04	
\enddata

\end{deluxetable}

\begin{deluxetable}{lrcccccc}

\tabletypesize{\footnotesize}

\tablewidth{0pt}
\tablenum{7}
\tablecaption{Subsample Statistical Difference Tests}
\tablecomments{P is the probability that the parameter being compared between two subsamples is drawn from the same parent distribution; low P values indicate that two subsamples are different.  The different statistical tests are: (1) Gehan generalized Wilcoxon test (with permuation variance); (2) Gehan generalized Wilcoxon test (with hypergeometric variance); (3) logrank test; (4) Peto \& Peto generalized Wilcoxon test; (5) Peto \& Prentice generalized Wilcoxon test.  If the average of the five statistical tests is less than 5\%, they are listed as ``different,'' in boldface.}

\tablehead{\colhead{Parameter} 	& \colhead{Subsamples} 		& \multicolumn{5}{c}{Statistical Difference Tests}  							 &\colhead{Different?}	\\
                  \colhead{ } 		& \colhead{Being Compared} 	& \colhead{P(1)} & \colhead{P(2)} & \colhead{P(3)} & \colhead{P(4)} & \colhead{P(5)} &\colhead{ } 	} 

\startdata

L [OI] 63 $\mu$m	&	Transitional vs. Full Disks	&	98.4\%	&	98.4\%	&	78.7\%	&	93.8\%	&	94.1\%	&	Not Different	\\
	&	Transitional vs. Outflow Disks	&	0.0\%	&	0.0\%	&	0.0\%	&	0.0\%	&	0.0\%	&	\textbf{Different}	\\
	&	Full vs. Outflow Disks	&	0.0\%	&	0.0\%	&	0.0\%	&	0.0\%	&	0.0\%	&	\textbf{Different}	\\
\hline															
L 63 $\mu$m	&	Transitional vs. Full Disks	&	9.8\%	&	10.2\%	&	4.8\%	&	9.8\%	&	10.0\%	&	Not Different	\\
	&	Transitional vs. Outflow Disks	&	2.1\%	&	1.4\%	&	5.9\%	&	5.9\%	& ---		&	\textbf{Different}	\\
	&	Full vs. Outflow Disks	&	0.1\%	&	0.0\%	&	0.0\%	&	0.1\%	&	0.0\%	&	\textbf{Different}	\\
\hline															
L [OI] 63 $\mu$m /  	&	Transitional vs. Full Disks	&	2.2\%	&	1.0\%	&	3.0\%	&	3.1\%	&	1.5\%	&	\textbf{Different}	\\
L 63 $\mu$m	&	Transitional vs. Outflow Disks	&	0.0\%	&	0.0\%	&	0.0\%	&	0.0\%	&	0.0\%	&	\textbf{Different}	\\
	&	Full vs. Outflow Disks	&	1.8\%	&	1.7\%	&	4.0\%	&	2.3\%	&	2.0\%	&	\textbf{Different}	\\
\hline														
T$_{eff}$	&	Transitional vs. Full Disks	&	78.9\%	&	79.1\%	&	50.1\%	&	50.1\%	&	---	&	Not Different	\\
	&	Transitional vs. Outflow Disks	&	30.8\%	&	31.8\%	&	45.2\%	&	45.2\%	&	---	&	Not Different	\\
	&	Full vs. Outflow Disks	&	11.4\%	&	11.5\%	&	10.3\%	&	10.3\%	&	---	&	Not Different	\\
\hline															
L$_{bol}$	&	Transitional vs. Full Disks	&	33.7\%	&	34.3\%	&	76.4\%	&	76.4\%	&	---	&	Not Different	\\
	&	Transitional vs. Outflow Disks	&	2.5\%	&	2.8\%	&	1.7\%	&	1.7\%	&	---	&	\textbf{Different}	\\
	&	Full vs. Outflow Disks	&	3.7\%	&	3.4\%	&	0.5\%	&	0.5\%	&	---	&	\textbf{Different}	\\
\hline															
L$_{X}$	&	Transitional vs. Full Disks	&	94.3\%	&	94.4\%	&	68.0\%	&	99.0\%	&	1.9\%	&	Not Different	\\
	&	Transitional vs. Outflow Disks	&	74.6\%	&	75.1\%	&	97.1\%	&	73.5\%	&	74.9\%	&	Not Different	\\
	&	Full vs. Outflow Disks	&	94.8\%	&	94.8\%	&	48.9\%	&	94.7\%	&	96.3\%	&	Not Different	\\
\hline															
L$_{acc}$	&	Transitional vs. Full Disks	&	68.8\%	&	69.1\%	&	96.2\%	&	68.9\%	&	69.1\%	&	Not Different	\\
	&	Transitional vs. Outflow Disks	&	1.9\%	&	1.0\%	&	5.4\%	&	1.9\%	&	1.3\%	&	\textbf{Different}	\\
	&	Full vs. Outflow Disks	&	1.4\%	&	0.9\%	&	8.5\%	&	1.4\%	&	0.9\%	&	\textbf{Different}	\\
\hline															
$\dot{M}$	&	Transitional vs. Full Disks	&	100.0\%	&	100.0\%	&	90.0\%	&	99.7\%	&	99.7\%	&	Not Different	\\
	&	Transitional vs. Outflow Disks	&	1.7\%	&	0.8\%	&	4.1\%	&	1.6\%	&	1.0\%	&	\textbf{Different}	\\
	&	Full vs. Outflow Disks	&	1.9\%	&	1.3\%	&	5.2\%	&	2.0\%	&	1.5\%	&	\textbf{Different}	\\
\hline															
m$_{disk}$	&	Transitional vs. Full Disks	&	59.6\%	&	59.8\%	&	42.6\%	&	53.6\%	&	54.5\%	&	Not Different	\\
	&	Transitional vs. Outflow Disks	&	50.7\%	&	50.5\%	&	50.7\%	&	48.3\%	&	48.7\%	&	\textbf{Different}	\\
	&	Full vs. Outflow Disks	&	75.4\%	&	75.4\%	&	99.8\%	&	75.4\%	&	75.3\%	&	\textbf{Different}	\\

\enddata

\end{deluxetable}

\begin{deluxetable}{llc}

\tabletypesize{\footnotesize}

\tablewidth{0pt}
\tablenum{8}
\tablecaption{Mean Parameter Values}
\tablecomments{The Kaplan-Meier estimator provides an estimate of the mean and standard deviation of the quantity measured, while taking data censoring into account.  See LaValley, Isobe, \& Feigelson, 1990 for more.}

\tablehead{\colhead{Parameter} & \colhead{Subsample} & \colhead{Kaplan-Meier Estimator}  \\
\colhead{ } & \colhead{ } &\colhead{Mean $\pm$ Standard Deviation} } 

\startdata

L [OI] 63 $\mu$m	&	Transitional Disks	&	-5.12	$\pm$	0.07	   $\,$ log  L$_{\Sun}$\\
	&	Full Disks	&	-5.09	$\pm$	0.07	  $\,$ log  L$_{\Sun}$\\
	&	Outflow Disks	&	-3.89	$\pm$	0.22	  $\,$ log  L$_{\Sun}$\\
\hline							
L 63 $\mu$m	&	Transitional Disks	&	-1.45	$\pm$	0.08	  $\,$ log  L$_{\Sun}$\\
	&	Full Disks	&	-1.86	$\pm$	0.17  $\,$ log  L$_{\Sun}$	\\
	&	Outflow Disks	&	-0.91	$\pm$	0.20	  $\,$ log  L$_{\Sun}$\\
\hline								
log( L [OI] 63 $\mu$m /  	&	Transitional Disks	&	-3.71	$\pm$	0.05  	\\
L 63 $\mu$m )	&	Full Disks	&	-3.39	$\pm$	0.08  	\\
	&	Outflow Disks	&	-2.99	$\pm$	0.10	 \\
\hline							
T$_{eff}$	&	Transitional Disks	&	4066	$\pm$	650	$\,$ K\\
	&	Full Disks	&	3976	$\pm$	399	$\,$ K\\
	&	Outflow Disks	&	4299	$\pm$	534	 $\,$ K\\
\hline							
L$_{bol}$	&	Transitional Disks	&	-0.43	$\pm$	0.12	  $\,$ log  L$_{\Sun}$\\
	&	Full Disks	&	-0.34	$\pm$	0.05	  $\,$ log  L$_{\Sun}$\\
	&	Outflow Disks	&	0.13	$\pm$	0.17  $\,$ log  L$_{\Sun}$	\\
\hline							
L$_{X}$	&	Transitional Disks	&	-3.89	$\pm$	0.18	  $\,$ log  L$_{\Sun}$\\
	&	Full Disks	&	-3.91	$\pm$	0.11	  $\,$ log  L$_{\Sun}$\\
	&	Outflow Disks	&	-3.78	$\pm$	0.21  $\,$ log  L$_{\Sun}$	\\
\hline							
L$_{acc}$	&	Transitional Disks	&	-1.51	$\pm$	0.22	  $\,$ log  L$_{\Sun}$\\
	&	Full Disks	&	-1.39	$\pm$	0.16	  $\,$ log  L$_{\Sun}$\\
	&	Outflow Disks	&	-0.34	$\pm$	0.32	  $\,$ log  L$_{\Sun}$\\
\hline								
$\dot{M}$	&	Transitional Disks	&	-8.60	$\pm$	0.17	  $\,$ log  M$_{\Sun}$/yr\\
	&	Full Disks	&	-8.42	$\pm$	0.18  $\,$ log  M$_{\Sun}$/yr	\\
	&	Outflow Disks	&	-7.39	$\pm$	0.36  $\,$ log  M$_{\Sun}$/yr	\\
\hline							
m$_{disk}$	&	Transitional Disks	&	10.38	$\pm$	3.09	  $\,$ M$_{Jupiter}$ \\
	&	Full Disks	&	12.21	$\pm$	3.19  $\,$ M$_{Jupiter}$	\\
	&	Outflow Disks	&	18.10	$\pm$	6.25	  $\,$ M$_{Jupiter}$\\
\enddata

\end{deluxetable}

\end{document}